\documentclass[subeqn,11pt,a4paper]{article}

\usepackage[margin=35mm]{geometry}

\usepackage{amssymb,amsmath,amsfonts,amsthm, amscd, mathrsfs, helvet, mathtools}
\usepackage{bm, stmaryrd}
\usepackage{graphicx, verbatim}
\usepackage[mathcal]{euscript}
\usepackage{color}
\usepackage[all]{xy}
\usepackage{relsize}

\usepackage{tikz}
\usetikzlibrary{cd}
\usetikzlibrary{matrix,calc}
\usetikzlibrary{decorations.markings,shapes.geometric,shapes.misc}
\usetikzlibrary{decorations.pathreplacing,positioning}
\usetikzlibrary{arrows}

\tikzset{cross/.style={cross out, draw=black, minimum size=2*(#1-\pgflinewidth), inner sep=0pt, outer sep=0pt}, cross/.default={1pt}}

\tikzset{cross/.style={cross out, draw=black, minimum size=2*(#1-\pgflinewidth), inner sep=0pt, outer sep=0pt}, cross/.default={1pt}}

\allowdisplaybreaks[1]

\usepackage{pict2e}

\usepackage{hyperref}
\hypersetup{
     colorlinks=false,         
     linkcolor=darkred,
     citecolor=blue,
}

\makeatletter
\DeclareFontFamily{OMX}{MnSymbolE}{}
\DeclareSymbolFont{MnLargeSymbols}{OMX}{MnSymbolE}{m}{n}
\SetSymbolFont{MnLargeSymbols}{bold}{OMX}{MnSymbolE}{b}{n}
\DeclareFontShape{OMX}{MnSymbolE}{m}{n}{
    <-6>  MnSymbolE5
   <6-7>  MnSymbolE6
   <7-8>  MnSymbolE7
   <8-9>  MnSymbolE8
   <9-10> MnSymbolE9
  <10-12> MnSymbolE10
  <12->   MnSymbolE12
}{}
\DeclareFontShape{OMX}{MnSymbolE}{b}{n}{
    <-6>  MnSymbolE-Bold5
   <6-7>  MnSymbolE-Bold6
   <7-8>  MnSymbolE-Bold7
   <8-9>  MnSymbolE-Bold8
   <9-10> MnSymbolE-Bold9
  <10-12> MnSymbolE-Bold10
  <12->   MnSymbolE-Bold12
}{}

\DeclareMathAlphabet{\mathscrbf}{OMS}{mdugm}{b}{n}

\let\llangle\@undefined
\let\rrangle\@undefined
\DeclareMathDelimiter{\llangle}{\mathopen}%
                     {MnLargeSymbols}{'164}{MnLargeSymbols}{'164}
\DeclareMathDelimiter{\rrangle}{\mathclose}%
                     {MnLargeSymbols}{'171}{MnLargeSymbols}{'171}
\makeatother



\definecolor{myPurple}{rgb}{0.5,0.1,0.6}
\definecolor{myOrange}{rgb}{1.0,0.5,0.0}
\definecolor{myRed}{rgb}{1.0,0.0,0.0}
\definecolor{myGreen}{rgb}{0.0,0.5,0.0}
\definecolor{LatexBlue}{rgb}{0.211765,0.227451,0.666667}
\definecolor{myBlue}{rgb}{0.0,0.0,1.0}
\definecolor{myBlack}{rgb}{0.0,0.0,0.0}
\definecolor{myGray}{rgb}{0.3,0.3,0.3}

\theoremstyle{plain}
\newtheorem{theorem}{Theorem}[section]
\newtheorem*{theorem*}{Theorem}
\newtheorem{proposition}[theorem]{Proposition}
\newtheorem*{proposition*}{Proposition}
\newtheorem{lemma}[theorem]{Lemma}

\theoremstyle{definition}
\newtheorem{definition}[theorem]{Definition}

\newtheorem{remarkx}[theorem]{Remark}

\newenvironment{remark}
  {\pushQED{\qed}\remarkx}
  {\popQED\endremarkx}

\DeclareMathOperator{\res}{res}
\DeclareMathOperator*{\lres}{res}	

\DeclareMathOperator{\End}{End}

\DeclareMathOperator{\im}{im}

\DeclareMathOperator{\Tr}{Tr}

\newcommand{\lda}{\lambda}

\newcommand{\g}{\mathfrak{g}}

\newcommand{\cD}{\mathcal{D}}
\newcommand{\CP}{\mathbb{P}^1}

\newcommand{\cL}{\mathcal{L}}

\newcommand{\Lag}{\mathscr{L}}
\newcommand{\bLag}{\mathscrbf{L}}

\def\b{\mathfrak{b}}
\def\gl{\mathfrak{gl}}
\def\sl{\mathfrak{sl}}
\def\id{\textup{id}}
\def\h{\mathfrak{h}}

\def\n{\mathfrak{n}}

\newcommand{\bb}[1]{\llbracket #1 \rrbracket}
\newcommand{\lau}[1]{(\kern-.2em( #1 )\kern-.2em)}

\newcommand{\ie}{{\it i.e.}\ }

\def\be{\begin{equation}}
\def\ee{\end{equation}}
\def\bea{\begin{eqnarray}}
\def\eea{\end{eqnarray}}

\def\A{\mathcal{A}}

\def\CC{\mathbb{C}}
\def\CP{\mathbb{C}P^1}

\def\RR{\mathbb{R}}

\def\ZZ{\mathbb{Z}}

\def\1{\bm{1}}

\numberwithin{equation}{section}

\begin{document}
	
\begin{center}
{\LARGE\bf Classical Yang-Baxter equation,}\\
\vspace{1.5mm}
{\LARGE\bf Lagrangian multiforms and}\\
\vspace{1.5mm}
{\LARGE\bf ultralocal integrable hierarchies}\\

\vspace{1.5em}
Vincent Caudrelier$^1$, Matteo Stoppato$^1$ and Beno\^{\i}t Vicedo$^2$

\vspace{1em}
\begingroup\itshape
$^1${\it School of Mathematics, University of Leeds, LS2 9JT, U.K.}\\
$^2${\it Department of Mathematics, University of York, York YO10 5DD, U.K.}
\par\endgroup
\vspace{1em}
\begingroup\ttfamily
v.caudrelier@leeds.ac.uk, stoppato.matteo@gmail.com, benoit.vicedo@gmail.com
\par\endgroup
\vspace{1.5em}
\end{center}

\begin{abstract}
We cast the classical Yang-Baxter equation (CYBE) in a variational context for the first time, by relating it 
to the theory of Lagrangian multiforms, a framework designed to capture integrability in a variational fashion. This provides a significant connection between Lagrangian multiforms and the CYBE, one of the most fundamental concepts of integrable systems. This is achieved by introducing a {\it generating} Lagrangian multiform which depends on a skew-symmetric classical $r$-matrix with spectral parameters. 
The multiform Euler-Lagrange equations produce a generating Lax equation which yields a generating zero curvature equation. 
 The CYBE plays a role at three levels: $1)$ It ensures the commutativity of the flows of the generating Lax equation; $2)$ It ensures that the generating zero curvature equation holds; $3)$ It implies the closure relation for the generating Lagrangian multiform. 
The specification of an integrable hierarchy is achieved by fixing certain data: a finite set $S\subset\CP$, a Lie algebra $\g$, a $\g$-valued rational function with poles in $S$ and an $r$-matrix. 
We show how our framework is able to generate a large class of {\it ultralocal} integrable hierarchies by providing several known and new examples pertaining to the rational or trigonometric class. These include the Ablowitz-Kaup-Newell-Segur hierarchy, the sine-Gordon (sG) hierarchy and various hierarchies related to Zakharov-Mikhailov type models which contain the Faddeev-Reshetikhin (FR) model and recently introduced deformed Gross-Neveu models as particular cases. The versatility of our method is illustrated by showing how to couple integrable hierarchies together to create new examples of integrable field theories and their hierarchies. We provide two examples: the coupling of the nonlinear Schr\"odinger system to the FR model and  the coupling of sG with the anisotropic FR model. 
\end{abstract}

\newpage

\tableofcontents

\newpage

\section{Introduction}

\subsection{Context and background}

\subsubsection{Integrability in the Hamiltonian framework}

A profound discovery in the modern theory of integrable systems was that the special partial differential equations originally treated in the seminal works \cite{GGKM,ZS}, using what is now known as the Inverse Scattering Method, were also infinite dimensional Hamiltonian systems \cite{G} for which an analog of the Liouville theorem for finite dimensional Hamiltonian systems could be established, \cite{ZF,ZMan}. This allows one, in particular, to see such systems as Hamiltonian field theories. The developments based on these early examples led to the beautiful theory of the classical $r$-matrix which captures the special Hamiltonian features of these models \cite{Dr,STS}. An important condition usually required of the $r$-matrix is that it satisfies the classical Yang-Baxter equation (CYBE)
\begin{equation} \label{CYBE}
\big[ r_{12}(\lambda, \mu), r_{13}(\lambda, \nu) \big] + \big[ r_{12}(\lambda, \mu),r_{23}(\mu, \nu) \big] -\big[r_{13}(\lambda, \nu), r_{32}(\nu, \mu) \big]=0.
\end{equation}
It ensures that a certain Poisson bracket defined using $r$ satisfies the Jacobi identity. Another important condition is to decide if $r$ is skew-symmetric or not, \ie whether or not it satisfies 
$$r_{12}(\lda,\mu)=-r_{21}(\mu,\lda)\,.$$  
This has deep mathematical and physical implications. If the $r$-matrix is skew-symmetric, the associated field theories are called ultralocal while they are non-ultralocal otherwise. In the present work, we restrict our attention to the ultralocal case.

A characteristic feature of integrable field theories is that their equations of motion come in hierarchies. Specifically, any given integrable Hamiltonian field theory has infinitely many conserved charges which can, themselves, be used as Hamiltonians to define flows with respect to the Poisson bracket. Because all the conserved charges Poisson commute amongst themselves, it is possible to impose all these flows simultaneously on the fields of the theory and thus treat the latter as depending on infinitely many times. The collection of equations of motion thus obtained is referred to as an integrable hierarchy. Schematically, for a scalar field theory with field $u$, there would be a countable number of conserved charges $H_j$, labelled by integers $j\ge 1$ say, in involution with respect to a given Poisson bracket, namely
\begin{equation*}
\{H_i,H_j\}=0
\end{equation*}
for every $i,j\geq 1$. The hierarchy would then consist of all the equations
\begin{equation} \label{Ham_hierarchy}
\partial_{t_j}u=\{H_j,u\},
\end{equation}
where we have introduced an infinite number of times $t_j$ for $j \geq 1$. Among all the conserved charges $H_j$, one of them can be taken to be the Hamiltonian of the integrable field theory one started with. Studying the hierarchy as a whole can reveal much more structure and properties of the initial model. This is of course not a new idea but here we depart from the established point of view in that we want to exploit this idea in a {\it Lagrangian} setting.

\subsubsection{Integrability in the Lagrangian framework}

When turning to the Lagrangian setting, one is immediately faced with the following question: how should integrable hierarchies be captured in the Lagrangian formalism? This question found an answer relatively recently in the theory of {\it Lagrangian multiforms} which was introduced in the seminal paper \cite{LN} and has rapidly developed in various direction. More recently, several works cast the original idea into the context of continuous integrable field theories, see \cite{SV,V,SNC,PV,SNC2,CS1,CS2,CS3} for examples of two-dimensional field theories (e.g. Korteweg-de Vries, sine-Gordon and nonlinear Schr\"odinger) and \cite{SNC2,SNC3} for a three-dimensional example (Kadomtsev--Petviashvili). For a two-dimensional field theory, the central object is a differential two-form
\begin{equation} \label{Lag multi intro}
\Lag[u]=\sum_{i,j}\Lag_{ij}[u]dt_i\wedge dt_j
\end{equation}
on an infinite-dimensional space $\RR^\infty$ parametrised by the infinite collection of times $t_i$ of the hierarchy. The coefficients $\Lag_{ij}[u]$ are Lagrangians depending on the fields of the theory, which are collectively denoted by $u$ here for simplicity (even though we are no longer restricting to the case of a single scalar field). For each Lagrangian coefficient $\Lag_{ij}[u]$ we can consider the associated action $S_{ij}[u] = \int_{\RR^2} \Lag_{ij}[u] dt_i \wedge dt_j$. Using the differential two-form \eqref{Lag multi intro} we can succinctly rewrite all these actions as $S_{ij}[u] = \int_{\sigma_{ij}} \Lag[u]$, where the integral here is over the two dimensional plane $\sigma_{ij} \simeq \RR^2$ spanned by the coordinates $t_i$ and $t_j$ in $\RR^\infty$. At this point, of course, there is no reason for the field theories described by the actions $S_{ij}[u]$ to belong to the same integrable hierarchy, let alone to produce equations of motion that are integrable! The key new ingredient is to impose a generalised variational principle on the more general action 
\begin{equation} \label{Action intro}
S[u,\sigma] = \int_\sigma \Lag[u],
\end{equation}
which now also depends on an arbitrary choice of two-dimensional smooth surface $\sigma$ in $\RR^\infty$. Note, in particular, that $S_{ij}[u] = S[u, \sigma_{ij}]$. The generalised variational principle which ties all these theories together is a least action principle for $S[u,\sigma]$ {\bf simultaneously for all smooth surfaces $\sigma$}. This results in what are called the {\bf multiform Euler--Lagrange (EL) equations}. These were first derived in \cite{SV} for the two-form case that we consider in this paper. It can be shown \cite{SV,SNC} that they can be written compactly as 
\be
\label{deltadL}
\delta d \Lag =0\,,
\ee 
where $d$ is the usual exterior derivative and $\delta$ denotes the variational derivative. In the Lagrangian multiform theory, the above generalised variational principle is complemented by another requirement: on critical points, one also requires that the action be stationary with respect to arbitrary local variations of $\sigma$. This gives us the important {\bf closure relation} on the equations of motion, i.e. {\it on shell} 
\be
\label{closure_relation}
d \Lag =0\,.
\ee  

Intuitively, requiring criticality of the action for an arbitrary surface is the new feature that encodes variationally the commutativity of the flows known to be a signature of integrability in the Hamiltonian world. Roughly speaking, the connection with \eqref{Ham_hierarchy} is that the Lagrangian coefficients $\Lag_{1j}$ correspond by a Legendre transform to the Hamiltonians $H_j$, with the understanding that the time $t_1$ plays some preferred role (the ``space'' variable) and the $t_j$, $j\ge 2$ are all the higher times of the hierarchy. The interpretation of all the other Lagrangian coefficients $\Lag_{ij}$ is best obtained by casting the hierarchy as a collection of compatible zero curvature equations involving Lax matrices $V_j(\lda)$, namely
\begin{equation}
\partial_{t_j}V_i(\lda)-\partial_{t_i}V_j(\lda)+\left[V_i(\lda) , V_j(\lda)\right]=0
\end{equation}
for $i, j \geq 1$. It is known that all these equations are in fact Hamiltonian, see \emph{e.g.} \cite{AC}, and the case $i=1$ corresponds to \eqref{Ham_hierarchy}. One of the main points of the present work is that they are also variational with Lagrangian $\Lag_{ij}$. It is important to realise that the multiform EL equations are largely overdetermined equations for the coefficients $\Lag_{ij}$. Part of these equations impose restrictions on the allowed coefficients, the idea being that they enforce the integrability of the corresponding theories. The rest consist of standard EL equations associated to these coefficients and give the equations of motion of the integrable hierarchy. 

\subsubsection{Motivating example: Ablowitz--Kaup--Newell--Segur hierarchy}\label{motivating_ex}

In \cite{CS3}, on the example of the Ablowitz--Kaup--Newell--Segur (AKNS) hierarchy, the notion of Lagrangian multiform was successfully combined with the idea of ``compounding hierarchies'' introduced in the Lagrangian framework in \cite{Nij} (itself inspired from the use of the generating formalism for integrable hierarchies, see e.g. \cite{Nij2}). This naturally leads to working with generating functions when dealing with hierarchies. The key object was what we can call a {\it generating Lagrangian multiform}. The simple idea is to organise the Lagrangian coefficients $\Lag_{ij}$ of the 2-form \eqref{Lag multi intro} into a generating series involving formal (spectral) parameters
\begin{equation} \label{gen Lag AKNS intro}
\Lag(\lda,\mu)=\sum_{i,j}\frac{\Lag_{ij}}{\lda^{i+1}\mu^{j+1}}.
\end{equation}
It is clear that there is a one-to-one corresponding between $\Lag[u]$ and $\Lag(\lda,\mu)$ where from the latter, one can extract the coefficients by the formula
\begin{equation*}
\Lag_{ij}=\res_\lambda \res_\mu \left(\lda^i\mu^j\Lag(\lda,\mu)\right),
\end{equation*}
where $\res_\lda$ returns the coefficient of $\lda^{-1}$ in the series expansion, and similarly for $\res_\mu$.
One advantage of working with generating series such as \eqref{gen Lag AKNS intro} stems from the usefulness of generating functions in general: properties of their coefficients are more easily studied and derived from those of the generating function. In our context, this means that we can manipulate an integrable hierarchy as a whole instead of studying each Lagrangian coefficient $\Lag_{ij}$ individually. Originally, the latter approach was used in the sense that only a given ``starting'' Lagrangian coefficient was known, say $\Lag_{12}$, and one would try to build the higher coefficients $\Lag_{ij}$ so as to obtain a consistent Lagrangian multiform. Methods to compute these coefficients were introduced for instance in \cite{V,SNC2}. Although the recursive algorithm could be applied in principle, in practice this is hard to implement beyond the first few coefficients. Moreover, the Lagrangians $\Lag_{ij}$ obtained in this way usually contain derivatives with respect to $t_1$ or $t_2$ (the times associated with $\Lag_{12}$). These are not natural times from the point of $\Lag_{ij}$: this is the so-called problem of ``alien derivatives'' which was identified and explained in \cite{V}. Having a generating Lagrangian multiform circumvents these issues. This will be elaborated upon in the examples. 

For the AKNS hierarchy, the generating Lagrangian multiform can be written as \cite{CS3}
\be
\label{ex_Lag_multiform}
\Lag(\lda,\mu)= i\Tr{\left( \phi(\mu)^{-1} \cD_\lambda \phi(\mu)\sigma_3 - \phi(\lambda) ^{-1} \cD_\mu \phi(\lambda)\sigma_3     \right) } -  \Tr{ \frac{Q(\lambda)Q(\mu)}{\mu-\lambda} }\,,
\ee
with $Q(\lda)=-i\phi(\lambda)\sigma_3\phi(\lambda) ^{-1}$, $\phi(\lda)$ being a group-valued formal series in $1/\lda$ with constant term equal to the identity and whose coefficients contain the dynamical variables. The operator $\displaystyle \cD_\lambda = \sum_{j \geq 0} \lambda^{-j-1} \partial_{t_j}$ is a formal collection of all the AKNS flows $\partial_{t_j}$, and similarly for $\cD_\mu$.
The generating Lagrangian multiform \eqref{ex_Lag_multiform} generates all the coefficients $\Lag_{ij}$ systematically and without the problem of alien derivatives, reproducing  the first few coefficients which had been constructed in \cite{SNC,SNC2,PV}, as it should. Its multiform EL equations yield the defining equations of the AKNS hierarchy as discussed by Flaschka--Newell--Ratiu (FNR) in \cite{FNR}, namely\footnote{The flow $t_0$ is the trivial linear flow but is included in the construction for convenience. In practice, one is interested in the nonlinear flows $t_j$, $j \ge 1$.} 
\begin{equation} \label{FNR intro}
\partial_{t_i} Q(\lambda) = [Q^{(i)}(\lambda), Q(\lambda)]\,,~~ i\ge 0\,,
\end{equation}
where $\displaystyle Q(\lda)=\sum_{j=0}^\infty Q_j\lda^{-j}$ and $\displaystyle Q^{(i)}(\lambda)=\sum_{j=0}^i Q_j\lda^{i-j}$ and $Q_0=-i\sigma_3$. More precisely, the multiform EL equations for \eqref{ex_Lag_multiform} produce the equations \eqref{FNR intro} in generating form
\begin{equation}
\label{AKNS_generating}
\cD_\mu Q(\lambda) = \frac{[Q(\mu),Q(\lambda)]}{ \mu - \lambda},
\end{equation}
where we used the formal series identity
\begin{equation}
\sum_{k=0}^\infty \frac{Q^{(k)}(\lambda)}{\mu^{k+1}} = \frac{Q(\mu)}{\mu - \lambda}\,.
\end{equation}

\subsection{Motivation, main results and plan}\label{motivation}

\paragraph{Motivation:} The present work is motivated by the following observations made on the generating Lagrangian multiform \eqref{ex_Lag_multiform} and the generating FNR equations \eqref{AKNS_generating}:
\begin{enumerate}
	\item The potential term in $\Lag(\lda,\mu)$ has a characteristic form which can be identified as the expression
	$$\Tr_{12}\left(r_{12}(\lda,\mu)Q_1(\lda)Q_2(\mu)\right)$$
	where $r_{12}(\lda,\mu)=\frac{P_{12}}{\mu-\lda}$ is the rational $r$-matrix, known to describe the Hamiltonian structure of the AKNS hierarchy. One could then imagine replacing this particular $r$-matrix with another skew-symmetric $r$-matrix. This leads to the question of whether the nice properties of the generating Lagrangian multiform still hold. One of our main results is that this is the case by virtue of the CYBE. Correspondingly, the RHS of \eqref{AKNS_generating} can also be written as $[\Tr_2 r_{12}(\lda,\mu) Q_2(\mu),Q_1(\lambda)]$ and the same generalisation can be contemplated.
	
	\item The choice of expanding all the objects as formal series in $1/\lda$ and $1/\mu$ is a sign that one is performing an expansion around the point at infinity. However, nothing would prevent us from considering other points in $\CP$.
	
	\item The Pauli matrix $\sigma_3$ appearing in \eqref{ex_Lag_multiform} is a special choice of constant element in the underlying loop algebra of $\sl_2$ and the form of $Q(\lda)$ indicates that one is building a phase space for the field theory as a (co)adjoint orbit around this particular element. One could consider other elements in the loop algebra to construct different phase spaces and hence different models. Moreover, one could also consider other Lie algebras than $\sl_2$.
\end{enumerate}
The careful implementation of these natural observations requires some machinery which is presented Section \ref{sec: adeles}. In a first instance, the reader may choose to read the rest of this introduction containing a summary of the formalism and results, and go directly to Section \ref{sec: Lagrangian}. 

The idea is to substitute the loop algebra of $\sl_2$ with a much more versatile structure: the Lie algebra of $\g$-valued ad\`eles associated with a Lie algebra $\g$. This algebra is presented in \cite{STS2} as the relevant structure to implement the second observation above. By doing so in our context, we build a ``universal'' generating Lagrangian multiform which is capable of describing a large class of ultralocal integrable  hierarchies  and we provide a large variety of examples.

In a nutshell, for a matrix Lie algebra $\g$, the Lie algebra of $\g$-valued ad\`eles is defined as
\begin{equation*}
	\bm\A_{\bm\lambda}(\g) \coloneqq \coprod_{a \in \CP} \g \otimes \CC\lau{\lambda_a}\,,
\end{equation*}
where $\lda_a=\lambda-a$ for $a\in\CC$ and $\lda_\infty=\frac{1}{\lda}$ are the local series expansion parameters. An element $\bm X(\bm\lambda) = (X^a(\lambda_a))_{a \in \CP}$ of this algebra consist of tuples with all but finitely many of the formal Laurent series $X^a(\lambda_a) \in \g \otimes \CC\lau{\lambda_a}$ being Taylor series in $\lambda_a$, \emph{i.e.} there exists a finite subset $S \subset \CP$ such that $X^a(\lambda_a) \in \g \otimes \CC\bb{\lambda_a}$ for every $a \in \CC \setminus S$. 
Let $R_\lambda(\g)$ denote the Lie algebra of $\g$-valued rational functions in the formal variable $\lambda$ and define the map 
\begin{equation} \label{iota map}
	\bm \iota_{\bm \lambda} : R_\lambda(\g) \longrightarrow \bm\A_{\bm\lambda}(\g), \qquad f \longmapsto (\iota_{\lambda_a} f)_{a \in \CP}
\end{equation}
where $\iota_{\lambda_a} f \in \g \otimes \CC\lau{\lambda_a}$ is the Laurent expansion  of $f\in R_\lambda(\g)$ at $a \in \CP$. Using certain solutions of the CYBE, it is possible to obtain a direct sum decomposition of this Lie algebra into two maximally isotropic Lie subalgebras
\begin{equation}
	\label{decomp}
	\bm\A_{\bm\lambda}(\g) = \bm\A^+_{\bm\lambda}(\g) \dotplus \bm \iota_{\bm\lambda} R_\lambda(\g)\,.
\end{equation}
We can also define a group $\bm\A^+_{\bm\lambda}(G)$ associated to $\bm\A^+_{\bm\lambda}(\g)$. If $\mu$ is another formal variable, we can work with double formal series locally in $\lda_a$ and $\mu_b$, $a,b\in\CP$ and consider tuples of the form $\bm X(\bm\lda,\bm\mu)=(X^{a,b}(\lda_a,\mu_b))_{a,b \in \CP}$.

Thanks to this ad\`elic framework, we can retain the power of the algebraic formulation of formal power series while working locally around arbitrary points in $\CP$. We introduce the following generalisation of \eqref{ex_Lag_multiform} which realises the above three observations
\begin{equation} \label{Lagrangian mult}
	\bLag(\bm \lambda, \bm \mu) \coloneqq \bm K(\bm \lambda,\bm \mu) - \bm U(\bm \lambda,\bm \mu)
\end{equation}
where the kinetic and potential terms are given by
\begin{subequations} \label{kin pot terms}
	\begin{align}
		\label{kinetic term} \bm K(\bm \lambda,\bm \mu) &\coloneqq \Tr\big( \bm \phi(\bm \lambda)^{-1} \cD_{\bm \mu} \bm \phi(\bm \lambda) (\bm \iota_{\bm \lambda} F(\lambda))_- \big)\\
		&\qquad\qquad\qquad - \Tr \big(\bm \phi(\bm \mu)^{-1} \cD_{\bm \lambda} \bm \phi(\bm \mu) (\bm \iota_{\bm \mu} F(\mu))_- \big), \notag\\
		\label{potential term} \bm U(\bm \lambda,\bm \mu) &\coloneqq \tfrac 12 \Tr_{12}\big( (\bm \iota_{\bm \lambda} \bm \iota_{\bm \mu} + \bm \iota_{\bm \mu} \bm \iota_{\bm \lambda})r_{12}(\lambda,\mu) \bm Q_1(\bm \lambda) \bm Q_2(\bm \mu)\big).
	\end{align}
\end{subequations}
Here $\bm \phi(\bm \lambda)$ is an element of the group $\bm\A^+_{\bm\lambda}(G)$, $\bm Q(\bm \lambda)= \bm \phi(\bm \lambda) \big( \bm \iota_{\bm \lambda} F(\lambda) \big)_- \bm \phi(\bm \lambda)^{-1}$ is an element of $\bm\A_{\bm \lambda}(\g)$, where $\big( \bm \iota_{\bm \lambda} F(\lambda) \big)_- = \big( F^a(\lambda_a)_- \big)_{a \in \CP}$ is the collection of principal parts of a $\g$-valued rational function $F(\lambda)\in R_\lambda(\g)$. In terms of components of the tuples, we have
\begin{eqnarray*}
	\Lag^{a,b}(\lambda_a, \mu_b) &=& \Tr\big( \phi^a(\lambda_a)^{-1} \cD_{\mu_b} \phi^a(\lambda_a) F^a(\lambda_a)_- \big)- \Tr \big(\phi^b(\mu_b)^{-1} \cD_{\lambda_a} \phi^b(\mu_b) F^b(\mu_b)_- \big)\\
&&- \tfrac 12 \Tr_{12}\big( (\iota_{\lambda_a} \iota_{\mu_b} + \iota_{\mu_b}\iota_{\lambda_a})r_{12}(\lambda,\mu)Q^a_1(\lambda_a)Q^b_2(\mu_b)\big),
\end{eqnarray*}
for every $a,b \in \CP$.
The operator $\cD_{\bm \lambda} \coloneqq (\cD_{\lambda_a})_{a \in \CP}$ denotes the $\CP$-tuple of formal operators $\cD_{\lambda_a}$ which contain the partial differential operators $\partial_{t^a_n}$ (see \eqref{def_gen_deriv}). The times $t_n^a$ will be the times of the integrable hierarchies we describe. The rational function $r_{12}(\lda,\mu)$ is the classical $r$-matrix defining the type of ultralocal hierarchies we consider (\emph{e.g.} rational or trigonometric) and corresponds to the $r$-matrix yielding the decomposition \eqref{decomp}.

\paragraph{Main results:}
\begin{enumerate}
			\item We show that the {\it generating Lax equation}
			\begin{equation}
				\cD_{\bm \mu} \bm Q_1(\bm \lambda) = \big[ \Tr_2 \big( \bm \iota_{\bm \lambda} \bm \iota_{\bm \mu} r_{12}(\lambda,\mu) \bm Q_2(\bm \mu) \big), \bm Q_1(\bm \lambda) \big]
			\end{equation} 
 is variational. It arises as the multiform EL equations associated to our generating Lagrangian multiform \eqref{Lagrangian mult}. This is the content of Theorem \ref{prop: eom Q}. This generalises the analogous result first obtained in \cite{SNC} in the context of the Zakharov-Mikhailov models \cite{ZM1}. The generating Lax equation plays here for field theories a role similar to the traditional Lax equation for finite dimensional systems. This is explained in Section \ref{sec: coadjoint}. We relate it to a {\it generating zero curvature equation} which is shown to hold as a consequence of the CYBE for the $r$-matrix appearing in \eqref{Lagrangian mult}.
 
	\item We relate for the first time the CYBE with the relatively recent notion of Lagrangian multiforms. The closure relation \eqref{closure_relation} in generating form, \ie the closure relation for \eqref{Lagrangian mult}, is shown to be a consequence of the CYBE for the $r$-matrix appearing in \eqref{Lagrangian mult}, see Theorem \ref{thm_closure}. On the one hand, this provides a variational interpretation of the CYBE, a fundamental equation that has only been introduced and studied from a Hamiltonian point of view so far. On the other hand, given the importance of the CYBE as a criterion for classical integrability, this further establishes the Lagrangian multiform approach as a variational criterion for integrability.

	\item Specialising the generating Lagrangian multiform \eqref{Lagrangian mult}, we recover known examples of integrable hierarchies and produce several new examples. We also introduce an easy method for coupling hierarchies together.
\end{enumerate}

\paragraph{Plan of the paper:}

In Section \ref{sec: adeles}, we introduce the Lie algebra of $\g$-valued ad\`eles and establish its decomposition into two complementary maximal isotropic Lie subalgebras which allows us to introduce the classical $r$-matrix of interest via the corresponding projectors onto the Lie subalgebras. This generalises to the ad\`eles case the well-known interpretation of a classical $r$-matrix as a difference of projectors. This is done explicitly for the rational and trigonometric cases. Section \ref{sec: Lagrangian} introduces the main elements of our framework: we state the generalisation of the generating FNR equations \eqref{AKNS_generating}, which we call the generating Lax equation, taking into account the above observations. Its properties are directly connected to the CYBE. Then we introduce the generating Lagrangian multiform that produces the generating Lax equation as its multiform EL equations. Again, its properties, in particular the closure relation, are shown to be a direct consequence of the CYBE. The subsequent Sections \ref{sec: AKNS hierarchy} to \ref{ZM_examples} are devoted to examples. Several were known previously, and these are used to show how our framework contains them naturally, \emph{e.g.} the AKNS hierarchy and the sine-Gordon hierarchy. For the latter example, we explain in detail how the first few known Lagrangian coefficients are recovered but without the problem of alien derivatives. Other examples, such as the trigonometric Zakharov-Mikhailov hierarchy, are new. For the recently introduced deformed Gross-Neveu models, the new feature brought in by our construction is that they are naturally embedded into an integrable hierarchy.
Various conclusions and discussions are presented in Section \ref{conclusion}. Finally, we recall in an appendix the relationship between the trigonometric $r$-matrix used in this paper and the more familiar $r$-matrix of the sine-Gordon model.

\section{Lie algebra of $\g$-valued ad\`eles} \label{sec: adeles}

\subsection{General setup} \label{sec: gen setup}

Let $N \in \ZZ_{\geq 1}$ and consider either the Lie algebra $\gl_N$ of all $N \times N$ matrices with complex entries or its Lie subalgebra $\sl_N$ of traceless matrices. We will treat both of these cases in parallel, using the common notation $\g$ throughout. The generalisation to other matrix Lie algebras is straightforward but for simplicity we shall restrict to these two cases. We also denote by $G$ the associated Lie group which corresponds either to the general linear group $GL_N$ of invertible $N \times N$ matrices or to its Lie subgroup $SL_N$ of matrices with determinant $1$.

We use the trace $\Tr : \gl_N \to \CC$ to endow the Lie algebra $\g$ with the non-degenerate invariant symmetric bilinear form $\g \times \g \to \CC$ given by $(X, Y) \mapsto \Tr(XY)$. 
Let $P_{12}$ be the tensor Casimir of $\g$ with the property that $\Tr_2(P_{12} X_2) = X$ for any $X \in \g$. Explicitly, for $\gl_N$ it is given by $P_{12} = \sum_{i,j=1}^N E_{ij} \otimes E_{ji}$ where $E_{ij}$ for $i,j=1, \ldots, N$ is the standard basis of $\gl_N$. Similarly, for $\sl_N$ we can write $P_{12} = \sum_a I_a \otimes I^a$ where $\{ I_a \}$ and $\{ I^a \}$ are dual bases of $\sl_N$ with respect to the above bilinear form.  For clarity, let us also recall that the notation $X_2$ means $\1\otimes X$ and the notation $\Tr_2(\dots)$ means that we apply the trace only in the second tensor factor.

Let $\lambda$ be a formal variable. For any $a \in \CC$ we define the formal local coordinate around $a$ as $\lambda_a \coloneqq \lambda - a$ and to the point at infinity we associate the formal local coordinate $\lambda_\infty \coloneqq \lambda^{-1}$. We consider the Lie algebra of $\g$-valued ad\`eles defined as
\begin{equation*}
\bm\A_{\bm\lambda}(\g) \coloneqq \coprod_{a \in \CP} \g \otimes \CC\lau{\lambda_a}.
\end{equation*}
Its elements consist of tuples $\bm X(\bm\lambda) = (X^a(\lambda_a))_{a \in \CP}$ with all but finitely many of the formal Laurent series $X^a(\lambda_a) \in \g \otimes \CC\lau{\lambda_a}$ being Taylor series, \emph{i.e.} there exists a finite subset $S \subset \CP$ such that $X^a(\lambda_a) \in \g \otimes \CC\bb{\lambda_a}$ for every $a \in \CC \setminus S$. 
The Lie bracket of two elements $\bm X(\bm\lambda) = (X^a(\lambda_a))_{a \in \CP}$ and $\bm Y(\bm\lambda) = (Y^a(\lambda_a))_{a \in \CP}$ in $\bm\A_{\bm\lambda}(\g)$ is defined component-wise, as
\begin{equation*}
[\bm X(\bm\lambda), \bm Y(\bm\lambda)] = \big( [X^a(\lambda_a), Y^a(\lambda_a)] \big)_{a \in \CP}.
\end{equation*}

Let $R_\lambda$ denote the algebra of rational functions in the formal variable $\lambda$. The Laurent expansion of a rational function $f \in R_\lambda$ at any $a \in \CP$ defines a homomorphism
\begin{equation} \label{rational expand}
\iota_{\lambda_a} : R_\lambda \longrightarrow \CC\lau{\lambda_a}, \qquad f \longmapsto \iota_{\lambda_a} f.
\end{equation}

We will consider two possible non-degenerate invariant bilinear forms on the Lie algebra $\bm\A_{\bm\lambda}(\g)$, namely
\begin{subequations} \label{bilinear form}
	\begin{equation}
	\langle\!\langle \cdot, \cdot \rangle\!\rangle_k : \bm\A_{\bm\lambda}(\g) \times \bm\A_{\bm\lambda}(\g) \longrightarrow \CC
	\end{equation}
	for $k=0$ and $k = -1$, defined as
	\begin{equation} \label{bilinear form b}
	\langle\!\langle \bm X(\bm\lambda), \bm Y(\bm \lambda) \rangle\!\rangle_k \coloneqq \sum_{a \in \CP} \res^\lambda_a \Tr \big( X^a(\lambda_a) Y^a(\lambda_a) \big) \lambda^k d\lambda,
	\end{equation}
\end{subequations}
for any $\bm X(\bm\lambda) = (X^a(\lambda_a))_{a \in \CP}$ and $\bm Y(\bm \lambda) = (Y^a(\lambda_a))_{a \in \CP}$. Strictly speaking, the rational function $\lambda^k$ on the right hand side of \eqref{bilinear form b} should be expanded at $a \in \CP$, namely we should write $\iota_{\lambda_a} \lambda^k$ instead of $\lambda^k$. In order to simplify the notation, such expansions will always be implicit when taking residues. Here, for any $a \in \CP$, the residue $\res^\lambda_a : \CC\lau{\lambda_a} d\lambda_a \to \CC$ returns the coefficient of $\lambda_a^{-1} d\lambda_a$. For computing the residue at infinity we note that $d\lambda = - \lambda_\infty^{-2} d\lambda_\infty$. 
Note that only finitely many terms in the sum in \eqref{bilinear form b} are non-zero by definition of $\bm\A_{\bm\lambda}(\g)$.

Let $R_\lambda(\g) \coloneqq \g \otimes R_\lambda$ denote the Lie algebra of $\g$-valued rational functions in the formal variable $\lambda$.
We have an embedding of Lie algebras
\begin{equation} \label{iota map}
\bm \iota_{\bm \lambda} : R_\lambda(\g) \longrightarrow \bm\A_{\bm\lambda}(\g), \qquad f \longmapsto (\iota_{\lambda_a} f)_{a \in \CP}
\end{equation}
where $\iota_{\lambda_a} f \in \g \otimes \CC\lau{\lambda_a}$ is the Laurent expansion of $f \in R_\lambda(\g)$ at $a \in \CP$ in the second tensor factor, as in \eqref{rational expand}. 
The Lie subalgebra $\bm \iota_{\bm \lambda} R_\lambda(\g) \subset \bm\A_{\bm\lambda}(\g)$ is maximally isotropic with respect to $\langle\!\langle \cdot, \cdot \rangle\!\rangle_k$, for any $k \in \ZZ$, by the strong residue theorem; see for instance \cite[Corollary 1]{Takh}.

In the remainder of this section we will describe two possible complementary Lie algebras to $\bm\iota_{\bm\lambda} R_\lambda(\g)$ in $\bm\A_{\bm\lambda}(\g)$, which are maximally isotropic with respect to $\langle\!\langle \cdot, \cdot \rangle\!\rangle_0$ and $\langle\!\langle \cdot, \cdot \rangle\!\rangle_{-1}$, respectively. These two main examples, which can be found for instance in \cite[Example 4]{Drinfeld}, correspond to the rational $r$-matrix and the trigonometric $r$-matrix, respectively.

\medskip

\paragraph{\bf Notation}
We will generally use boldface to denote $\CP$-tuples. For instance, given any $n \in \ZZ$ we will write $\bm \lambda^n \bm X(\bm \lambda)$ for the element $(\lambda_a^n X^a(\lambda_a))_{a \in \CP} \in \bm\A_{\bm\lambda}(\g)$ of the Lie algebra of $\g$-valued ad\`eles. More generally, we would write $\bm \lambda^n \bm X(\bm \lambda) d \bm \lambda$ as a shorthand for the $\CP$-tuple $(\lambda_a^n X^a(\lambda_a) d\lambda_a)_{a \in \CP}$. Note, crucially, that although $d\lambda_a = d\lambda$ for all $a \in \CC$, we have $d\lambda_\infty = - \lambda^{-2} d\lambda$ for the point at infinity. Therefore the two expressions $\bm \lambda^n \bm X(\bm \lambda) d \bm \lambda$ and $\bm \lambda^n \bm X(\bm \lambda) d \lambda$ subtly differ only in the component at infinity. If $\mu$ is another formal variable then $\bm \mu$ will denote a separate $\CP$-tuple carrying an independent label $b \in \CP$. For instance, we would have
\begin{equation*}
[\bm X(\bm \lambda), \bm Y(\bm \mu)] = \big( \delta_{ab}[X^a(\lambda_a), Y^b(\mu_b)] \big)_{a,b \in \CP}
\end{equation*}
for any $\bm X(\bm \lambda) \in \bm\A_{\bm \lambda}(\g)$ and $\bm Y(\bm \mu) \in \bm\A_{\bm \mu}(\g)$.
We will make use of such notation with multiple formal variables extensively from Section \ref{sec: Lagrangian} onwards.

\subsection{Rational $r$-matrix} \label{sec: rational r}

Throughout this section we fix the choice $k=0$ in the bilinear form \eqref{bilinear form}. Consider the Lie subalgebra of $\g$-valued integral ad\`eles
\begin{equation} \label{Ap glN def}
\bm\A^{\rm rat}_{\bm\lambda}(\g) \coloneqq \g \otimes \lambda_\infty \CC\bb{\lambda_\infty} \times \coprod_{a \in \CC} \g \otimes \CC\bb{\lambda_a}.
\end{equation}
Note that we have excluded the constant term from the Taylor series at infinity.
We shall also need the corresponding group
\begin{equation} \label{Ap GLN def}
\bm\A^{\rm rat}_{\bm\lambda}(G) \coloneqq \widehat{G}_\infty \times \coprod_{a \in \CC} \widehat{G}_a, 
\end{equation}
where in the $GL_N$ case $\widehat{G}_a$ consists of all invertible $N \times N$ matrices with entries in $\CC\bb{\lambda_a}$ while $\widehat{G}_\infty$ consists of all invertible $N \times N$ matrices with off-diagonal entries in $\lambda_\infty \CC\bb{\lambda_\infty}$ and diagonal entries in $1 + \lambda_\infty \CC\bb{\lambda_\infty}$. In the $SL_N$ case the groups $\widehat{G}_a$ for all $a \in \CP$ are defined in the same way but with the added condition that the matrices are of determinant $1$.

For later practical purposes, it is convenient to collect the following notations in a definition.
\begin{definition}
Let $a \in \CC$ and $X^a(\lambda_a) \in \g \otimes \CC\lau{\lambda_a}$ be a Laurent series in $\lda_a$ with coefficients in $\g$. We shall use the notation
\begin{subequations} \label{rat pole parts}
	\begin{equation} \label{rat pole part a}
	X^a(\lambda_a)^{\rm rat}_- \in \g \otimes \lambda_a^{-1} \CC[\lambda_a^{-1}]
	\end{equation}
	to represent the pole part of $X^a(\lambda_a)$. Similarly, for $X^\infty(\lambda_\infty)  \in \g \otimes \CC\lau{\lambda_\infty} = \g \otimes \CC\lau{\lambda^{-1}}$, we denote by
	\begin{equation} \label{rat pole part inf}
	X^\infty(\lambda_\infty)^{\rm rat}_- \in \g \otimes \CC[\lambda_\infty^{-1}] = \g \otimes \CC[\lambda]
	\end{equation}
	the pole part of $X^\infty(\lambda_\infty)$. Note that the constant term in $\lda_\infty$ is included around infinity.
\end{subequations}
	
\end{definition}

The Lie subalgebra $\bm\A^{\rm rat}_{\bm\lambda}(\g) \subset \bm\A_{\bm\lambda}(\g)$ is clearly maximally isotropic with respect to the bilinear form $\langle\!\langle \cdot, \cdot \rangle\!\rangle_0$ defined in \eqref{bilinear form}. Here we made use of the fact that the constant term was excluded from the Taylor series at infinity in the definition \eqref{Ap glN def}. It follows that the Lie algebra $\bm\A_{\bm\lambda}(\g)$ decomposes as a direct sum of vector spaces
\begin{equation} \label{rational decomp}
\bm\A_{\bm\lambda}(\g) = \bm\A^{\rm rat}_{\bm\lambda}(\g) \dotplus \bm \iota_{\bm\lambda} R_\lambda(\g)
\end{equation}
into complementary Lagrangian (\emph{i.e.} maximal isotropic) Lie subalgebras. Let $\pi^{\rm rat}_\pm$ denote the projections onto $\bm\A^{\rm rat}_{\bm\lambda}(\g)$ and $\bm \iota_{\bm\lambda} R_\lambda(\g)$, respectively, relative to \eqref{rational decomp}.
\begin{definition}[{\bf Rational $r$-matrix}]
Recall the notation $P_{12}$ for the tensor Casimir of $\g$ from Section \ref{sec: gen setup}. The rational $r$-matrix is defined as the following rational function of the formal variables $\lda$ and $\mu$:
\be
\label{def_r_rat}
r_{12}^{\rm rat}(\lda,\mu)=\frac{P_{12}}{\mu-\lda}\,.
\ee	
\end{definition}
As is well-known, it is skew-symmetric: $r_{12}^{\rm rat}(\lda,\mu)=-r_{21}^{\rm rat}(\mu,\lda)$. The following result shows that its known connection to projectors associated to the decomposition of a Lie algebra into isotropic Lie subalgebras extends to the present ad\`eles setting.

\begin{proposition} \label{prop: Ppm kernels}
	For any $\bm X(\bm \lambda) \in \bm\A_{\bm\lambda}(\g)$, its projections onto the complementary subalgebras $\bm\A^{\rm rat}_{\bm\lambda}(\g)$ and $\bm \iota_{\bm\lambda} R_\lambda(\g)$ relative to the direct sum decomposition \eqref{rational decomp} are given respectively by $\pi^{\rm rat}_+ \bm X(\bm \lambda) = \big( (\pi^{\rm rat}_+ X)^a(\lambda_a) \big)_{a \in \CP}$ and $\pi^{\rm rat}_- \bm X(\bm \lambda) = \big( (\pi^{\rm rat}_- X)^a(\lambda_a) \big)_{a \in \CP}$ where
	\begin{subequations} \label{P+ P- expressions}
		\begin{align}
		\label{P+ expression}
		(\pi^{\rm rat}_+ X)^a(\lambda_a) &= \sum_{b \in \CP} \res^\mu_b \Tr_2 \bigg( \iota_{\mu_b} \iota_{\lambda_a} r_{12}^{\rm rat}(\lda,\mu) X^b(\mu_b)_2 \bigg) d\mu,\\
		\label{P- expression}
		(\pi^{\rm rat}_- X)^a(\lambda_a) &= - \sum_{b \in \CP} \res^\mu_b \Tr_2 \bigg( \iota_{\lambda_a} \iota_{\mu_b} r_{12}^{\rm rat}(\lda,\mu) X^b(\mu_b)_2 \bigg) d\mu.
		\end{align}
	\end{subequations}
	\begin{proof}
		Let $\bm X(\bm \lambda) \in \bm\A_{\bm\lambda}(\g)$. We consider, to begin with, its projection onto $\bm \iota_{\bm\lambda} R_\lambda(\g)$. The $\g$-valued rational function in $R_\lambda(\g)$ constructed out of the pole parts of the collection of Laurent series in $\bm X(\bm \lambda)$ is given by
		\begin{align*}
		&\sum_{b \in \CP} \res^\mu_b \Tr_2 \bigg( \iota_{\mu_b} \frac{P_{12}}{\lambda - \mu} X^b(\mu_b)_2 \bigg) d\mu = \sum_{b \in \CC} \sum_{n = 0}^\infty \res^\mu_b \frac{\mu_b^n}{\lambda_b^{n+1}} X^b(\mu_b) d\mu\\
		&\qquad\qquad\qquad - \sum_{n = 0}^\infty \res^\mu_\infty \frac{\lambda^n}{\mu^{n+1}} X^\infty(\mu_\infty) d\mu = \sum_{b \in \CP} X^b(\lambda_b)^{\rm rat}_-
		\end{align*}
		where in the first equality we took the trace and split the term at $b=\infty$ from the rest of the sum over $b \in \CP$.
		The expression \eqref{P- expression} is then obtained by taking the Laurent series expansion of this rational function at each $a \in \CP$, corresponding to applying the map \eqref{iota map}.
		
		Consider now the projection of $\bm X(\bm \lambda) \in \bm\A_{\bm\lambda}(\g)$ onto $\bm\A^{\rm rat}_{\bm\lambda}(\g)$. Note that for any $a \in \CC$ we have
		\begin{align*}
		&\sum_{b \in \CP} \res^\mu_b \Tr_2 \bigg( \iota_{\mu_b} \iota_{\lambda_a} \frac{P_{12}}{\mu - \lambda} X^b(\mu_b)_2 \bigg) d\mu
		= \sum_{n=0}^\infty \lambda_a^n \sum_{b \in \CP} \res^\mu_b \iota_{\mu_b} \mu_a^{-n-1} X^b(\mu_b) d\mu.
		\end{align*}
		If $\bm X(\bm \lambda) \in \bm \iota_{\bm\lambda} R_\lambda(\g)$, say $\bm X(\bm \lambda) = \bm \iota_{\bm \lambda} f(\lambda)$ for some $f(\lambda) \in R_\lambda(\g)$, then the above vanishes at each order in the $\lambda_a$-expansion by the residue theorem. Indeed, the coefficient of $\lambda_a^n$ is given by the sum of all the residues of $(\mu - a)^{-n-1} f(\mu) d\mu$. On the other hand, if $\bm X(\bm \lambda) \in \bm\A^{\rm rat}_{\bm\lambda}(\g)$ then the only term contributing to the sum over $b \in \CP$ is the term for $b=a$ which is equal to $X^a(\lambda_a)$. The same statements hold for $a = \infty$ and hence the result follows.
	\end{proof}
\end{proposition}

Define the linear operator $r^{\rm rat} \coloneqq \pi^{\rm rat}_+ - \pi^{\rm rat}_-$. It follows from Proposition \ref{prop: Ppm kernels} that its kernel is given by
\begin{equation}
\label{def_rat_kernel}
\bigg( (\iota_{\mu_b} \iota_{\lambda_a} + \iota_{\lambda_a} \iota_{\mu_b}) \frac{P_{12}}{\mu - \lambda} \bigg)_{a, b \in \CP}.
\end{equation}
The kernel of the identity operator $\id = \pi^{\rm rat}_+ + \pi^{\rm rat}_-$ is similarly given by an expansion of zero (see e.g. \cite[Chap. 2]{LL}) since
\begin{equation}
	\label{delta_from_r_matrix1}
\bigg( (\iota_{\mu_b} \iota_{\lambda_a} - \iota_{\lambda_a} \iota_{\mu_b}) \frac{P_{12}}{\mu - \lambda} d\mu \bigg)_{a, b \in \CP} = \big( P_{12} \delta_{ab} \delta(\lambda_a, \mu_a) d\mu_a \big)_{a, b \in \CP}
\end{equation}
where we defined 
\be
\label{def_delta}
\delta(\lambda_a, \mu_a) \coloneqq \sum_{n \in \ZZ} \lambda_a^n \mu_a^{-n-1}\,.
\ee
\begin{lemma}\label{lemma_expansion}
	Let $\bm X(\bm \mu) = \big( X^a(\mu_a) \big)_{a \in \CP} \in \bm \A_{\bm \mu}(\g)$ with $\displaystyle X^a(\mu_a) = \sum_{n=-N_a}^\infty X^a_n \mu_a^n$ for some $N_a \in \ZZ$, where $N_a > 0$ for finitely many $a \in \CP$. For any $a \in \CC$ we have
	\begin{equation*}
	\iota_{\mu_a} \frac{X^a(\mu_a)}{\mu - \lambda} = - \sum_{r=-N_a}^\infty \mu_a^r \big( \lambda_a^{-r-1} X^a(\lambda_a) \big)^{\rm rat}_-
	\end{equation*}
	and at infinity we have
	\begin{equation*}
	\iota_{\mu_\infty} \frac{X^\infty(\mu_\infty)}{\mu - \lambda} = \sum_{r=-N_\infty}^\infty \mu_\infty^{r+1} \big( \lambda_\infty^{-r} X^\infty(\lambda_\infty) \big)^{\rm rat}_-.
	\end{equation*}
	\begin{proof}
		First, let $a \in \CC$. Then we have
		\begin{align*}
		\iota_{\mu_a} \frac{X^a(\mu_a)}{\mu - \lambda} &= - \sum_{n= - N_a}^\infty X^a_n \mu_a^n \sum_{s=0}^\infty \mu_a^s \lambda_a^{-s-1}
		= - \sum_{n=-N_a}^\infty \sum_{r=n}^\infty X^a_n \mu_a^r \lambda_a^{n-r-1}\\
		&= - \sum_{r=-N_a}^\infty \mu_a^r \sum_{n=-N_a}^r X^a_n \lambda_a^{n-r-1}
		= - \sum_{r=-N_a}^\infty \mu_a^r \big( \lambda_a^{-r-1} X^a(\lambda_a) \big)^{\rm rat}_-
		\end{align*}
		where in the second equality we changed variables from $s$ to $r = s+n$ in the second sum and in the second line we changed the order of the sums.
		
		Consider now the point at infinity. We have
		\begin{align*}
		\iota_{\mu_\infty} \frac{X^\infty(\mu_\infty)}{\mu - \lambda} &= \sum_{n=-N_\infty}^\infty X^\infty_n \mu_\infty^n \sum_{s=0}^\infty \mu_\infty^{s+1} \lambda_\infty^{-s}
		= \sum_{n=-N_\infty}^\infty \sum_{r=n}^\infty X^\infty_n \mu_\infty^{r+1} \lambda_\infty^{n-r}\\
		&= \sum_{r=-N_\infty}^\infty \mu_\infty^{r+1} \sum_{n=-N_\infty}^r X^\infty_n \lambda_\infty^{n-r}
		= \sum_{r=-N_\infty}^\infty \mu_\infty^{r+1} \big( \lambda_\infty^{-r} X^\infty(\lambda_\infty) \big)^{\rm rat}_-
		\end{align*}
		where in the second equality we changed variables $s = r-n$ as before and in the second line we changed the order of the sums.
	\end{proof}
\end{lemma}

\subsection{Trigonometric $r$-matrix} \label{sec: trigonometric r}

Throughout this section we will choose $k=-1$ in the bilinear form \eqref{bilinear form}.
We shall also make use of the standard nilpotent subalgebras $\n_\pm$ and Borel subalgebras $\b_\pm$ of $\g$. Explicitly, $\n_+$ (resp. $\n_-$) is spanned by $E_{ij}$ for $i < j$ (resp. $i > j$). In the $\gl_N$ case $\b_+$ (resp. $\b_-$) is spanned by $E_{ij}$ for $i \leq j$ (resp. $i \geq j$) while in the $\sl_N$ case $\b_+$ (resp. $\b_-$) is spanned by $E_{ij}$ for $i < j$ (resp. $i > j$) together with $E_{ii} - E_{jj}$ for $i < j$. The Cartan subalgebra $\h$ is spanned by $E_{ii}$ for $i = 1, \ldots, N$ in the $\gl_N$ case and by $E_{ii} - E_{jj}$ for $i < j$ in the $\sl_N$ case. We have the direct sum decompositions $\b_\pm = \h \oplus \n_\pm$. We shall also make use of the corresponding subgroups $N_\pm$, $B_\pm$ and $H$ in $G$. For $GL_N$ these are the groups of unipotent upper/lower-triangular $N \times N$ matrices, invertible upper/lower-triangular $N \times N$ matrices and invertible diagonal $N \times N$ matrices, respectively. For $SL_N$ we add the condition that the matrices are of determinant $1$.

Recall the notation $P_{12}$ for the tensor Casimir of $\g$ from Section \ref{sec: gen setup}. We can split this into three parts as $P_{12} = P^-_{12} + P^0_{12} + P^+_{12}$ where $P^\pm_{12} \in \n_\pm \otimes \n_\mp$ and $P^0_{12} \in \h \otimes \h$. Explicitly, in the $\gl_N$ case these read
\begin{equation*}
P^+_{12} = \sum_{\substack{i,j=1\\i < j}}^N E_{ij} \otimes E_{ji}, \qquad
P^0_{12} = \sum_{i=1}^N E_{ii} \otimes E_{ii}, \qquad
P^-_{12} = \sum_{\substack{i,j=1\\i < j}}^N E_{ji} \otimes E_{ij}.
\end{equation*}
For $\sl_N$ the expression for $P^0_{12}$ is given in terms of dual bases $\{ u^i \}$ and $\{ u_i \}$ of the Cartan subalgebra $\h$ with respect to the trace bilinear form as $P^0_{12} = \sum_{i=1}^{N-1} u^i \otimes u_i$. We note that $P^+_{21} = P^-_{12}$, $P^0_{21} = P^0_{12}$ and $P_{21} = P_{12}$. We also define the corresponding projectors $P^\pm : \g \to \n_\pm$ and $P^0 : \g \to \h$ onto the nilpotent Lie subalgebras $\n_\pm$ and the Cartan subalgebra $\h$, respectively, given for any $X \in \g$ as
\begin{equation*}
P^\pm X \coloneqq \Tr_2(P^\pm_{12} X_2), \qquad P^0 X \coloneqq \Tr_2(P^0_{12} X_2),
\end{equation*}
so that $\id_{\g} = P^- + P^0 + P^+$.

In the trigonometric setting, the role of the Lie subalgebra $\bm\A^{\rm rat}_{\bm\lambda}(\g) \subset \bm\A_{\bm\lambda}(\g)$ in \eqref{Ap glN def} will be played by the following alternative Lie subalgebra
\begin{equation} \label{Ap glN def trig}
\bm\A^{\rm trig}_{\bm\lambda}(\g) \coloneqq \bm{\mathcal B}_{\bm \lambda}^{0, \infty}(\g) \times \coprod_{a \in \CC^\times} \g \otimes \CC\bb{\lambda_a}
\end{equation}
where $\CC^\times \coloneqq \CC \setminus \{0\}$ and
\begin{equation*}
\bm{\mathcal B}_{\bm \lambda}^{0, \infty}(\g) \subset \big( \b_+ \oplus \g \otimes \lambda \CC\bb{\lambda} \big) \times \big( \b_- \oplus \g \otimes \lambda_\infty \CC\bb{\lambda_\infty} \big)
\end{equation*}
is the Lie subalgebra consisting of pairs of Taylor series $X^0(\lambda) = \sum_{n=0}^\infty X^0_n \lambda^n$ and $X^\infty(\lambda_\infty) = \sum_{n=0}^\infty X^\infty_n \lambda_\infty^n$ with $X^0_n, X^\infty_n \in \g$ for all $n \geq 1$ but with $X^0_0 \in \b_+$ and $X^\infty_0 \in \b_-$ subject to the constraint $P^0 X^0_0 = - P^0 X^\infty_0$.
We shall also need the corresponding group $\bm\A^{\rm trig}_{\bm\lambda}(G)$ defined as follows.

In the $GL_N$ case we let $\widehat{B}_+$ denote the group of all invertible $N \times N$ matrices with entries below the diagonal in $\lambda \CC\bb{\lambda}$ and entries on or above the diagonal in $\CC\bb{\lambda}$. Likewise, we let $\widehat{B}_-$ be the group of all invertible $N \times N$ matrices with entries on or below the diagonal in $\CC\bb{\lambda_\infty}$ and entries above the diagonal in $\lambda_\infty \CC\bb{\lambda_\infty}$. Concretely, an element of $\widehat{B}_+$ can be expanded as a Taylor series $g(\lambda) = \sum_{n=0}^\infty g_n \lambda^n$ with $g_0$ upper triangular and $g_n \in \gl_N$ for $n \geq 1$, while an element of $\widehat{B}_-$ is a Taylor series $h(\lambda_\infty) = \sum_{n=0}^\infty h_n \lambda_\infty^n$ with $h_0$ lower triangular and $h_n \in \gl_N$ for $n \geq 1$. As usual, in the $SL_N$ case we define the subgroups $\widehat{B}_\pm$ as in the $GL_N$ case but with the added condition that the matrices are of determinant $1$.
We then set
\begin{equation} \label{Ap GLN def trig}
\bm\A^{\rm trig}_{\bm\lambda}(G) \coloneqq \bm{\mathcal B}_{\bm \lambda}^{0, \infty}(G) \times \coprod_{a \in \CC} \widehat{G}_a
\end{equation}
where the first factor is the subgroup
$\bm{\mathcal B}_{\bm \lambda}^{0, \infty}(GL_N) \subset \widehat{B}_+ \times \widehat{B}_-$
consisting of pairs of Taylor series $g^0(\lambda) = \sum_{n=0}^\infty g^0_n \lambda^n$ and $g^\infty(\lambda_\infty) = \sum_{n=0}^\infty g^\infty_n \lambda_\infty^n$ with $g^0_n, g^\infty_n \in \gl_N$ for all $n \geq 1$ but where the upper triangular matrix $g^0_0$ and the lower triangular matrix $g^\infty_0$ are subject to the constraint $P^0 g^0_0 = (P^0 g^\infty_0)^{-1}$.

Note that for consistency we should really keep denoting the local coordinate at the origin as $\lambda_0$, following the general notation introduced in Section \ref{sec: gen setup}. However, since $\lambda_0$ is nothing but $\lambda$, we will most often prefer to write the local coordinate at the origin simply as $\lambda$, rather than use the more cumbersome notation $\lambda_0$.

It will be convenient in what follows to introduce slightly different notions of pole parts of Laurent series at the origin and infinity in the trigonometric case. As they are important in practical calculations, we gather them in the following definition. 
\begin{definition}
Given any $\displaystyle X^0(\lambda) = \sum_{n=-N_0}^\infty X^0_n \lambda^n \in \g \otimes \CC\lau{\lambda}$ we define
	\begin{subequations} \label{trig pole parts}
		\begin{equation} \label{trig pole part 0}
		X^0(\lambda)^{\rm trig}_- \coloneqq \big( P^- + \tfrac 12 P^0 \big) X^0_0 + X^0(\lambda)^{\rm rat}_-
		\in \b_- \oplus \g \otimes \lambda^{-1} \CC[\lambda^{-1}].
		\end{equation}
		Similarly, for any $\displaystyle X^\infty(\lambda^{-1}) = \sum_{n=-N_\infty}^\infty X^\infty_n \lambda^{-n} \in \g \otimes \CC\lau{\lambda^{-1}}$ we define
		\begin{equation} \label{trig pole part inf}
		X^\infty(\lambda_\infty)^{\rm trig}_- \coloneqq \big( P^+ + \tfrac 12 P^0 \big) X^\infty_0 + \sum_{n=-N_\infty}^{-1} X^\infty_n \lambda_\infty^n \in \b_+ \oplus \g \otimes \lambda_\infty^{-1} \CC[\lambda_\infty^{-1}].
		\end{equation}
		Furthermore, for a Laurent series $\displaystyle X^b(\lambda_b) = \sum_{n=-N_b}^\infty X^b_n \lambda^n_b \in \g \otimes \CC\lau{\lambda_b}$ at any other point $b \in \CC^\times$ we set
		\begin{equation} \label{trig pole part b}
		X^b(\lambda_b)^{\rm trig}_- \coloneqq - \big( P^- + \tfrac 12 P^0 \big) X^b(-b)^{\rm rat}_- + X^b(\lambda_b)^{\rm rat}_- \in \b_- \oplus \g \otimes \lambda_b^{-1} \CC[\lambda_b^{-1}],
		\end{equation}
		where in the first term $X^b(-b)^{\rm rat}_-$ is the pole part $X^b(\lambda_b)^{\rm rat}_-$ at $b$ evaluated at $\lambda = 0$.
		In particular, as compared to the pole part $X^b(\lambda_b)^{\rm rat}_- \in \g \otimes \lambda_b^{-1} \CC[\lambda_b^{-1}]$ introduced in \eqref{rat pole part a}, we note that the pole part $X^b(\lambda_b)^{\rm trig}_-$ includes a constant term (provided that $( P^- + \tfrac 12 P^0) X^b(-b)^{\rm rat}_- \neq 0$) which, moreover, is valued in $\b_-$.
	\end{subequations}
\end{definition}

\begin{proposition} \label{prop: Atrig complementary}
	The Lie subalgebra $\bm\A^{\rm trig}_{\bm\lambda}(\g) \subset \bm\A_{\bm\lambda}(\g)$ is maximally isotropic with respect to
	$\langle\!\langle \cdot, \cdot \rangle\!\rangle_{-1}$. Moreover, we have a direct sum of vector spaces
	\begin{equation} \label{trig decomp}
	\bm\A_{\bm\lambda}(\g) = \bm\A^{\rm trig}_{\bm\lambda}(\g) \dotplus \bm \iota_{\bm\lambda} R_\lambda(\g)
	\end{equation}
	into complementary Lagrangian (\emph{i.e.} maximal isotropic) Lie subalgebras.
	\begin{proof}
		To see that $\bm\A^{\rm trig}_{\bm\lambda}(\g)$ is isotropic with respect to the bilinear form $\langle\!\langle \cdot, \cdot \rangle\!\rangle_{-1}$, let $\bm X(\bm \lambda), \bm Y(\bm \lambda) \in \bm\A^{\rm trig}_{\bm\lambda}(\g)$ be arbitrary and consider the pairing $\langle\!\langle \bm X(\bm \lambda), \bm Y(\bm \lambda) \rangle\!\rangle_{-1}$ as given in \eqref{bilinear form b}. There are no contributions from any $a \in \CC^\times$. The only contributions come from $0$ and $\infty$, which read
		\begin{align*}
		&\res^\lambda_0 \Tr \big( X^0(\lambda) Y^0(\lambda) \big) \lambda^{-1} d\lambda + \res^\lambda_\infty \Tr \big( X^\infty(\lambda_\infty) Y^\infty(\lambda_\infty) \big) \lambda^{-1} d\lambda\\
		&\qquad = \Tr(X^0_0 Y^0_0) - \Tr(X^\infty_0 Y^\infty_0)\\
		&\qquad = \Tr\big( P^0(X^0_0) P^0(Y^0_0) \big) - \Tr\big( P^0(X^\infty_0) P^0(Y^\infty_0) \big) = 0.
		\end{align*}
		In the first equality we wrote $X^0(\lambda) = \sum_{n=0}^\infty X^0_n \lambda^n$, $X^\infty(\lambda_\infty) = \sum_{n=0}^\infty X^\infty_n \lambda_\infty^n$ and similarly for $Y^0(\lambda)$ and $Y^\infty(\lambda_\infty)$. The second equality above follows from the fact that $X^0_0, Y^0_0 \in \b_+$ and $X^\infty_0, Y^\infty_0 \in \b_-$ and the last step makes use of the conditions in the definition of $\bm{\mathcal B}_{\bm \lambda}^{0, \infty}(\g)$ that $P^0 X^0_0 = - P^0 X^\infty_0$ and $P^0 Y^0_0 = - P^0 Y^\infty_0$.
		In order to show that $\bm\A^{\rm trig}_{\bm\lambda}(\g)$ is maximally isotropic it suffices to prove the second statement, namely that we have the direct sum decomposition of vector spaces as in \eqref{trig decomp}.
		
		To any $\bm X(\bm \lambda) \in \bm\A_{\bm\lambda}(\g)$ we associate the rational function
		\begin{equation} \label{fX expression}
		f_X(\lambda) = X^0(\lambda)^{\rm trig}_- + \sum_{b \in \CC^\times} X^b(\lambda_b)^{\rm trig}_- + X^\infty(\lambda_\infty)^{\rm trig}_-
		\end{equation}
		in $R_\lambda(\g)$. Consider the element $\widetilde{\bm X}(\bm \lambda) = (\widetilde X^a(\lambda_a))_{a \in \CP}$ defined by
		\begin{equation*}
		\widetilde X^a(\lambda_a) \coloneqq X^a(\lambda_a) - \iota_{\lambda_a} f_X(\lambda)
		\end{equation*}
		for every $a \in \CP$. We have $\widetilde X^a(\lambda_a) \in \g \otimes \CC\bb{\lambda_a}$ for every $a \in \CP$. But more precisely, noting that
		\begin{equation*}
		X^b(\lambda_b)^{\rm trig}_- \big|_{\lambda = 0} = \big( P^+ + \tfrac 12 P^0 \big) X^b(-b)^{\rm rat}_-, \qquad
		X^\infty(\lambda_\infty)^{\rm trig}_- \big|_{\lambda = 0}  = \big( P^+ + \tfrac 12 P^0 \big) X^\infty_0
		\end{equation*}
		for every $b \in \CC^\times$, we have, in fact, $\widetilde X^0(\lambda_0) \in \b_+ \oplus \g \otimes \lambda \CC\bb{\lambda}$ whose leading term in $\b_+$ is given by
		\begin{equation} \label{const piece at 0}
		\big( P^+ + \tfrac 12 P^0 \big) \big( X^0_0 - X^\infty_0 - X^b(-b)^{\rm rat}_- \big) \in \b_+.
		\end{equation}
		Likewise, we have
		\begin{align*}
		X^b(\lambda_b)^{\rm trig}_- \big|_{\lambda = \infty} &= - \big( P^- + \tfrac 12 P^0 \big) X^b(-b)^{\rm rat}_-, \\
		X^0(\lambda)^{\rm trig}_- \big|_{\lambda = \infty}  &= \big( P^- + \tfrac 12 P^0 \big) X^0_0
		\end{align*}
		from which it follows that $\widetilde X^\infty(\lambda_\infty) \in \b_- \oplus \g \otimes \lambda_\infty \CC\bb{\lambda_\infty}$ with leading coefficient in $\b_-$ given by
		\begin{equation} \label{const piece at inf}
		\big( P^- + \tfrac 12 P^0 \big) \big( - X^0_0 + X^\infty_0 + X^b(-b)^{\rm rat}_- \big) \in \b_-.
		\end{equation}
		Moreover, comparing the Cartan components of \eqref{const piece at 0} and \eqref{const piece at inf} we see that these are opposite. Hence we conclude that $\widetilde{\bm X}(\bm \lambda) \in \bm\A^{\rm trig}_{\bm\lambda}(\g)$. In other words,
		\begin{equation*}
		\bm X(\bm \lambda) = \widetilde{\bm X}(\bm \lambda) + \bm \iota_{\bm\lambda} f_X(\lambda)
		\end{equation*}
		gives the desired decomposition of a general element $\bm X(\bm \lambda) \in \bm\A_{\bm\lambda}(\g)$ as in \eqref{trig decomp}.
		
		This decomposition is unique since any element which belongs to both $\bm\A^{\rm trig}_{\bm\lambda}(\g)$ and $\bm \iota_{\bm\lambda} R_\lambda(\g)$ must vanish. Indeed, suppose $f(\lambda) \in R_\lambda(\g)$ is such that $\bm \iota_{\bm \lambda} f(\lambda) \in \bm\A^{\rm trig}_{\bm\lambda}(\g)$. Then it is clear from the definition of $\bm\A^{\rm trig}_{\bm \lambda}(\g)$ in \eqref{Ap glN def trig} that $f(\lambda)$ cannot be singular at any point in $\CP$ and so must be constant. But then it follows from the definition of $\bm{\mathcal B}_{\bm \lambda}^{0, \infty}(\g)$ that this constant must in fact be zero.
	\end{proof}
\end{proposition}
\begin{definition}[{\bf Trigonometric $r$-matrix}]
The trigonometric $r$-matrix is defined as the following function of two formal variables $\lambda$ and $\mu$:
\begin{equation} \label{trig kernel}
r^{\rm trig}_{12}(\lambda, \mu) = \frac{1}{2} \bigg( P^+_{12} - P^-_{12} + \frac{\mu + \lambda}{\mu - \lambda} P_{12} \bigg) = \frac{\mu P_{12}}{\mu - \lambda} - P^-_{12} - \tfrac 12 P^0_{12}.
\end{equation}	
\end{definition}
It is skew-symmetric, namely we have $r^{\rm trig}_{21}(\mu, \lambda) = - r^{\rm trig}_{12}(\lambda, \mu)$.
It provides the trigonometric counterpart of the kernel \eqref{def_rat_kernel} for the choice of complement \eqref{Ap glN def trig}. Indeed, we have the following analogue of Proposition \ref{prop: Ppm kernels} in the trigonometric case. 

\begin{proposition} \label{prop: Ppm kernels trig}
	For any $\bm X(\bm \lambda) \in \bm\A_{\bm\lambda}(\g)$, its projections onto the complementary subalgebras $\bm\A^{\rm trig}_{\bm\lambda}(\g)$ and $\bm \iota_{\bm\lambda} R_\lambda(\g)$ relative to the direct sum decomposition \eqref{trig decomp} are given respectively by $\pi^{\rm trig}_+ \bm X(\bm \lambda) = \big( (\pi^{\rm trig}_+ X)^a(\lambda_a) \big)_{a \in \CP}$ and $\pi^{\rm trig}_- \bm X(\bm \lambda) = \big( (\pi^{\rm trig}_- X)^a(\lambda_a) \big)_{a \in \CP}$ where
	\begin{subequations} \label{P+ P- expressions}
		\begin{align}
		\label{pi+ expression}
		(\pi^{\rm trig}_+ X)^a(\lambda_a) &= \sum_{b \in \CP} \res^\mu_b \Tr_2 \big( \iota_{\mu_b} \iota_{\lambda_a} r^{\rm trig}_{12}(\lambda, \mu) X^b(\mu_b)_2 \big) \mu^{-1} d\mu,\\
		\label{pi- expression}
		(\pi^{\rm trig}_- X)^a(\lambda_a) &= - \sum_{b \in \CP} \res^\mu_b \Tr_2 \big( \iota_{\lambda_a} \iota_{\mu_b} r^{\rm trig}_{12}(\lambda, \mu) X^b(\mu_b)_2 \big) \mu^{-1} d\mu.
		\end{align}
	\end{subequations}
	\begin{proof}
				We first describe the image of $\pi^{\rm trig}_-$ explicitly. Then we show that $\pi^{\rm trig}_-$ sends $\bm\A^{\rm trig}_{\bm\lambda}(\g)$ to zero and that it acts as the identity on $\bm \iota_{\bm\lambda} R_\lambda(\g)$. Hence, $\pi^{\rm trig}_-$ is the projection onto $\bm \iota_{\bm\lambda} R_\lambda(\g)$ along $\bm\A^{\rm trig}_{\bm\lambda}(\g)$. Similarly, we prove that $\pi^{\rm trig}_+$ is the projection onto $\bm\A^{\rm trig}_{\bm\lambda}(\g)$ along $\bm \iota_{\bm\lambda} R_\lambda(\g)$. 
				
				Given any $\bm X(\bm \lambda) \in \bm\A_{\bm\lambda}(\g)$,
		consider the $\g$-valued rational function
		\begin{align*}
		f_X(\lambda) &= - \sum_{b \in \CP} \res^\mu_b \Tr_2 \big( \iota_{\mu_b} r^{\rm trig}_{12}(\lambda, \mu) X^b(\mu_b)_2 \big) \mu^{-1} d\mu\\
		&= \sum_{b \in \CP} \res^\mu_b \bigg( \iota_{\mu_b} \mu^{-1} \big( P^- + \tfrac 12 P^0\big)(X^b(\mu_b)) + \iota_{\mu_b} \frac{1}{\lambda - \mu} X^b(\mu_b) \bigg) d\mu.
		\end{align*}
		We compute the residues at each $b \in \CC^\times$ and then at the origin and infinity. Firstly, for the residue at $b \in \CC^\times$ we find $X^b(\lambda_b)^{\rm trig}_-$.
		For the residue at the origin we find
		$X^0(\lambda)^{\rm trig}_-$
		and, likewise, for the residue at infinity we find
		\begin{equation*}
		- \big( P^- + \tfrac 12 P^0 \big) X^\infty_0 + X^\infty(\lambda_\infty)^{\rm rat}_-
		= X^\infty(\lambda_\infty)^{\rm trig}_-,
		\end{equation*}
		where in the first expression we are using the pole part $X^\infty(\lambda_\infty)^{\rm rat}_- \in \g \otimes \CC[\lambda_\infty^{-1}]$ defined in \eqref{rat pole part inf} of a Laurent series at infinity, and in the second expression we are using the other notion of pole part $X^\infty(\lambda_\infty)^{\rm trig}_-$ introduced above in \eqref{trig pole part inf}. Putting the above together we conclude that $f_X(\lambda)$ is the rational function \eqref{fX expression} used in the proof of Proposition \ref{prop: Atrig complementary}.
		By construction we have $(\pi^{\rm trig}_- X)^a(\lambda_a) = \iota_{\lambda_a} f_X(\lambda)$ for every $a \in \CP$.
		
		Now suppose $\bm X(\bm \lambda) \in \bm\A^{\rm trig}_{\bm\lambda}(\g)$. Clearly $X^b(\lambda_b)^{\rm rat}_- = 0$, hence also $X^b(\lambda_b)^{\rm trig}_- = 0$ using the definition \eqref{trig pole part b}, so the sum over $b \in \CC^\times$ on the right hand side of \eqref{fX expression} vanishes. On the other hand, $X^0(\lambda)^{\rm trig}_- = \tfrac 12 P^0 X^0_0$ and $X^\infty(\lambda_\infty)^{\rm trig}_- = \tfrac 12 P^0 X^\infty_0$. But since $P^0 X^0_0 = - P^0 X^\infty_0$ by definition of $\bm X(\bm \lambda)$ belonging to $\bm\A^{\rm trig}_{\bm\lambda}(\g)$, it follows that the remaining two terms in \eqref{fX expression} cancel. So we have shown that $\pi^{\rm trig}_- \bm X(\bm \lambda) = 0$ for any $\bm X(\bm \lambda) \in \bm\A^{\rm trig}_{\bm\lambda}(\g)$.
		
		On the other hand, suppose now that $\bm X(\bm \lambda) = \bm \iota_{\bm\lambda} f(\lambda)$ for some $f(\lambda) \in R_\lambda(\g)$. If the latter has a pole at some $a \in \CC^\times$ then its pole part there is given by $X^a(\lambda_a)^{\rm rat}_-$. If it has a pole at the origin then its pole part there is equal to
		\begin{equation} \label{pole part at 0 trig}
		X^0(\lambda)^{\rm trig}_- - \sum_{b \in \CC^\times} \big( P^- + \tfrac 12 P^0 \big) X^b(-b)^{\rm rat}_- - \big( P^- + \tfrac 12 P^0 \big) X^\infty_0,
		\end{equation}
		where $X^\infty_0$ is the constant term in the expansion of $f(\lambda)$ at infinity. Indeed, recall from \eqref{trig pole part 0} that $X^0(\lambda)^{\rm trig}_-$ is given by the pole part $X^0(\lambda)^{\rm rat}_-$ at the origin plus $(P^- + \tfrac 12 P^0) X^0_0$ where $X^0_0$ is given here by the value at the origin of all the other pole parts of $f(\lambda)$. This is why we must subtract the latter from $X^0(\lambda)^{\rm trig}_-$ in \eqref{pole part at 0 trig} to be left only with the desired pole part at the origin. Finally, the pole part of $f(\lambda)$ at infinity is given by
		\begin{equation} \label{pole part at inf trig}
		X^\infty(\lambda_\infty)^{\rm trig}_- + \big( P^- + \tfrac 12 P^0 \big) X^\infty_0.
		\end{equation}
		Indeed, the pole part at infinity should contain the constant term but $X^\infty(\lambda_\infty)^{\rm trig}_-$ only contains part of it. The remaining part is precisely the piece added in \eqref{pole part at inf trig}. It now follows that the expression on the right hand side of \eqref{fX expression} built from $\bm X(\bm \lambda) = \bm \iota_{\bm\lambda} f(\lambda)$ coincides exactly with the partial fraction decomposition of $f(\lambda)$. This establishes that $\pi^{\rm trig}_- \bm X(\bm \lambda) = \bm X(\bm \lambda)$ for any $\bm X(\bm \lambda) \in \bm \iota_{\bm\lambda} f(\lambda)$. In other words, we have therefore shown that $\pi^{\rm trig}_-$ is indeed the projection onto $\bm \iota_{\bm\lambda} R_\lambda(\g)$ along $\bm\A^{\rm trig}_{\bm\lambda}(\g)$.
		
		It remains to consider $\pi^{\rm trig}_+$. For any $\bm X(\bm \lambda) \in \bm\A_{\bm\lambda}(\g)$ and $a \in \CC$ we have
		\begin{align} \label{pi plus expression trig}
		(\pi^{\rm trig}_+ X)^a(\lambda_a) &= - \sum_{b \in \CP} \res^\mu_b \Big( \iota_{\mu_b} \mu^{-1} \big( P^- + \tfrac 12 P^0\big)(X^b(\mu_b))  \Big) d\mu \notag\\
		&\qquad\qquad + \sum_{n=0}^\infty \lambda_a^n \sum_{b \in \CP} \res^\mu_b \iota_{\mu_b} \mu_a^{-n-1} X^b(\mu_b) d\mu.
		\end{align}
		If $\bm X(\bm \lambda) \in \bm \iota_{\bm\lambda} R_\lambda(\g)$ then the first term on the right hand side vanishes by the residue theorem and the second term likewise at each order in the $\lambda_a$-expansion. If instead we consider $a = \infty$ then
		\begin{align} \label{pi plus inf expression trig}
		(\pi^{\rm trig}_+ X)^\infty(\lambda_\infty) &= - \sum_{b \in \CP} \res^\mu_b \Big( \iota_{\mu_b} \mu^{-1} \big( P^- + \tfrac 12 P^0\big)(X^b(\mu_b))  \Big) d\mu \notag\\
		&\qquad\qquad - \sum_{n=0}^\infty \lambda^{-n-1} \sum_{b \in \CP} \res^\mu_b \iota_{\mu_b} \mu^n X^b(\mu_b) d\mu,
		\end{align}
		but both terms vanish once again by the residue theorem if $\bm X(\bm \lambda) \in \bm \iota_{\bm\lambda} R_\lambda(\g)$. So we deduce that $\pi^{\rm trig}_+ \bm X(\bm \lambda) = 0$ for every $\bm X(\bm \lambda) \in \bm \iota_{\bm \lambda} R_\lambda(\g)$.
		
		Suppose now that $\bm X(\bm \lambda) \in \bm\A^{\rm trig}_{\bm\lambda}(\g)$. The first term on the right hand side of \eqref{pi plus expression trig} gets a contribution only from the terms $b=0$ and $b=\infty$, which read
		\begin{align*}
		&- \res^\mu_0 \Big( \mu^{-1} \big( P^- + \tfrac 12 P^0\big)(X^0(\mu))  \Big) d\mu - \res^\mu_\infty \Big( \mu^{-1} \big( P^- + \tfrac 12 P^0\big)(X^\infty(\mu_\infty))  \Big) d\mu\\
		&\qquad = - \big( P^- + \tfrac 12 P^0\big) X^0_0 + \big( P^- + \tfrac 12 P^0\big) X^\infty_0 = \big( P^- + P^0\big) X^\infty_0 = X^\infty_0
		\end{align*}
		where we wrote $X^0(\mu) = \sum_{n=0}^\infty X^0_n \mu^n$ and $X^\infty(\mu_\infty) = \sum_{n=0}^\infty X^\infty_n \mu_\infty^n$. The second equality above follows since by assumption we have $X^0_0 \in \b_+$ so that $P^- X^0_0 = 0$ and also $P^0 X^0_0 = - P^0 X^\infty_0$. The third equality also follows since by assumption $X^\infty_0 \in \b_-$. The second sum in \eqref{pi plus expression trig} is just as in the rational case, however since the series at infinity now contains a constant term we get a contribution to the sum over $b \in \CP$ from both $b=a$ and $b = \infty$, yielding $X^a(\lambda_a) - X^\infty_0$. So in total, we deduce that $(\pi^{\rm trig}_+ X)^a(\lambda_a) = X^a(\lambda_a)$ for every $a \in \CC$.
		
		Consider now the case $a = \infty$. The first term on the right hand side of \eqref{pi plus inf expression trig} is again equal to $X^\infty_0$ while the second term gives $\displaystyle \sum_{n=1}^\infty X^\infty_n \lambda_\infty^n = X^\infty(\lambda_\infty) - X^\infty_0$. Putting these together we deduce that $(\pi^{\rm trig}_+ X)^\infty(\lambda_\infty) = X^\infty(\lambda_\infty)$. In conclusion, we have shown that $\pi^{\rm trig}_+ \bm X(\bm \lambda) = \bm X(\bm \lambda)$ for all $\bm X(\bm \lambda) \in \bm\A^{\rm trig}_{\bm\lambda}(\g)$ so that $\pi^{\rm trig}_+$ is the projection onto $\bm\A^{\rm trig}_{\bm\lambda}(\g)$ along $\bm \iota_{\bm\lambda} R_\lambda(\g)$, as claimed.
	\end{proof}
\end{proposition}

We can now define the linear operator $r^{\rm trig} \coloneqq \pi^{\rm trig}_+ - \pi^{\rm trig}_-$. It follows from Proposition \ref{prop: Ppm kernels trig} that its kernel reads
\begin{equation}
\big( (\iota_{\mu_b} \iota_{\lambda_a} + \iota_{\lambda_a} \iota_{\mu_b}) r^{\rm trig}_{12}(\lambda, \mu) \big)_{a, b \in \CP}.
\end{equation}
Moreover, the kernel of the identity operator $\id = \pi^{\rm trig}_+ + \pi^{\rm trig}_-$ is similarly given by an expansion of zero since
\begin{equation}
	\label{delta_from_r_matrix2}
\big( (\iota_{\mu_b} \iota_{\lambda_a} - \iota_{\lambda_a} \iota_{\mu_b}) r^{\rm trig}_{12}(\lambda, \mu) \mu^{-1} d\mu \big)_{a, b \in \CP} = \big( P_{12} \delta_{ab} \delta(\lambda_a, \mu_a) d\mu_a \big)_{a, b \in \CP}
\end{equation}
using the same notation $\delta_{ab}$ and $\delta(\lambda, \mu)$ as in the rational case.

The following is the analogue of Lemma \ref{lemma_expansion} in the trigonometric case.
\begin{lemma}\label{lemma_expansion trig}
	Let $\bm X(\bm \mu) = \big( X^a(\mu_a) \big)_{a \in \CP} \in \bm \A_{\bm \mu}(\g)$ with $X^a(\mu_a) = \sum_{n=-N_a}^\infty X^a_n \mu_a^n$ for some $N_a \in \ZZ$, where $N_a > 0$ for finitely many $a \in \CP$. For any $a \in \CC$ we have
	\begin{equation*}
	\iota_{\mu_a} \Tr_2 \big( r^{\rm trig}_{12}(\lambda, \mu) X^a(\mu_a)_2 \big) = - \sum_{r=-N_a}^\infty \mu_a^r \big( (\lambda_a^{-r} + a \lambda_a^{-r-1}) X^a(\lambda_a) \big)^{\rm trig}_-,
	\end{equation*}
	while at infinity we have
	\begin{equation*}
	\iota_{\mu_\infty} \Tr_2 \big( r^{\rm trig}_{12}(\lambda, \mu) X^\infty(\mu_\infty)_2 \big) = \sum_{r=-N_\infty}^\infty \mu_\infty^r \big( \lambda_\infty^{-r} X^\infty(\lambda_\infty) \big)^{\rm trig}_-.
	\end{equation*}
	\begin{proof}
		First, let $a \in \CC^\times$. We have
		\begin{align*}
		&\iota_{\mu_a} \Tr_2 \big( r^{\rm trig}_{12}(\lambda, \mu) X^a(\mu_a)_2 \big) = - \big( P^- + \tfrac 12 P^0\big)(X^a(\mu_a)) - \iota_{\mu_a} \frac{\mu}{\lambda - \mu} X^a(\mu_a)\\
		&\qquad\quad = - \sum_{r=-N_a}^\infty \mu_a^r (P^- + \tfrac 12 P^0) X_r^a - \sum_{n=-N_a}^\infty \sum_{s=0}^\infty (\mu_a + a) \frac{\mu_a^{n+s}}{\lambda_a^{s+1}} X_n^a\\
		&\qquad\quad = - \sum_{r=-N_a}^\infty \mu_a^r \bigg( (P^- + \tfrac 12 P^0) X_r^a + \sum_{n=-N_a}^{r-1} \lambda_a^{n-r} X^a_n + a \sum_{n=-N_a}^r \lambda_a^{n-r-1} X^a_n \bigg).
		\end{align*}
		In the third equality we split the double sum into two terms, containing $\mu_a$ and $a$ respectively from the first factor. We changed variable from $s$ to $r = s+n+1$ in the first and from $s$ to $r = s+n$ in the second, and then changed the order of the two double sums. It remains to note that
		\begin{equation*}
		\sum_{n=-N_a}^{r-1} \lambda_a^{n-r} X^a_n + a \sum_{n=-N_a}^r \lambda_a^{n-r-1} X^a_n = \big( (\lambda_a^{-r} + a \lambda_a^{-r-1}) X^a(\lambda_a) \big)^{\rm rat}_-
		\end{equation*}
		and that this is equal to $- X^a_r$ when evaluated at $\lambda = 0$. The result at $a \in \CC^\times$ now follows by definition \eqref{trig pole part b} of the pole part at $a$.
		
		At the origin we have
		\begin{align*}
		&\iota_\mu \Tr_2 \big( r^{\rm trig}_{12}(\lambda, \mu) X^0(\mu)_2 \big)
		= - \sum_{r=-N_0}^\infty \mu^r (P^- + \tfrac 12 P^0) X^0_r - \sum_{n=-N_0}^\infty \sum_{s=1}^\infty \mu^{n+s} \lambda^{-s} X^0_n\\
		&\quad = - \sum_{r=-N_0}^\infty \mu^r \bigg( (P^- + \tfrac 12 P^0) X^0_r + \sum_{n=-N_0}^{r-1} \lambda^{n-r} X^0_n \bigg) = - \sum_{r=-N_0}^\infty \mu^r \big( \lambda^{-r} X^0(\lambda) \big)^{\rm trig}_-.
		\end{align*}
		In the second equality we changed variable in the double sum from $s$ to $r = n+s$ and then changed the order of the two sums. We have also added the term $r=-N_0$ in this double sum since this term vanishes due to the range in the sum over $n$ being empty. The last equality uses the definition \eqref{trig pole part 0}. Note that the result at the origin coincides with the result obtained above for $a \in \CC^\times$ but taken at $a=0$. This is not completely obvious since the definitions of the pole parts \eqref{trig pole part 0} and \eqref{trig pole part b} at $0$ and a generic point $a \in \CC^\times$ are different.
		Likewise, at infinity we have
		\begin{eqnarray*}
		\iota_{\mu_\infty} \Tr_2 \big( r^{\rm trig}_{12}(\lambda, \mu) X^\infty(\mu_\infty)_2 \big)&=& - \sum_{r=-N_\infty}^\infty \mu_\infty^r (P^- + \tfrac 12 P^0) X^\infty_r + \sum_{n=-N_\infty}^\infty \sum_{s=0}^\infty \mu_\infty^{n+s} \lambda_\infty^{-s} X^\infty_n\\
	&=& \sum_{r=-N_\infty}^\infty \mu_\infty^r \bigg( - (P^- + \tfrac 12 P^0) X^\infty_r + \sum_{n=-N_\infty}^r \lambda_\infty^{n-r} X^\infty_n \bigg)\\
		&=& \sum_{r=-N_\infty}^\infty \mu_\infty^r \big( \lambda_\infty^{-r} X^\infty(\lambda_\infty) \big)^{\rm trig}_-.
		\end{eqnarray*}
		In the second equality we changed variable in the double sum from $s$ to $r = n+s$ and then changed the order of the two sums. The last equality uses \eqref{trig pole part inf}.
	\end{proof}
\end{lemma}

\section{Generating Lagrangian multiform and CYBE} \label{sec: Lagrangian}

In this section we will treat uniformly both the rational and trigonometric cases discussed in Section \ref{sec: rational r} and Section \ref{sec: trigonometric r}, respectively. More precisely, we shall work with the Lie algebra of $\g$-valued ad\`eles $\bm\A_\lambda(\g)$ equipped with the bilinear form \eqref{bilinear form} with either $k=0$ or $k=-1$. The corresponding vector space direct sum decompositions \eqref{rational decomp} and \eqref{trig decomp} will be denoted by
\begin{equation*}
\bm\A_{\bm\lambda}(\g) = \bm\A^+_{\bm\lambda}(\g) \dotplus \bm \iota_{\bm\lambda} R_\lambda(\g)
\end{equation*}
where $\bm\A^+_{\bm\lambda}(\g)$ stands for the rational Lie subalgebra $\bm\A^{\rm rat}_{\bm\lambda}(\g)$ when $k=0$ and the trigonometric Lie subalgebra $\bm\A^{\rm trig}_{\bm\lambda}(\g)$ when $k=-1$. Correspondingly, we shall use the common notation $\bm\A^+_{\bm\lambda}(G)$ for the groups $\bm\A^{\rm rat}_{\bm\lambda}(G)$ in the rational case and $\bm\A^{\rm trig}_{\bm\lambda}(G)$ in the trigonometric case.

Given a general element $\bm X(\bm \lambda) = (X^a(\lambda_a))_{a \in \CP} \in \bm\A_{\bm \lambda}(\g)$ of the Lie algebra of $\g$-valued ad\`eles, we will also denote by $X^a(\lambda_a)_-$ the principal part of the formal Laurent series $X^a(\lambda_a)$, which stands for $X^a(\lambda_a)^{\rm rat}_-$ in the rational case, see \eqref{rat pole parts}, or for $X^a(\lambda_a)^{\rm trig}_-$ in the trigonometric case, see \eqref{trig pole parts}.

\subsection{Dynamical equations} \label{sec: coadjoint}

We will describe integrable field theories by taking the point of view that the spatial coordinate $x$, on which all the Hamiltonian fields are usually taken to depend, should be treated on an equal footing to all the other times in the hierarchy. To explain this new perspective on integrable hierarchies it is useful to begin by recalling the traditional point of view.

The dynamical equations of different integrable field theories in the same hierarchy are usually described as zero curvature equations
\begin{equation} \label{traditional ZC}
\partial_x V^a_n(\lambda) - \partial_{t^a_n} U(\lambda) + [V^a_n(\lambda), U(\lambda)] = 0
\end{equation}
where the Lax matrix $U(\lambda) \in R_\lambda(\g)$ is a coadjoint orbit of $\bm\A^+_{\bm \lambda}(G)$ in $R_\lambda(\g)$ which encodes the finite collection of \emph{fields} of the hierarchy. The $V^a_n(\lambda) \in R_\lambda(\g)$, associated to the times $t^a_n$ for some labels $a \in \CC$ and $n \in \ZZ$ to be specified below and which we also refer to as Lax matrices, are coadjoint orbits of $\bm\A^+_{\bm \lambda}(G)$ in $R_\lambda(\g)$ built out of differential polynomials in the fields. From this traditional point of view, \eqref{traditional ZC} represents a set of equations which is seen as a natural extension of the Lax equations $\partial_{t^a_n} L(\lambda) = [M^a_n(\lambda), L(\lambda)]$, used to describe finite-dimensional systems, to the field theory case where every degree of freedom now depends on $x$. In particular, $U(\lambda)$ is usually treated as the fundamental object since the $V^a_n(\lambda)$ can all be built out of it and as such it is seen as the natural analogue of the Lax matrix $L(\lambda)$ in the field theory case.

The crucial point is that the particular flow $\partial_x$ can, and from our point of view \emph{should}, be thought of as a linear combination of some of the {\it elementary} time flows $\partial_{t^a_n}$. But if we are to treat the coadjoint orbit $U(\lambda) \in R_\lambda(\g)$ on an equal footing to all the other coadjoint orbits $V^a_n(\lambda) \in R_\lambda(\g)$ then we should also abandon the idea that each $V^a_n(\lambda)$ is parametrised by differential polynomials with respect to $x$ of the finite collection of fields contained in $U(\lambda)$. Instead, we should treat all the coadjoint orbits $V^a_n(\lambda) \in R_\lambda(\g)$ as truly independent. We shall see, in a sense which is much closer in spirit to the Lax formalism for finite-dimensional systems, that all the Lax matrices $V^a_n(\lambda)$ can be derived from a single object $\bm Q(\bm \lambda) \in \bm \A_{\bm \lambda}(\g)$, a certain \emph{adjoint} orbit of $\bm\A^+_{\bm \lambda}(G)$ in the full space of ad\`eles $\bm \A_{\bm \lambda}(\g)$. In particular, the latter will satisfy a Lax equation (see \eqref{Lax 1 reformulate V} below)
\begin{equation*}
\partial_{t^a_n} \bm Q(\bm \lambda) = [ \bm \iota_{\bm \lambda} V^a_n(\lambda), \bm Q(\bm \lambda) ]
\end{equation*}
with respect to all the times $t^a_n$. As such, in our approach to hierarchies of integrable field theories, $\bm Q(\bm \lambda)$ will play a very similar role to that of the usual Lax matrix $L(\lambda)$ for finite-dimensional systems. For us, the fundamental object will therefore be $\bm Q(\bm \lambda)$ rather than $U(\lambda)$. The relationship between these two objects, and in particular the connection between our approach to hierarchies of integrable field theories and the usual one recalled above, comes from fixing a particular linear combination of time flows as our choice of spatial derivative $\partial_x$. We discuss this in detail in Section \ref{sec: connection to IFT}, together with what we call the FNR procedure. 

Since there is a close parallel between our treatment of integrable field theories and various familiar constructions in the theory of finite-dimensional integrable systems, we will draw the comparison throughout this section in a series of remarks.

\subsubsection{Adjoint orbit}

Let $\bm \phi(\bm \lambda) = (\phi^a(\lambda_a))_{a \in \CP} \in \bm\A^+_{\bm \lambda}(G)$. We regard the entries of the matrix coefficients in the expansions
\begin{equation*}
\phi^a(\lambda_a) = \sum_{n=0}^\infty \phi^a_n \lambda_a^n
\end{equation*}
for all $a \in \CP$ as an infinite collection of dynamical variables.  In general, these are not all independent. For instance, $\phi^a(\lambda_a)$ should be invertible in the $GL_N$ case, which means that the first term $\phi^a_0$ should be invertible, or $\phi^a(\lambda_a)$ should have determinant $1$ in the $SL_N$ case which will impose non-trivial relations between the coefficients at each order in $\lambda_a$. The infinitely many degrees of freedom contained in $\bm \phi(\bm \lambda)$, or equivalently in $\bm Q(\bm{\lda})$ defined in \eqref{Q def} below, will be used to describe infinitely many different integrable hierarchies of integrable field theories. We will refer to these as {\it group or algebra coordinates} (respectively): they represent the dependent variables and are the fields satisfying the equations of motion of a hierarchies.

A particular integrable hierarchy will be determined by a choice of non-dynamical rational function, with poles in a finite subset $S \subset \CP$, which we can write using a partial fraction decomposition as
\begin{equation*}
F(\lambda) = \sum_{a \in S} F^a(\lambda_a)_- \in R_\lambda(\g)
\end{equation*}
where $F^a(\lambda_a)_- \in \g \otimes \CC[\lambda_a^{-1}]$ are (rational or trigonometric, depending on the case) principal parts at each $a \in S$. In particular, $F^a(\lambda_a)_- d\lambda$ has a pole of order $N_a > 0$ at any $a \in S \cap \CC$ and a pole of order $N_\infty + 2 \geq 2$ at infinity if $\infty \in S$.
Its expansion at all of the points $a \in \CP$ defines an element $\bm \iota_{\bm \lambda} F(\lambda) = (\iota_{\lambda_a} F(\lambda))_{a \in \CP} \in \bm\A_{\bm \lambda}(\g)$ of the $\g$-valued ad\`eles, via the embedding \eqref{iota map}. By design, we have $(\iota_{\lambda_a} F(\lambda))_- = F^a(\lambda_a)_-$ for each $a \in S$ and $(\iota_{\lambda_a} F(\lambda))_- = 0$ for every other points $a \in \CP \setminus S$. The element of the $\g$-valued ad\`eles with these components, which we can denote by
\begin{equation} \label{iota F}
\big( \bm \iota_{\bm \lambda} F(\lambda) \big)_- \coloneqq \big( F^a(\lambda_a)_- \big)_{a \in \CP} \in \bm \A_{\bm \lambda}(\g),
\end{equation}
is just a finite collection of principal parts. We consider its adjoint orbit under the group element $\bm \phi(\bm \lambda) \in \bm\A^+_{\bm \lambda}(G)$ introduced above, namely
\begin{equation} \label{Q def}
\bm Q(\bm \lambda) \coloneqq \bm \phi(\bm \lambda) \big( \bm \iota_{\bm \lambda} F(\lambda) \big)_- \bm \phi(\bm \lambda)^{-1} \in \bm\A_{\bm \lambda}(\g).
\end{equation}
Explicitly, its component at any pole $a \in S$ is $Q^a(\lambda_a)=\phi^a(\lambda_a) F^a(\lambda_a)_- \phi^a(\lambda_a)^{-1}$ while the component at any other $a \in \CP \setminus S$ vanishes. We can further expand the latter as a Laurent series in $\lambda_a$, namely
\begin{equation} \label{Qan def}
Q^a(\lambda_a) = \sum_{n=-N_a}^\infty Q^a_n \lambda_a^n,
\end{equation}
for some $Q^a_n \in \g$, where $N_a > 0$ is the order of the pole of $F(\lambda)$ at $a \in S \cap \CC$. For the point at infinity we can have $N_\infty \geq 0$.

\begin{remark} \label{rem: Q vs Lax}
	The adjoint orbit \eqref{Q def} within the full Lie algebra of $\g$-valued ad\`eles $\bm\A_{\bm \lambda}(\g)$ will play the role of the Lax matrix in the present infinite-dimensional setting. For comparison, it is useful to recall that in the finite-dimensional setting the Lax matrix is given by a coadjoint orbit
	\begin{equation} \label{Lax coadjoint orbit}
	L(\lambda) = \Pi_- \Big( \bm \phi(\bm \lambda) \big( \bm \iota_{\bm \lambda} F(\lambda) \big)_- \bm \phi(\bm \lambda)^{-1} \Big) \in R_\lambda(\g)
	\end{equation}
	where $\Pi_-$ denotes either $\pi_-^{\rm rat}$ or $\pi_-^{\rm trig}$, depending on whether we are in the rational or trigonometric setting, but without applying the expansion $\bm\iota_{\bm \lambda}$ so that we obtain an element of $R_\lambda(\g)$ rather than $\bm \iota_{\bm \lambda} R_\lambda(\g)$. In particular, the rational function $L(\lambda)$ depends only on finitely many dynamical variables in $\bm \phi(\bm \lambda) \in \bm\A_{\bm \lambda}(\g)$.
\end{remark}

\subsubsection{Generating Lax equation}\label{sec_gen_Lax}

As our aim is to work with hierarchies of equations of motion, to each point $a\in \CP$ we attach an infinite family of time coordinates $t^a_n$ for $n\in\ZZ$. Related to each time is the usual partial derivative $\partial_{t_n^a}$ (meant as a total derivative when acting on functions of the fields). For our purposes, let us define the following generating operators 
\be
\label{def_gen_deriv}
\cD_{\lambda_a} \coloneqq \sum_{n\in\ZZ} \lambda_a^n \partial_{t^a_n}\,,~~a\in\CC\,, \qquad \cD_{\lambda_\infty} \coloneqq \sum_{n\in\ZZ} \lambda_\infty^{n+k+1} \partial_{t^\infty_n}
\ee
with $k = 0$ in the rational case and $k=-1$ in the trigonometric case. We let $\cD_{\bm \lambda} \coloneqq (\cD_{\lambda_a})_{a \in \CP}$ denote the $\CP$-tuple of these differential operators.
Then, if $\mu$ is another formal variable we will use the notation
\begin{equation}
\label{def_D_on_Q}
\cD_{\bm \lambda} \bm Q(\bm \mu) = \bigg( \cD_{\lambda_a}Q^b(\mu_b) \bigg)_{a, b \in \CP}
\end{equation}
which encodes the flows $\partial_{t^a_m} Q^b_n$ of all the dynamical variables $Q^b_n$ with respect to all the times $t^a_m$ for each pair of points $a, b \in \CP$.

Following the first observation in Section \ref{motivation} of the introduction, we want to declare the evolution of $\bm Q(\bm \lambda) \in \bm \A_{\bm \lambda}(\g)$ with respect to the above infinite family of times $t^a_n$ to be governed by the following general Lax equation in $r$-matrix and generating form
\begin{equation} \label{Lax 1}
\cD_{\bm \mu} \bm Q_1(\bm \lambda) = \big[ \Tr_2 \big( \bm \iota_{\bm \lambda} \bm \iota_{\bm \mu} r_{12}(\lambda,\mu) \bm Q_2(\bm \mu) \big), \bm Q_1(\bm \lambda) \big].
\end{equation}
However, a few comments and precautions are necessary. First, writing such an equation with the understanding that $\cD_{\bm \mu}$ is the $\CP$-tuple of commuting differential operators defined in \eqref{def_gen_deriv} assumes that the vector fields on the right-hand side commute, if we want to be able to interpret the times $t_n^a$ as coordinates on a manifold. In other words, defining the generating vector ${\cal X}_{\bm \mu}$ acting on $\bm \A_{\bm \lambda}(\g)$ by 
\be
\label{def_gen_vf}
{\cal X}_{\bm \mu} \bm Q_1(\bm \lambda) = \big[ \Tr_2 \big( \bm \iota_{\bm \lambda} \bm \iota_{\bm \mu} r_{12}(\lambda,\mu) \bm Q_2(\bm \mu) \big), \bm Q_1(\bm \lambda) \big]\,,
\ee 
we must first prove that $[{\cal X}_{\bm \mu},{\cal X}_{\bm \nu}]=0$. Only then can we set ${\cal X}_{\bm \mu}=\cD_{\bm \mu}$ and view the generating Lax equation \eqref{Lax 1} as describing compatible time flows on $\bm \A_{\bm \lambda}(\g)$.
This is shown below in Proposition \ref{commuting_vf} and is a beautiful consequence of the CYBE for $r$.

Second, note that the right-hand side of \eqref{Lax 1} lives in $\coprod_{a, b \in \CP} \g \otimes \lambda^{-N_a} \mu^{-N_b} \CC\bb{\lambda, \mu}$.  Indeed, at $b \in \CP$ the power of $\mu_b$ is bounded below by $-N_b$ since $\iota_{\lambda_a} \iota_{\mu_b} r_{12}(\lambda, \mu)$ is a Taylor series in $\mu_b$ while $Q^b_2(\mu_b)$ is a Laurent series with leading term of order $\mu_b^{-N_b}$ by definition \eqref{Qan def}. By the following lemma we then also deduce that at $a \in \CP$ the power of $\lambda_a$ on the right hand side of \eqref{Lax 1} is bounded below by $-N_a$. For the left hand side, this means that the flow with respect to the times $t_m^b$, with $m<-N_b$ are trivial: $\bm Q(\bm \lambda)$ does not depend on those times and for all practical purposes related to a hierarchy of field theories, they can be ignored. 

\begin{lemma} \label{lem: iota reorder}
	We have
	\begin{equation*}
	\big[ \Tr_2 \big( \bm \iota_{\bm \lambda} \bm \iota_{\bm \mu} r_{12}(\lambda,\mu) \bm Q_2(\bm \mu) \big),\bm Q_1(\bm \lambda) \big] = \big[ \Tr_2 \big( \bm \iota_{\bm \mu} \bm \iota_{\bm \lambda} r_{12}(\lambda,\mu) \bm Q_2(\bm \mu) \big),\bm Q_1(\bm \lambda) \big].
	\end{equation*}
	\begin{proof}
		 Using the identity \eqref{delta_from_r_matrix1} (or \eqref{delta_from_r_matrix2} in the trigonometric case) we deduce that for any $a, b \in \CP$ we have
		\begin{equation*}
		\big[ \Tr_2 \big( (\iota_{\lambda_a} \iota_{\mu_b} - \iota_{\mu_b} \iota_{\lambda_a}) r_{12}(\lambda,\mu) Q^b_2(\mu_b) \big),Q^a_1(\lambda_a) \big] \propto \delta(\lambda_a, \mu_a) [Q^a(\lambda_a), Q^a(\mu_a)].
		\end{equation*}
		Since $[Q^a(\lambda_a), Q^a(\mu_a)]$ vanishes when $\lambda_a = \mu_a$ it is proportional to $\lambda_a - \mu_a$ and so it follows that the right hand side above vanishes, as required.
	\end{proof}
\end{lemma}

\begin{remark} \label{rem: Lax eq}
	The Lax equation \eqref{Lax 1} is to be compared with the Lax equation in the usual finite-dimensional setting for the evolution of the Lax matrix with respect to the times associated with the coefficients in the partial fraction decomposition of the quadratic Hamiltonian
	\begin{equation*}
	H(\mu) = \tfrac 12 \Tr\big( L(\mu)^2 \big) = \sum_{a \in S} \sum_{n=0}^{n_a-1} \frac{H^a_n}{(\mu - a)^{n+1}}.
	\end{equation*}
	If we gather together the flows $\partial_{t^a_n} = \{ H^a_n, \cdot \}$ associated with the Hamiltonians $H^a_n$ by defining the differential operator valued rational function
	\begin{equation*}
	\cD_\mu = \sum_{a \in S} \sum_{n=0}^{n_a-1} \frac{\partial_{t^a_n}}{(\mu - a)^{n+1}},
	\end{equation*}
	which is to be compared with the ad\`elic object \eqref{def_gen_deriv} in the present infinite-dimensional setting, then the Lax equations in the finite-dimensional setting take the form
	\begin{equation} \label{Lax fin dim}
	\cD_\mu L_1(\lambda) = \big[ \Tr_2 \big( r_{12}(\lambda,\mu) L_2(\mu) \big), L_1(\lambda) \big].
	\end{equation}
	Both sides of this equation are $\g$-valued rational functions in both $\lambda$ and $\mu$ with poles in $\lambda$ and $\mu$ at each $a \in S$ of order at most $N_a$.
\end{remark}

\begin{proposition}\label{commuting_vf}
	The flows \eqref{Lax 1} are compatible as a consequence of the commutativity of the corresponding vector fields, \emph{i.e.} for any three formal variables $\lambda$, $\mu$ and $\nu$ we have
	\begin{equation} \label{flows_commute}
	{\cal X}_{\bm \nu} {\cal X}_{\bm \mu} \bm Q(\bm \lambda) = {\cal X}_{\bm \mu} {\cal X}_{\bm \nu} \bm Q(\bm \lambda).
	\end{equation}
	\begin{proof}
We have
		\begin{align} \label{Dmu Dnu Qlambda}
		{\cal X}_{\bm \nu} {\cal X}_{\bm \mu} \bm Q_1(\bm \lambda) &= \big[ \Tr_2 \big( \bm \iota_{\bm \lambda} \bm \iota_{\bm \mu} r_{12}(\lambda,\mu) {\cal X}_{\bm \nu} \bm Q_2(\bm \mu) \big),\bm Q_1(\bm \lambda) \big] \notag\\
		&\qquad + \big[ \Tr_2 \big( \bm \iota_{\bm \lambda} \bm \iota_{\bm \mu} r_{12}(\lambda,\mu) \bm Q_2(\bm \mu) \big), {\cal X}_{\bm \nu} \bm Q_1(\bm \lambda) \big] \notag\\
		&= \Tr_{23} \Big[ \bm \iota_{\bm \lambda} \bm \iota_{\bm \mu} r_{12}(\lambda,\mu) \big[ \bm \iota_{\bm \mu} \bm \iota_{\bm \nu} r_{23}(\mu,\nu) \bm Q_3(\bm \nu),\bm Q_2(\bm \mu) \big],\bm Q_1(\bm \lambda) \Big] \notag\\
		&\qquad + \Tr_{23} \Big[ \bm \iota_{\bm \lambda} \bm \iota_{\bm \mu} r_{12}(\lambda,\mu) \bm Q_2(\bm \mu), \big[ \bm \iota_{\bm \lambda} \bm \iota_{\bm \nu} r_{13}(\lambda,\nu) \bm Q_3(\bm \nu),\bm Q_1(\bm \lambda) \big] \Big].
		\end{align}
		By using the cyclicity of the trace over space $2$ in the first term on the right hand side and the Jacobi identity on the last term, this can be rewritten as
		\begin{align*}
		{\cal X}_{\bm \nu} {\cal X}_{\bm \mu} \bm Q_1(\bm \lambda) &= \Tr_{23} \Big[ \bm \iota_{\bm \lambda} \bm \iota_{\bm \mu} \bm \iota_{\bm \nu} \big[ r_{12}(\lambda,\mu), r_{23}(\mu,\nu) \big] \bm Q_2(\bm \mu) \bm Q_3(\bm \nu),\bm Q_1(\bm \lambda) \Big]\\
		&\qquad + \Tr_{23} \Big[ \bm \iota_{\bm \lambda} \bm \iota_{\bm \mu} \bm \iota_{\bm \nu} \big[ r_{12}(\lambda,\mu), r_{13}(\lambda,\nu) \big] \bm Q_2(\bm \mu) \bm Q_3(\bm \nu),\bm Q_1(\bm \lambda) \Big]\\
		&\qquad + \Tr_{23} \Big[ \bm \iota_{\bm \lambda} \bm \iota_{\bm \nu} r_{13}(\lambda,\nu) \bm Q_3(\bm \nu),\big[ \bm \iota_{\bm \lambda} \bm \iota_{\bm \mu} r_{12}(\lambda,\mu) \bm Q_2(\bm \mu), \bm Q_1(\bm \lambda) \big] \Big].
		\end{align*}
		Likewise, exchanging $\mu \leftrightarrow \nu$ in \eqref{Dmu Dnu Qlambda} we obtain
		\begin{align*}
		{\cal X}_{\bm \mu} {\cal X}_{\bm \nu} \bm Q_1(\bm \lambda) &= \Tr_{23} \Big[ \bm \iota_{\bm \lambda} \bm \iota_{\bm \nu} r_{13}(\lambda,\nu) \big[ \bm \iota_{\bm \nu} \bm \iota_{\bm \mu} r_{32}(\nu,\mu) \bm Q_2(\bm \mu),\bm Q_3(\bm \nu) \big],\bm Q_1(\bm \lambda) \Big]\\
		&\qquad + \Tr_{23} \Big[ \bm \iota_{\bm \lambda} \bm \iota_{\bm \nu} r_{13}(\lambda,\nu) \bm Q_3(\bm \nu), \big[ \bm \iota_{\bm \lambda} \bm \iota_{\bm \mu} r_{12}(\lambda,\mu) \bm Q_2(\bm \mu),\bm Q_1(\bm \lambda) \big] \Big]\\
		&= \Tr_{23} \Big[ \bm \iota_{\bm \lambda} \bm \iota_{\bm \mu} \bm \iota_{\bm \nu} \big[ r_{13}(\lambda,\nu), r_{32}(\nu,\mu) \big] \bm Q_2(\bm \mu) \bm Q_3(\bm \nu),\bm Q_1(\bm \lambda) \Big]\\
		&\qquad + \Tr_{23} \Big[ \bm \iota_{\bm \lambda} \bm \iota_{\bm \nu} r_{13}(\lambda,\nu) \bm Q_3(\bm \nu), \big[ \bm \iota_{\bm \lambda} \bm \iota_{\bm \mu} r_{12}(\lambda,\mu) \bm Q_2(\bm \mu),\bm Q_1(\bm \lambda) \big] \Big],
		\end{align*}
		where in the second equality we used Lemma \ref{lem: iota reorder} to swap the order of $\bm \iota_{\bm \nu}$ and $\bm \iota_{\bm \mu}$ in the first term, along with the cyclicity of the trace over space $3$.
		Thus $\big[{\cal X}_{\bm \nu}, {\cal X}_{\bm \mu}\big]\bm Q_1(\bm \lambda)$ equals 
\begin{align*}
&\Tr_{23}\Big[\bm \iota_{\bm \lambda} \bm \iota_{\bm \mu} \bm \iota_{\bm \nu} \Big( [ r_{12}(\lambda, \mu), r_{13}(\lambda, \nu)] + [r_{12}(\lambda, \mu), r_{23}(\mu, \nu)]\\
&\qquad\qquad\qquad\qquad\qquad\qquad\qquad\qquad - [ r_{13}(\lambda,\nu), r_{32}(\nu, \mu)] \Big) \bm Q_2(\bm \mu) \bm Q_3(\bm \nu) ,\bm Q_1(\bm \lambda) \Big]
\end{align*}
which vanishes as a consequence of the CYBE \eqref{CYBE}.
	\end{proof}
\end{proposition}

\subsubsection{Generating zero curvature equation}\label{sec:gen_ZC}

In the context of integrable field theories the role of the Lax equation, cf. \eqref{Lax fin dim}, is replaced by the zero curvature equation for a Lax connection. Therefore, as a first step towards relating the present formalism to integrable \emph{field} theories, we now associate with each time $t^a_n$, for any $a \in S$ and $n \geq -N_a$, a rational Lax matrix $V^a_n(\lambda) \in R_\lambda(\g)$ such that any pair of these satisfies a zero curvature equation.

The equations of motion \eqref{Lax 1} can be written succinctly as
\begin{equation} \label{Lax 1 reformulate}
\cD_{\bm \mu} \bm Q(\bm \lambda) = [ \bm \iota_{\bm \lambda} \bm V(\lambda; \bm \mu), \bm Q(\bm \lambda) ]
\end{equation}
where we have introduced
\begin{equation} \label{V def}
\bm V(\lambda; \bm \mu) \coloneqq \Tr_2 \big( \bm \iota_{\bm \mu} r_{12}(\lambda,\mu) \bm Q_2(\bm \mu) \big). \end{equation}
Note that in \eqref{V def} we do \emph{not} expand the right hand side in powers of $\lambda_a$ for $a \in \CP$, \emph{i.e.} we do not apply the homomorphism $\bm \iota_{\bm \lambda}$. Instead, this expansion is taken explicitly in \eqref{Lax 1 reformulate}.
In particular, the semi-colon in the notation $\bm V(\lambda; \bm \mu)$ is used to emphasise that $\lambda$ is just a formal variable whereas $\bm \mu$ is the usual boldface notation used as a shorthand for a collection $\big( V^b(\lambda; \mu_b) \big)_{b \in \CP}$ where
\begin{subequations} \label{Vb expansion}
	\begin{align}
	V^b(\lambda; \mu_b) &= \sum_{n=-N_b}^\infty V^b_n(\lambda) \mu_b^n\,,~~b\in\CC\,, \\
	V^\infty(\lambda; \mu_\infty) &= \sum_{n=-N_\infty}^\infty V^\infty_n(\lambda) \mu_\infty^{n+k+1}\,.
	\end{align}
\end{subequations}
As usual, we take $k = 0$ in the rational case and $k=-1$ in the trigonometric case.
Here $V^b_n(\lambda) \in R_\lambda(\g)$ are $\g$-valued rational functions in $\lambda$ with a pole at $\lambda = b$.
Unpacking the notation in \eqref{Lax 1 reformulate} slightly, recalling the definition of the operators $\cD_{\bm \mu}$ and \eqref{def_gen_deriv}, we see that the flow of $\bm Q(\bm \lambda)$ with respect to the time $t^a_n$ is controlled by $V^a_n(\lambda)$, namely we have the Lax equation
\begin{equation} \label{Lax 1 reformulate V}
\partial_{t^a_n} \bm Q(\bm \lambda) = [ \bm \iota_{\bm \lambda} V^a_n(\lambda), \bm Q(\bm \lambda) ].
\end{equation}
Moreover, by the following proposition $V^b(\lambda; \mu_b)$ can be seen as a generating series in $\mu_b$ of a hierarchy of Lax matrices $V^b_n(\lambda)$ associated with the times $t^b_n$.

\begin{proposition} \label{prop: ZC generating}
	We have the zero curvature equation in generating form
	\begin{equation} \label{zc generating}
	\cD_{\bm \nu} \bm V(\lambda; \bm \mu)- \cD_{\bm \mu} \bm V(\lambda; \bm \nu)+ \big[\bm V(\lambda; \bm \mu), \bm V(\lambda; \bm \nu)\big]=0.
	\end{equation}
	Equivalently, in components we have the zero curvature equation
	\begin{equation*}
	\partial_{t^b_n} V^a_m(\lambda) - \partial_{t^a_m} V^b_n(\lambda) + \big[V^a_m(\lambda), V^b_n(\lambda) \big]=0
	\end{equation*}
	for every $a, b \in \CP$ and $m \geq - N_a$ and $n \geq - N_b$.
	\begin{proof}
		Using the Lax equation \eqref{Lax 1} we find
		\begin{align*}
		\cD_{\bm \nu} \bm V(\lambda; \bm \mu) &= \Tr_2 \big( \bm \iota_{\bm \mu} r_{12}(\lambda,\mu) \cD_{\bm \nu} \bm Q_2(\bm \mu) \big)\\
		&= \Tr_{23} \big( \bm \iota_{\bm \mu} r_{12}(\lambda,\mu) \big[ \bm \iota_{\bm \mu} \bm \iota_{\bm \nu} r_{23}(\mu,\nu) \bm Q_3(\bm \nu),\bm Q_2(\bm \mu) \big] \big)\\
		&= \Tr_{23} \big( \bm \iota_{\bm \mu} \bm \iota_{\bm \nu} \big[ r_{12}(\lambda,\mu), r_{23}(\mu,\nu) \big] \bm Q_2(\bm \mu) \bm Q_3(\bm \nu) \big),
		\end{align*}
		where in the last equality we used the cyclicity of the trace in space $2$.
		Likewise, we also have
		\begin{align*}
		\cD_{\bm \mu} \bm V(\lambda; \bm \nu) &= \Tr_3 \big( \bm \iota_{\bm \nu} r_{13}(\lambda,\nu) \cD_{\bm \mu} \bm Q_3(\bm \nu) \big)\\
		&= \Tr_{23} \big( \bm \iota_{\bm \nu} r_{13}(\lambda,\nu) \big[ \bm \iota_{\bm \nu} \bm \iota_{\bm \mu} r_{32}(\nu,\mu) \bm Q_2(\bm \mu),\bm Q_3(\bm \nu) \big] \big)\\
		&= \Tr_{23} \big( \bm \iota_{\bm \mu} \bm \iota_{\bm \nu} \big[ r_{13}(\lambda,\nu), r_{32}(\nu,\mu) \big] \bm Q_2(\bm \mu) \bm Q_3(\bm \nu) \big)
		\end{align*}
		where in the final step we used Lemma \ref{lem: iota reorder} to swap the order of $\bm \iota_{\bm \nu}$ and $\bm \iota_{\bm \mu}$, before using the cyclicity of the trace in space $3$. Finally, we have
		\begin{align*}
		\big[\bm V(\lambda; \bm \mu), \bm V(\lambda; \bm \nu)\big] &= \Tr_{23} \big[ \bm \iota_{\bm \mu} r_{12}(\lambda,\mu) \bm Q_2(\bm \mu), \bm \iota_{\bm \lambda} \bm \iota_{\bm \nu} r_{13}(\lambda,\nu) \bm Q_3(\bm \nu) \big]\\
		&= \Tr_{23} \big( \bm \iota_{\bm \mu} \bm \iota_{\bm \nu} \big[ r_{12}(\lambda,\mu), r_{13}(\lambda,\nu) \big] \bm Q_2(\bm \mu) \bm Q_3(\bm \nu) \big).
		\end{align*}
		The result now follows by the classical Yang-Baxter equation \eqref{CYBE}.
	\end{proof}
\end{proposition}

\begin{remark} \label{rem: V vs M}
	Note the clear resemblance between the generating series \eqref{V def} for the hierarchy of Lax matrices $V^a_n(\lambda)$ and the usual generating rational function
	\begin{equation*}
	M(\lambda; \mu) = \Tr_2 \big( r_{12}(\lambda,\mu) L_2(\mu) \big)
	\end{equation*}
	in the finite-dimensional case. The coefficients in the partial fraction decomposition of the latter with respect to $\mu$ are $\g$-valued rational matrices $M^a_n(\lambda)$ which control the flow of the Lax matrix $L(\lambda)$ with respect to the associated time $t^a_n$ via the Lax equation
	$\partial_{t^a_n} L(\lambda) = [ M^a_n(\lambda), L(\lambda) ]$,
	which is to be compared with \eqref{Lax 1 reformulate V}.
	
	In the finite-dimensional case, however, one may also need to consider the more general generating rational function
	\begin{equation*}
	M^{(n)}(\lambda; \mu) = \Tr_2 \big( r_{12}(\lambda,\mu) L_2(\mu)^{n-1} \big)
	\end{equation*}
	for integers $n \geq 2$.
	Indeed, the Lax equation in \eqref{Lax fin dim} involves $M(\lambda; \mu) = M^{(2)}(\lambda; \mu)$ which is only associated with the quadratic Hamiltonians $H(\mu) = \frac 12 \Tr(L(\mu)^2)$. But in the finite-dimensional setting one should equally consider the Lax equations where $M^{(n)}(\lambda; \mu)$ replaces $M^{(2)}(\lambda; \mu)$ since these describe the flows of the Lax matrix $L(\lambda)$ with respect to the higher order Hamiltonians built from $\frac 1n \Tr(L(\mu)^n)$.
	
In the present infinite-dimensional context, we observe that the generating series \eqref{V def} is sufficient to produce the infinite number of Lax matrices $V^a_n(\lambda)$ associated with the infinite number of times $t^a_n$ in the hierarchy that one expects from the traditional examples of the AKNS or the sine-Gordon hierarchies (see below). It is not clear to us what an appropriate analog of taking higher powers of $L(\lambda)$ is in terms of $\bm Q(\bm \lambda)$ and whether the resulting Lax matrices and commuting flows would be independent of those obtained already.

\end{remark}

\begin{remark}
	It is instructive to compare the generating series \eqref{V def} for the hierarchy of Lax matrices $V^a_n(\lambda)$ with formulas for similar generating series of Lax matrices obtained in the more traditional approach to integrable field theories which involves the monodromy matrix associated to a given auxiliary equation $\partial_x\Psi=U\Psi$. For example, in \cite[pp. 203-204]{FT}, it is shown that the object
	\be
	\label{FT_gen_V}
	V(x,\lda,\mu)=\frac{1}{2(\lda-\mu)}(\1+W(x,\mu))(-i\sigma_3)(\1+W(x,\mu))^{-1}
	\ee
	``is the generating series of the Lax matrices $V_n(x,\lda)$ appearing in the zero curvature equation representation of the higher NS equations''. The expansion is to be understood as
	\be
	V(x,\lda,\mu)=\sum_{n=1}^{\infty}V_n(x,\lda)\mu^{-n}\,.
	\ee
	The point is that \eqref{FT_gen_V} can be rewritten as
		\be
			\label{FT_gen_V2}
	V(x,\lda,\mu)=-\frac{1}{2}\Tr_2\left(\iota_{\mu_\infty}r_{12(\lda,\mu)}(\1+W(x,\mu))_2(-i\sigma_3)_2(\1+W(x,\mu))_2^{-1}\right).
	\ee
	We note the explicit dependence of the preferred variable $x$, indicative of the fact that this object has been built from a particular, preferred time $x$ associated to the Lax matrix denoted $U(x,\lda)$, which is nothing but $V_1(x,\lda)$, as it should be. Other than this dependence, formula \eqref{FT_gen_V} has exactly the same structure as our formula \eqref{V def} when specialised to the AKNS hierarchy, see Section \ref{sec: AKNS hierarchy}. Indeed, in that case the only pole to consider is at infinity and the function $F(\lda)$ is taken to be $-i\sigma_3$. Hence the only non zero element in the tuple \eqref{V def} is 
	\begin{equation}
	\label{our_gen_V}
V^\infty(\lambda;  \mu_\infty) = \Tr_2 \big( \iota_{ \mu_\infty} r_{12}(\lambda,\mu) \phi^\infty_2( \mu_\infty)(-i\sigma_3)_2\phi^\infty_2( \mu_\infty)^{-1} \big). 
	\end{equation}
	To complete the comparison, note that the term $(\1+W(x,\mu))$ in \eqref{FT_gen_V2} comes from writing the monodromy matrix $T(x,y,\lda)$ on the finite interval $[y,x]$, associated to $U(x,\lda)$, as
	\be
	T(x,y,\mu)=(\1+W(x,\mu))e^{Z(x,y,\mu)}(\1+W(y,\mu))^{-1}
	\ee
	where $Z$ is a diagonal matrix and both $Z$ and $W$ are Taylor series in $1/\mu$ (with no constant term for $W$). We refer the curious reader to \cite{FT} for more details about $Z$ and $W$ which are not of importance for our discussion here. Considering for instance the case of fast decaying fields as $|x|\to\infty$, we can work with the monodromy matrix on $(-\infty,x)$
	\be
	\label{form_T}
	T^-(x,\mu)=(\1+W(x,\mu))e^{Z^-(x,\mu)}\,.
	\ee
	This is the object that plays the role of our group element $\phi^\infty( \mu_\infty)$. Indeed, formally plugging  $T^-(x,\mu)$ into \eqref{our_gen_V} in place of $\phi^\infty( \mu_\infty)$, and remembering that $e^{Z^-(x,\mu)}$ commutes with $\sigma_3$, we see that we get \eqref{FT_gen_V2} (up to an irrelevant factor $-1/2$ which comes from a different choice of normalisation between us and \cite{FT}).
	
	In \cite{ACDK,AC}, the argument from \cite{FT} was generalised to obtain the analog of formula \eqref{FT_gen_V2} but where one now builds it from the monodromy matrix associated to the time $t_k$ and Lax matrix $V_k(t_k,\lda)$, for an arbitrary but fixed $k\ge 1$. This represented the first step towards providing a generating function of Lax matrices that treats all times in the AKNS hierarchy equally. Our formula \eqref{V def} achieves this fully in that it makes no reference to a preferred time and an associated monodromy matrix as a starting point. It is also valid well beyond the realm of AKNS only, as our various examples below demonstrate. 
\end{remark}

It will be useful, in view of applying our general framework to construct explicit examples in the next few sections, to be more explicit about the form of the Lax matrices $V^a_n(\lambda)$. This can be done using Lemma \ref{lemma_expansion} in the rational case or Lemma \ref{lemma_expansion trig} in the trigonometric case.

\begin{proposition} \label{prop: Va explicit form}
	In the rational case, for every $a \in \CC$ and $n \geq - N_a$, we have
	\begin{equation*}
	V^a_n(\lambda) = - \big( \lambda_a^{-n-1} Q^a(\lambda_a) \big)^{\rm rat}_-,
	\end{equation*}
	while at infinity, for any $n \geq - N_\infty$ we have
	\begin{equation*}
	V^\infty_n(\lambda) = \big( \lambda_\infty^{-n} Q^\infty(\lambda_\infty) \big)^{\rm rat}_-.
	\end{equation*}
	
	In the trigonometric case, for every $a \in \CC$ and $n \geq -N_a$ we have
	\begin{equation*}
	V^a_n(\lambda) = - \big( (\lambda_a^{-n} + a \lambda_a^{-n-1}) Q^a(\lambda_a) \big)^{\rm trig}_-,
	\end{equation*}
	which at the origin simply reads $V^0_n(\lambda) = - \big( \lambda^{-n} Q^0(\lambda) \big)^{\rm trig}_-$, while at infinity we have, for every $n \geq - N_\infty$,
	\begin{equation*}
	V^\infty_n(\lambda) = \big( \lambda^n Q^\infty(\lambda^{-1}) \big)^{\rm trig}_-.
	\end{equation*}
	\begin{proof}
		In the rational (resp. trigonometric) case this is a direct consequence of the definition \eqref{Vb expansion} together with Lemma \ref{lemma_expansion} (resp. Lemma \ref{lemma_expansion trig}).
	\end{proof}
\end{proposition}

Recall that $Q^a(\lambda_a) = 0$ if $a \in \CP \setminus S$ so that, in fact, $V^a_n(\lambda) = 0$ unless $a \in S$.
By construction each Lax matrix $V^a_n(\lambda) \in R_\lambda(\g)$ for any $a \in S$ and $n \geq -N_a$, or rather their embedding in $\bm \A_{\bm \lambda}(\g)$ via \eqref{iota map}, is a coadjoint orbit in $\bm \iota_{\bm \lambda} R_\lambda(\g)$. For instance, in the rational case for $a \in \CC \cap S$ we have
\begin{equation*}
V^a_n(\lambda) = - \Big( \phi^a(\lambda_a) \lambda_a^{-n-1} F^a(\lambda_a)_- \phi^a(\lambda_a)^{-1} \Big)^{\rm rat}_- \in \g \otimes \CC[\lambda_a^{-1}] \subset R_\lambda(\g).
\end{equation*}

\subsubsection{Connection to integrable field theory and FNR procedure} \label{sec: connection to IFT}

Up to this point, the framework we have been discussing is very similar to the one used to describe finite-dimensional integrable systems, as emphasised in Remarks \ref{rem: Q vs Lax}, \ref{rem: Lax eq} and \ref{rem: V vs M}.
However, as we will see explicitly in all the examples discussed in later sections, our formalism encodes entire hierarchies of integrable field theories!

\paragraph{The FNR procedure.}

One way to make explicit contact with the traditional approach to integrable field theory is to choose a preferred coordinate, denote it by $x$ and set it as a particular combination of the fundamental times $t^a_n$ for $a \in S$ and $n \geq -N_a$. Quite generally, we can choose some finite subsets $T_a \subset \ZZ_{\geq -N_a}$ for each $a \in S$ and define $\displaystyle \partial_x \coloneqq \sum_{a \in S} \sum_{n \in T_a} r^a_n \partial_{t^a_n}$ for some $r^a_n \in \CC^\ast$. The Lax matrix associated with the coordinate $x$ is then given by
\begin{equation} \label{U def}
U(\lambda) \coloneqq \sum_{a \in S} \sum_{n \in T_a} r^a_n V^a_n(\lambda) \in R_\lambda(\g).
\end{equation}
As explained above, $\bm \iota_{\bm \lambda} U(\lambda)$ is then a coadjoint orbit in the dual space $\bm \iota_{\bm \lambda} R_\lambda(\g)$ of $\bm \A^+_{\bm \lambda}(\g)$. This is the coadjoint orbit alluded to at the very start of this section which encodes the finite collection of fields of our integrable hierarchy.
It follows from \eqref{Lax 1 reformulate}, or even more directly from \eqref{Lax 1 reformulate V}, that the spatial dependence of $\bm Q(\bm \lambda)$ is governed by the Lax equation
\begin{equation} \label{Lax 1 spatial}
\partial_x \bm Q(\bm \lambda) = [ \bm \iota_{\bm \lambda} U(\lambda), \bm Q(\bm \lambda) ].
\end{equation}
As we will see on examples, the equation \eqref{Lax 1 spatial} can be solved recursively to express the coefficients $Q^a_n$ of $Q^a(\lambda_a)$, cf. \eqref{Qan def}, as differential polynomials in the fields, \emph{i.e.} the variables contained in the Lax matrix $U(\lambda)$. All other Lax matrices $V^a_n(\lambda)$ associated to the fundamental times $t^a_n$ will then have components expressed as differential polynomials of the fields.

\medskip

We will outline below how \eqref{Lax 1 spatial} can, in principle, be solved recursively for each $Q^a_n$. Since certain details of the recursive procedure depend on the model considered, we will only illustrate here the part of the construction which applies universally to all models in Lemma \ref{lem: Q in terms of U} below. We will see later on examples how to apply this construction to specific models. 

To state the lemma, we first need to make a few observations and definitions. Since the Laurent expansions $\iota_{\lambda_a} V^a_n(\lambda)$ each have a non-zero principal part, it follows from the definition \eqref{U def} that we can write
\begin{equation} \label{series U at a}
\iota_{\lambda_a} U(\lambda) = \sum_{p = - n_a}^\infty U^a_p \lambda_a^p
\end{equation}
for some $n_a \geq 1$ and non-zero leading coefficient $U^a_{-n_a} \in \g$.
By definition \eqref{Q def} we have that $Q^a(\lambda_a)=\phi^a(\lambda_a) F^a(\lambda_a)_- \phi^a(\lambda_a)^{-1}$.
It thus follows from the relationship between each $\iota_{\lambda_a} V^a_n(\lambda)$ and $Q^a(\lambda_a)$, as described explicitly in Proposition \ref{prop: Va explicit form}, that the coefficients of the most singular terms in the formal Laurent series $Q^a(\lambda_a)$ and $\iota_{\lambda_a} U(\lambda)$, given in \eqref{Qan def} and \eqref{series U at a} respectively, are proportional. In other words, we have $Q^a_{-N_a} = c\, U^a_{-n_a}$ for some $c \in \CC^\ast$. Explicitly, it can be seen from Proposition \ref{prop: Va explicit form} that $c$ is given up to a sign by the coefficient $r^a_n$ in \eqref{U def} with $n = \max T_a$. We can thus write
\begin{equation} \label{Q series first term}
\lambda_a^{N_a} Q^a(\lambda_a) = c\, U^a_{-n_a} + \sum_{r=1}^\infty Q^a_{-N_a+r} \lambda_a^r.
\end{equation}
Now let $\mathfrak{k} \coloneqq \ker(\textup{ad}\, U^a_{-n_a})$ and $\mathfrak{i} \coloneqq \im(\textup{ad}\, U^a_{-n_a})$. We fix any complements $\mathfrak{k}'$ of $\mathfrak{k}$ and $\mathfrak{i}'$ of $\mathfrak{i}$ in $\g$ so that we have the direct sum decompositions
\begin{equation} \label{decompositions glN}
\g = \mathfrak{k} \oplus \mathfrak{k}' = \mathfrak{i} \oplus \mathfrak{i}'.
\end{equation}
Let $\pi_{\mathfrak{k}} : \g \to \mathfrak{k}$ and $\pi_{\mathfrak{k}'} : \g \to \mathfrak{k}'$ denote the projections onto $\mathfrak{k}$ and $\mathfrak{k}'$ relative to the first decomposition. Likewise, let $\pi_{\mathfrak{i}} : \g \to \mathfrak{i}$ and $\pi_{\mathfrak{i}'} : \g \to \mathfrak{i}'$ denote the projections onto $\mathfrak{i}$ and $\mathfrak{i}'$ relative to the second decomposition in \eqref{decompositions glN}.

\begin{lemma} \label{lem: Q in terms of U}
For any $r \geq 1$, $\pi_{\mathfrak{k}'}(Q^a_{-N_a+r})$ is expressible as a differential polynomial in $x$ of the elements $Q^a_{-N_a+s}$ for $s < r$.
\begin{proof}
Using the explicit forms \eqref{Qan def} and \eqref{series U at a} for the Laurent series of $Q^a(\lambda_a)$ and $\iota_{\lambda_a} U(\lambda)$, we may rewrite the component of \eqref{Lax 1 spatial} at $a \in S$ more explicitly as
\begin{equation*}
\sum_{n=-N_a}^\infty \lambda_a^n \partial_x Q^a_n = \sum_{m=-N_a}^\infty \sum_{p = - n_a}^\infty \lambda_a^{m+p} [ U^a_p, Q^a_m] = \sum_{n=-N_a-n_a}^\infty \lambda_a^n \sum_{p = - n_a}^\infty [ U^a_p, Q^a_{n-p}].
\end{equation*}
In the second equality we have changed variables in the double sum from $m \geq -N_a$ to $n \coloneqq m+p \geq -N_a - n_a$. Comparing the coefficients of $\lambda_a^{-N_a - n_a + r}$ on both sides of the above equation for all $r \geq 0$ we find the following. For every $0 \leq r \leq n_a - 1$,
\begin{subequations} \label{recurrence for Q}
\begin{equation} \label{recurrence for Q a}
[U^a_{-n_a}, Q^a_{-N_a + r}] = - \sum_{q=1}^r [U^a_{-n_a+q}, Q^a_{-N_a+r-q}]
\end{equation}
where we changed variables in the sum from $p$ to $q \coloneqq p+n_a$. Notice that for $r=0$ this gives $[U^a_{-n_a}, Q^a_{-N_a}] = 0$ which is consistent with the observation in \eqref{Q series first term} that $Q^a_{-N_a}$ is proportional to $U^a_{-n_a}$.
On the other hand, for $r \geq n_a$ we have
\begin{equation} \label{recurrence for Q b}
[ U^a_{-n_a}, Q^a_{-N_a+r}] = \partial_x Q^a_{-N_a -n_a+ r} - \sum_{q=1}^r [ U^a_{-n_a+q}, Q^a_{-N_a+r-q}].
\end{equation}
\end{subequations}
Denoting the right hand side of the equations \eqref{recurrence for Q} by $B_r$, for each $r \geq 0$ we can rewrite all of them more uniformly as
\begin{equation} \label{recur for Q}
[ U^a_{-n_a}, Q^a_{-N_a+r}] = B_r
\end{equation}
for $r \geq 0$. 
Since the left hand side of \eqref{recur for Q} lies in $\mathfrak{i}$ we have, for every $r \geq 0$,
\begin{equation} \label{recur for Q split}
\big[ U^a_{-n_a}, \pi_{\mathfrak{k}'}(Q^a_{-N_a+r}) \big] = \pi_{\mathfrak{i}}(B_r), \qquad
0 = \pi_{\mathfrak{i}'}(B_r),
\end{equation}
where in the first equation we have also decomposed $Q^a_{-N_a+r}$ relative to the first decomposition in \eqref{decompositions glN} and used the fact that $\pi_{\mathfrak{k}}(Q^a_{-N_a+r})$ commutes with $U^a_{-n_a}$.
		
Now the linear map $\textup{ad}\, U^a_{-n_a} : \mathfrak{k}' \to \mathfrak{i}$ is a bijection. Indeed, it is clearly surjective by definition of $\mathfrak{i}$. To see that it is injective, note that if $[U^a_{-n_a}, X] = [U^a_{-n_a}, Y]$ for any $X, Y \in \mathfrak{k}'$ then $X-Y \in \mathfrak{k}$ and hence $X - Y = 0$, as required. It follows that $\pi_{\mathfrak{k}'}(Q^a_{-N_a+r})$ is uniquely determined in terms of $\pi_{\mathfrak{i}}(B_r)$ for every $r \geq 0$ by the first equation in \eqref{recur for Q split}.
The result now follows.
\end{proof}
\end{lemma}

In order to completely determine the coefficients $Q^a_{-N_a+r}$ for $r \geq 0$, 
it remains to show that the $\pi_{\mathfrak{k}}(Q^a_{-N_a+r})$ for every $r \geq 0$ can also be determined recursively.
This is the part which will typically depend on the model considered. Here we will show, generalising an argument for the ZS-AKNS $n \times n$ hierarchy given in \cite[Theoerem 2.2]{TerngUhlenbeck}, see also \cite{Sattinger}, how this can be done under the assumption that there is a polynomial $P_a$ with coefficients in $\CC[\lambda_a]$ such that $P_a\big( \lambda_a^{N_a} F^a(\lambda_a)_- \big) = 0$ and $P_a'(c\, U^a_{-n_a}) \in \CC[\lambda_a]$ is invertible in $\CC\bb{\lambda_a}$. Recalling \eqref{Q series first term}, we have the identity
\begin{equation*}
P_a \bigg( \! c\, U^a_{-n_a} + \sum_{r=1}^\infty Q^a_{-N_a+r} \lambda_a^r \bigg) = 0.
\end{equation*}
And using the fact that each $\pi_{\mathfrak{k}}(Q^a_{-N_a+r})$ commutes with $U^a_{-n_a}$, by definition of $\mathfrak{k}$, we can then rewrite the above in the form
\begin{equation} \label{identity poly P}
P_a(c\, U^a_{-n_a}) + P'_a(c\, U^a_{-n_a}) \sum_{r=1}^\infty \pi_{\mathfrak{k}}(Q^a_{-N_a+r}) \lambda_a^r = \mathcal{R} \Big( \big\{ Q^a_{-N_a+r} \big\}_{r=1}^\infty \Big)
\end{equation}
where the right hand side is a sum of terms, each of which contains either higher powers of $\sum_{r=1}^\infty \pi_{\mathfrak{k}}(Q^a_{-N_a+r}) \lambda_a^r$ or at least one factor of $\sum_{r=1}^\infty \pi_{\mathfrak{k}'}(Q^a_{-N_a+r}) \lambda_a^r$. Since we are assuming that $P'_a(c\, U^a_{-n_a}) \in \CC[\lambda_a]$ is invertible, it follows by comparing powers of $\lambda_a^r$ on both sides of \eqref{identity poly P} that $\pi_{\mathfrak{k}}(Q^a_{-N_a+r})$ can be expressed as a finite sum of terms involving only $\pi_{\mathfrak{k}}(Q^a_{-N_a+s})$ for $s < r$ or $\pi_{\mathfrak{k}'}(Q^a_{-N_a+s})$ for $s \leq r$.
		
In conjunction with Lemma \ref{lem: Q in terms of U}, this shows that each $\pi_{\mathfrak{k}}(Q^a_{-N_a+r})$ and $\pi_{\mathfrak{k}'}(Q^a_{-N_a+r})$, and therefore $Q^a_{-N_a+r}$ itself, can be determined recursively for each $r \geq 0$. In particular, all the coefficients $Q^a_n$, $n \geq - N_a$ of the Laurent series $Q^a(\lambda_a)$ in \eqref{Qan def} can be expressed as differential polynomials in $x$ of the coefficients of the rational function $U(\lambda)$. The same conclusion still holds even when there is no polynomial $P_a$ with the above properties, as will be shown on the example of the sine-Gordon hierarchy in Section \ref{sec:sG}.

\medskip

It is important to observe that our choice of `spatial' coordinate $x$ defined by the linear combination $\partial_x = \sum_{a \in S} \sum_{n \in T_a} r^a_n \partial_{t^a_n}$ and its associated Lax matrix in \eqref{U def} was completely arbitrary. Indeed, one of the main advantages of working with the adjoint orbit $\bm Q(\bm \lambda)$ in $\bm \A_{\bm \lambda}(\g)$ rather than the coadjoint orbit $U(\lambda)$ in $R_\lambda(\g)$ is that it keeps all the times on an equal footing by not singling out a particular (linear combination of) time as `space'.

\paragraph{On the redundancy of the FNR procedure.} 
The previous discussion casts in the present framework the original idea of \cite{FNR} whereby one should first solve for the coordinates in $\bm Q(\bm \lambda)$ in terms of the finite collection of fields contained in a given Lax matrix $U(\lda)$, now interpreted as fields depending on a preferred space variable $x$. The other times in the hierarchies are viewed as (compatible) time flows imposed on this finite collection of fields and define a preferred field theory alongside its higher symmetries. 

Here we want to elaborate on a point of view originally advocated in \cite{CS3} whereby the above ``traditional'' approach is not needed at all and, in fact, represents a conceptual obstruction to the formalism we want to put forward in this work: we treat all the {\it times} in a hierarchy {\bf as well as} all the {\it (algebra or group) coordinates} (\ie the dependent variables contained in $\bm Q(\bm \lda)$ or $\bm \phi(\bm \lambda)$ respectively) on the same footing. From this point of view, one should consider the entirety of the Lax equations contained in the generating Lax equation \eqref{Lax 1}, or equivalently, the collection of zero curvature equations \eqref{prop: ZC generating}. The point is that the latter implement the FNR procedure anyway but they present the advantage of being amenable to a covariant Hamiltonian formulation, which was one of the main results of \cite{CS1,CS3}. This aspect is beyond the scope of the present work but remains one motivation for it. The fact that the zero curvature equations contain the equations of the FNR procedure was already observed and used in the particular example of the AKNS hierarchy in \cite{AC}. For convenience, let us sketch the argument here in the simplest case of a single pole $a\in\CC$, with a collection of times $t_n^a$, $n\ge -N_a$. Suppose we fix $n\ge -N_a$ and we want to solve 
\begin{equation}
\label{FNR1}
\partial_{t_n^a}  Q^a(\lambda_a) = [ \iota_{ \lambda_a} V_n^a(\lambda),  Q^a( \lambda_a) ]\,,
\end{equation}
given $Q^a_{-N_a}$, along the lines of Lemma \ref{lem: Q in terms of U} and the discussion after it. Without loss of generality, shifting the power of $\lda_a$ by $N_a$, we can always assume for simplicity that $N_a=0$. Then, \eqref{FNR1} amounts to the collection of equations
\begin{equation}
\label{FNR2}
\partial_{t_n^a}  Q^a_j = \sum_{p =0}^n[ Q_{j+n-p+1}^a,  Q^a_p ]\,,~~j\ge 0\,.
\end{equation}
As discussed above, in certain cases (which include the AKNS hierarchy and the sG hierarchy as we show explicitly in Section \ref{sec:sG}), this allows one to express all the algebra coordinates in $Q_j$, $j\ge n$ as differential polynomials with respect to $t_n^a$ in the coordinates contained in $Q_k$, $k=0,\dots,n$. Now consider the zero curvature equations, for $m\ge n+1$,
\begin{equation}
\label{ZC_m}
\partial_{t_n^a} V_m^a(\lambda) - \partial_{t_m^a} V_n^a(\lambda)+ [  V_n^a(\lambda),  V_m^a(\lambda) ]=0\,.
\end{equation}
Looking at the coefficient of $1/\lda^j$, for $j=n+2,\dots,m+1$, we find that they contain the equations
\begin{equation}
\partial_{t_n^a}  Q^a_{m+1-k} = \sum_{p =0}^n[ Q_{m+n+2-k-p}^a,  Q^a_p ]\,,~~k=n+2,\dots,m+1\,.
\end{equation}
If we set $j=m+1-k$, these become
\begin{equation}
\partial_{t_n^a}  Q^a_{j} = \sum_{p =0}^n[ Q_{j+n+1-p}^a,  Q^a_p ]\,,~~j=0,\dots,m-n-1\,.
\end{equation}
So the collection of zero curvature equations \eqref{ZC_m} for $m\ge n+1$ produces exactly the set of FNR equations \eqref{FNR2}. Hence, there is no point in implementing the FNR procedure a priori to determine the ``fields'' and then impose the zero curvature equations to determine their equations of motion. The latter suffices. With this in mind, we will come back to this point in certain examples below to illustrate our position and show how abandoning the FNR procedure allows us to eliminate the problem of alien derivatives mentioned in the introduction.

\subsection{Generating Lagrangian multiform}

In this section, we introduce the main object of this paper, the generating Lagrangian multiform \eqref{Lagrangian mult}-\eqref{kin pot terms coord}, and we show that the Lax equation \eqref{Lax 1} as it derives from $\bLag(\bm \lambda, \bm \mu)$. Although the equations of motion \eqref{Lax 1} can be written in terms of $\bm Q(\bm \lambda) \in \bm \A_{\bm \lambda}(\g)$ alone, in order to write $\bLag(\bm \lambda, \bm \mu)$ we need the group-valued element $\bm \phi(\bm \lambda) \in \bm \A^+_{\bm \lambda}(G)$. This is very reminiscent of the fact that writing down the Zakharov-Mikhailov action describing the Zakharov-Shabat equations of motion requires introducing a group valued field \cite{ZM1}.
Recall the definition of $\bm Q(\bm \lambda) \in \bm \A_{\bm \lambda}(\g)$ in \eqref{Q def} as an adjoint orbit of the element $(\bm \iota_{\bm \lambda} F(\lambda))_- \in \bm \A_{\bm \lambda}(\g)$, defined in \eqref{iota F}, under the action of $\bm \phi(\bm \lambda) \in \bm \A^+_{\bm \lambda}(G)$.

We consider the following generating Lagrangian multiform
\begin{equation} \label{Lagrangian mult2}
\bLag(\bm \lambda, \bm \mu) \coloneqq \bm K(\bm \lambda,\bm \mu) - \bm U(\bm \lambda,\bm \mu)
\end{equation}
where the kinetic and potential terms are given by
\begin{subequations} \label{kin pot terms2}
	\begin{align}
	\label{kinetic term} \bm K(\bm \lambda,\bm \mu) &\coloneqq \Tr\big( \bm \phi(\bm \lambda)^{-1} \cD_{\bm \mu} \bm \phi(\bm \lambda) (\bm \iota_{\bm \lambda} F(\lambda))_- \big)\\
	&\qquad\qquad\qquad - \Tr \big(\bm \phi(\bm \mu)^{-1} \cD_{\bm \lambda} \bm \phi(\bm \mu) (\bm \iota_{\bm \mu} F(\mu))_- \big), \notag\\
	\label{potential term2} \bm U(\bm \lambda,\bm \mu) &\coloneqq \tfrac 12 \Tr_{12}\big( (\bm \iota_{\bm \lambda} \bm \iota_{\bm \mu} + \bm \iota_{\bm \mu} \bm \iota_{\bm \lambda})r_{12}(\lambda,\mu) \bm Q_1(\bm \lambda) \bm Q_2(\bm \mu)\big).
	\end{align}
\end{subequations}
As mentioned at the end of Section \ref{sec: gen setup}, the boldface notation \eqref{Lagrangian mult} is used as a shorthand for an equality of components
\begin{equation*}
\Lag^{a,b}(\lambda_a, \mu_b) = K^{a,b}(\lambda_a, \mu_b) - U^{a,b}(\lambda_a, \mu_b)
\end{equation*}
for every $a,b \in \CP$, and the kinetic and potential terms \eqref{kin pot terms} in components are given explicitly by
\begin{subequations} \label{kin pot terms coord}
	\begin{align}
	\label{kinetic term coord} K^{a,b}(\lambda_a,\mu_b) &= \Tr\big( \phi^a(\lambda_a)^{-1} \cD_{\mu_b} \phi^a(\lambda_a) F^a(\lambda_a)_- \big)\\
	&\qquad\qquad - \Tr \big(\phi^b(\mu_b)^{-1} \cD_{\lambda_a} \phi^b(\mu_b) F^b(\mu_b)_- \big), \notag\\
	\label{potential term coord} U^{a,b}(\lambda_a,\mu_b) &= \tfrac 12 \Tr_{12}\big( (\iota_{\lambda_a} \iota_{\mu_b} + \iota_{\mu_b} \iota_{\lambda_a})r_{12}(\lambda,\mu)Q^a_1(\lambda_a)Q^b_2(\mu_b)\big).
	\end{align}
\end{subequations}
The kinetic term \eqref{kinetic term} is clearly skew-symmetric under the exchange $\bm \lda\leftrightarrow \bm \mu$, so the skew-symmetry of $\bLag(\bm \lambda, \bm\mu)$ is equivalent to the skew-symmetry of the potential term \eqref{potential term}, namely
\begin{align*}
&\Tr_{12}\big( (\bm \iota_{\bm \lambda} \bm \iota_{\bm \mu} + \bm \iota_{\bm \mu} \bm \iota_{\bm \lambda}) r_{12}(\lambda,\mu)\bm Q_1(\bm \lambda) \bm Q_2(\bm \mu)\big)\\
&\qquad\qquad = - \Tr_{12}\big( (\bm \iota_{\bm \lambda} \bm \iota_{\bm \mu} + \bm \iota_{\bm \mu} \bm \iota_{\bm \lambda}) r_{21}(\mu,\lambda) \bm Q_1(\bm \lambda) \bm Q_2(\bm \mu)\big).
\end{align*}
This holds since $r$ is skew-symmetric. 

\subsubsection{Extracting Lagrangians and Lagrangian multiforms}

We have been using the generating formalism efficiently so far. Here, we spend some time discussing the connection of our generating Lagrangian multiform with Lagrangians and Lagrangian multiforms. This will be useful to reformulate the multiform EL equations and the closure relation in generating form, allowing to continue to take advantage of this for general computations. 

From the definition of the generating Lagrangian multiform \eqref{Lagrangian mult}, we see that the kinetic term $K^{a,b}(\lambda_a, \mu_b)$ given by \eqref{kinetic term coord} is a Laurent series in both $\lambda_a$ and $\mu_b$, with powers bounded below by $-N_a$ and $-N_b$, respectively. In particular, for any $m, n \in \ZZ$ the coefficient of $\lambda_a^m \mu_b^n$ is well defined. The same is true for the potential term \eqref{potential term coord} by the following lemma.
\begin{lemma}
	For any $m, n \in \ZZ$ and any $a, b \in \CP$, the coefficient of $\lambda_a^m \mu_b^n$ in the potential term $U^{a,b}(\lambda_a, \mu_b)$ given by \eqref{potential term coord} is a well defined expression which is quadratic in the coefficients of $Q^a(\lambda_a)$ and $Q^b(\mu_b)$.
	\begin{proof}
		If $b \neq a$ then $(\iota_{\lambda_a} \iota_{\mu_b} + \iota_{\mu_b} \iota_{\lambda_a})r_{12}(\lambda,\mu)$ is valued in $(\g \otimes \g) \otimes \CC\bb{\lambda_a, \mu_b}$. Since by definition \eqref{Qan def} we have $Q^a(\lambda_a) \in \g \otimes \lambda_a^{-N_a} \CC\bb{\lambda_a}$ and $Q^b(\mu_b) \in \g \otimes \mu_b^{-N_b} \CC\bb{\mu_b}$, it follows that $U^{a,b}(\lambda_a, \mu_b)$ is a Laurent series in both $\lambda_a$ and $\mu_b$, with powers bounded below by $-N_a$ and $-N_b$, respectively.
		
		If $b=a$ then $(\iota_{\lambda_a} \iota_{\mu_a} + \iota_{\mu_a} \iota_{\lambda_a})r_{12}(\lambda,\mu)$ contains a doubly infinite Laurent series in $\lambda_a \mu_a^{-1}$ coming from the expansion of $1/(\lambda - \mu)$, possibly also multiplied by some polynomial in $\lambda_a$ and $\mu_a$ depending on the precise form of the $r$-matrix. Multiplying this by the Laurent series $Q^a(\lambda_a) \in \g \otimes \lambda_a^{-N_a} \CC\bb{\lambda_a}$ and $Q^a(\mu_a) \in \g \otimes \mu_a^{-N_a} \CC\bb{\mu_a}$, we produce terms of the form $\lambda_a^{r+j+p} \mu_a^{s-j+q}$ with $r, s \geq -N_a$, $j \in \ZZ$ and $p, q$ ranging over finitely many possible values. In order to form a term proportional to $\lambda_a^m \mu_a^n$ we need $m = r+j+p$ and $n=s-j+q$. But then $m-j-p = r \geq -N_a$ so that $j \leq m +N_a - p$ and also $n+j-q = s \geq - N_a$ so that $j \geq -n -N_a+q$. In other words, $j \in \ZZ$ must be bounded from above and below so that it ranges only over finitely many values. Hence, there are only finitely many terms contributing to the coefficient of $\lambda_a^m \mu_a^n$ and the result follows.
	\end{proof}
\end{lemma}
As a consequence, for any $a,b \in \CP$ and $m, n \in \ZZ$ with $m \geq - N_a$ and $n \geq - N_b$, we may now extract the following Lagrangian coefficients associated to the times $t_n^a$ and $t_m^b$:
\begin{definition}[{\bf Elementary Lagrangians}]
\begin{equation}
\label{def_elem_lag}
\Lag^{a,b}_{m,n} \coloneqq \res^\lambda_a \res^\mu_b \Lag^{a,b}(\lambda_a, \mu_b)\lambda^{-m-1}d\lda \,\mu^{-n-1}d{\mu}\,.
\end{equation}	
Recall the notational convention explained after \eqref{bilinear form b}, in particular for residues computed at infinity. In short, Definition \eqref{def_elem_lag} means that $\Lag^{a,b}_{m,n}$ is the coefficient of $\lda_a^m\mu_b^n$ in the expansion of $\Lag^{a,b}(\lambda_a, \mu_b)$, as one would want. This is what we use to compute elementary Lagrangians in all our examples.
\end{definition}
 As explained below, when building a hierarchy, one chooses a finite set $S\in\CP$ and all but a finite number of the elementary Lagrangians $\Lag^{a,b}_{m,n}$ vanish (those for which $a$ and/or $b$ is in $\CP\setminus S$). 
The Lagrangian multiform of the hierarchy is then given by
\begin{equation}
\label{def_lag_multiform_S}
\Lag^{\rm S} \coloneqq \tfrac{1}{2} \sum_{a,b \in S}\sum_{m,n}\Lag^{a,b}_{m,n}\,d{t^a_m} \wedge d{t^b_n}=\sum_{(m,a)<(n,b)}\Lag^{a,b}_{m,n}\,d{t^a_m} \wedge d{t^b_n}\,.
\end{equation}
Note that we introduced an order on the pairs $(m,a)\in\ZZ\times S$ in the last equality (recall that $\Lag^{a,b}_{m,n}=-\Lag^{b,a}_{n,m}$). With $S=\{a_1,\dots,a_n\}$, it is defined by
$$(m,a_i)< (n,a_j)\Leftrightarrow i<j ~\text{or}~ (i=j ~\text{and}~ m< n)\,.$$

These definitions generalise the correspondence explained in the introductory section \ref{motivating_ex} between $\Lag[u]$ and $\Lag(\lda,\mu)$ for the AKNS hierarchy. As we will see in detail in Section \ref{sec: AKNS hierarchy}, the latter indeed corresponds to the case where $S=\{\infty\}$. In practice, one calculates the elementary Lagrangians \eqref{def_elem_lag} directly by computing the appropriate Laurent series expansion of $\Lag^{a,b}(\lambda_a, \mu_b)$. The corresponding Lagrangian multiform is easily obtained as in \eqref{def_lag_multiform_S}.

The essential point of the present discussion is to identify the generating form of the two main equations of the theory of Lagrangian multiforms: the multiform EL equations $\delta d \Lag^{\rm S}=0$ and the closure relation $d \Lag^{\rm S}=0$ which should hold on solutions of the multiform EL equations. We see that the key object to translate in generating form is therefore 
$d \Lag^{\rm S}$.
In view of \eqref{def_lag_multiform_S}, $d \Lag^{\rm S}$ has the form
$$d \Lag^{\rm S}=\sum_{(k,c)<(m,a)<(n,b)}\left(\partial_{t_k^c}\Lag^{a,b}_{m,n}+\partial_{t_n^b}\Lag^{c,a}_{k,m}+\partial_{t_m^a}\Lag^{b,c}_{n,k}  \right)\,dt_k^c\wedge d{t^a_m} \wedge d{t^b_n}\,. $$
The generating function corresponding to the coefficient $\partial_{t_k^c}\Lag^{a,b}_{m,n}+\partial_{t_n^b}\Lag^{c,a}_{k,m}+\partial_{t_m^a}\Lag^{b,c}_{n,k}$ is
$$\cD_{\nu_c}\Lag^{a,b}(\lambda_a, \mu_b)+\cD_{\mu_b}\Lag^{c,a}(\nu_c,\lambda_a )+\cD_{\lambda_a}\Lag^{b,c}(\mu_b,\nu_c)\,.$$ 
Summarizing our discussion, the set $S$ was fixed but arbitrary, so going back to the ad\'elic setting, we will be working compactly with 
$$\delta \cD_{\bm \nu}\Lag(\bm \lambda, \bm \mu)+\delta \cD_{ \bm \mu}\Lag(\bm \nu,\bm \lambda)+\delta\cD_{\bm \lambda}\Lag( \bm \mu,\bm \nu)$$
when deriving the multiform EL equations in generating form, and with
$$\cD_{\bm \nu}\Lag(\bm \lambda, \bm \mu)+\cD_{ \bm \mu}\Lag(\bm \nu,\bm \lambda)+\cD_{\bm \lambda}\Lag( \bm \mu,\bm \nu)$$
when studying the closure relation.

\subsubsection{Generating multiform Euler-Lagrange equations}

Having introduced the main object of our framework, we proceed to derive the associated multiform EL equations (in generating form) and show that they give the generating Lax equation \eqref{Lax 1}.

\begin{theorem}
 \label{prop: eom Q}
The generating Lax equation \eqref{Lax 1} is variational: the multiform EL equations deriving from the generating Lagrangian multiform $\bLag(\bm \lambda, \bm \mu)$ take the form
$$\cD_{\bm \mu} \bm Q_1(\bm \lambda) = \big[ \Tr_2 \big( \bm \iota_{\bm \lambda} \bm \iota_{\bm \mu} r_{12}(\lambda,\mu) \bm Q_2(\bm \mu) \big), \bm Q_1(\bm \lambda) \big].$$
\begin{proof}
We derive the equations induced by the requirement $\delta d\Lag=0$	in generating form. This means that we compute $\delta \cD_{\bm \nu} \bm\Lag(\bm\lambda,\bm\mu) + \circlearrowleft=0$, where $\circlearrowleft$ means cyclic permutations of $\lda,\mu,\nu$, and set the independent coefficients to zero. 
	We start with the kinetic terms.
	\bea
	\cD_{\bm \nu} \bm K(\bm\lambda,\bm\mu) =& \Tr\Big(- \bm\phi^{-1}(\bm\lambda) \cD_{\bm\nu} \bm\phi(\bm\lambda) \bm\phi^{-1}(\bm\lambda) \cD_{\bm\mu} \bm\phi(\bm\lambda) (\bm\iota_{\bm\lda}F(\lambda))_-
	+ \bm\phi^{-1}(\bm\lambda) \cD_{\bm\nu} \cD_{\bm\mu} \bm\phi(\bm\lambda) (\bm\iota_{\bm\lda}F(\lambda))_- \nonumber \\
	&+ \bm\phi^{-1}(\bm\mu)\cD_{\bm\nu} \bm\phi(\bm\mu) \bm\phi^{-1}(\bm\mu) \cD_{\bm\lambda} \bm\phi(\bm\mu) (\bm\iota_{\bm\mu}F(\mu))_- - \bm\phi^{-1}(\bm\mu)\cD_{\bm\nu} \cD_{\bm\lambda} \bm\phi(\bm\mu)(\bm\iota_{\bm\mu}F(\mu))_- \Big) \nonumber
	\eea
	so that $\cD_{\bm\nu} \bm K(\bm\lambda,\bm\mu) + \circlearrowleft$ is equal to 
	\bea
	&\Tr\Big(\left[\bm\phi^{-1}(\lambda) \cD_{\bm\mu}\bm\phi(\bm\lambda) ,\bm\phi^{-1}(\bm\lambda) \cD_{\bm\nu} \bm\phi(\bm\lambda)\right] (\bm\iota_{\bm\lda}F(\lambda))_- \Big)+ \circlearrowleft\,.
	\eea
	After we apply the $\delta$ differential we get
	\bea
	&&\delta \cD_{\bm\nu} \bm K(\bm\lambda,\bm\mu) + \circlearrowleft=\Tr\Big(  \cD_{\bm\nu} \bm\phi(\bm\lambda)\bm\phi^{-1}(\bm\lambda) \cD_{\bm\mu} \bm Q(\bm\lambda) - \cD_{\bm\mu} \bm\phi(\bm\lambda)\bm\phi^{-1}(\bm\lambda) \cD_{\bm\nu} \bm Q(\bm\lambda)\Big)\delta \bm\phi(\lambda)\bm\phi^{-1}(\bm\lambda)\nonumber\\
	&&+\Tr\Big(\bm\phi^{-1}(\bm\lambda) \cD_{\bm\nu} \bm Q(\bm\lambda)\delta \cD_{\bm\mu} \bm\phi(\bm\lambda) - \bm\phi^{-1}(\bm\lambda) \cD_{\bm\mu} \bm Q(\bm\lambda) \delta \cD_{\bm\nu} \bm\phi(\bm\lambda) \Big) + \circlearrowleft\,.\nonumber
	\eea
	We now turn to the the potential term
	\be
	\bm U(\bm\lambda,\bm\mu) =\tfrac 12 \Tr_{12}\big( (\bm \iota_{\bm \lambda} \bm \iota_{\bm \mu} + \bm \iota_{\bm \mu} \bm \iota_{\bm \lambda})r_{12}(\lambda,\mu) \bm Q_1(\bm \lambda) \bm Q_2(\bm \mu)\big)\,.
	\ee
	We drop $\bm\lambda$ and $\bm\mu$ in $\bm \phi$ and $\bm Q$ for conciseness since they follow the spaces $1$ and $2$ consistently. Let us also denote $(\bm \iota_{\bm \lambda} \bm \iota_{\bm \mu} + \bm \iota_{\bm \mu} \bm \iota_{\bm \lambda})r_{12}(\lambda,\mu)$ by $\bm r_{12}$. We compute
	\be
	\cD_{\bm \nu} \bm U(\bm\lambda,\bm\mu) = \tfrac 12 \Tr_{12}\big( \bm r_{12} \left(\cD_{\bm\nu} \bm Q_1 \bm Q_2+ \bm Q_1  \cD_{\bm\nu} \bm Q_2\right)\big)
	\ee	
	and after applying the $\delta$-differential we get
	\be
	\delta \cD_{\bm \nu} \bm U(\bm\lambda,\bm\mu)= \tfrac 12 \Tr_{12}\big( \bm r_{12} \left(\delta \cD_{\bm\nu} \bm Q_1 \bm Q_2+ \cD_{\bm\nu} \bm Q_1 \delta\bm Q_2+\delta \bm Q_1  \cD_{\bm\nu} \bm Q_2+\bm Q_1 \delta \cD_{\bm\nu} \bm Q_2\right)\big)
	\ee
	and similarly for the cyclic permutations. 
	We use the following identities
	\begin{eqnarray*}
		\Tr_{12} \bm r_{12} \delta \cD_{\bm\nu} \bm Q_1\bm Q_2 &=& \Tr_{12}(-\bm Q_2 r_{12} \bm \cD_{\bm \nu} \bm  Q_1 - \bm  Q_1 \cD_{\bm \nu} \bm \phi_1 \bm \phi^{-1}_1 \bm  Q_2 r_{12} \nonumber\\
		&&+ \cD_{\bm \nu}\bm  \phi_1\bm  \phi^{-1}_1 \bm Q_2 \bm r_{12} \bm Q_1 + \bm \phi_1 \cD_{\bm \nu} X_1 \bm \phi^{-1}_1 \bm Q_2 r_{12})\delta \bm \phi_1 \bm \phi^{-1}_1\\
		&&+\Tr_{12} [\bm Q_1, \bm r_{12} \bm Q_2] \delta \cD_{\bm \nu}\bm \phi_1 \bm \phi^{-1}_1\nonumber\\
		&=& \Tr_{12}( [\cD_{\bm \nu}\bm  Q_1, \bm r_{12}\bm Q_2] - \cD_{\bm \nu} \bm \phi_1 \bm \phi^{-1}_1 [\bm Q_1, \bm r_{12}\bm Q_2])\delta \bm \phi_1 \bm \phi^{-1}_1\\
		&& +\Tr_{12} [\bm Q_1, \bm r_{12} \bm Q_2] \delta \cD_{\bm \nu}\bm \phi_1 \bm \phi^{-1}_1\,,
	\end{eqnarray*}
	\begin{eqnarray*}
		\Tr_{12} \bm r_{12}\bm Q_1 \delta \cD_{\bm \nu} \bm Q_2 &=& \Tr_{12}( - \bm r_{12} \bm Q_1 \cD_{\bm \nu} \bm Q_2 - \bm Q_2 \cD_{\bm \nu} \bm \phi_2 \bm \phi^{-1}_2 \bm r_{12} \bm Q_1 \nonumber\\
		&&+ \cD_{\bm \nu} \bm \phi_2 \bm \phi^{-1}_2 \bm r_{12} \bm Q_1 \bm Q_2 + \bm \phi_2 \cD_{\bm \nu} X_2 \bm \phi^{-1}_2 \bm r_{12} \bm Q_1) \delta \bm \phi_2 \bm \phi^{-1}_2\\
		&&+\Tr_{12} [\bm Q_2,\bm r_{12}\bm Q_1]\delta \cD_{\bm \nu} \bm \phi_2 \bm \phi^{-1}_2\nonumber\\
		&=&\Tr_{12} ([\cD_{\bm \nu} \bm Q_2, \bm r_{12}\bm Q_1] - \cD_{\bm \nu} \bm \phi_2 \bm \phi^{-1}_2[\bm Q_2,\bm r_{12} \bm Q_1]) \delta \bm \phi_2 \bm \phi^{-1}_2\\
		&&+\Tr_{12} [\bm Q_2,\bm r_{12}\bm Q_1]\delta \cD_{\bm \nu} \bm \phi_2 \bm \phi^{-1}_2\,,\nonumber
	\end{eqnarray*}
	and
	\begin{eqnarray*}	
		&&\Tr_{12} \bm r_{12} \delta \bm Q_1 \cD_{\bm \nu} \bm Q_2 = \Tr_{12} [\bm Q_1,\bm r_{12} \cD_{\bm \nu} \bm Q_2] \delta \bm \phi_1 \bm \phi^{-1}_1\,,\\
		&&\Tr_{12} \bm r_{12} \cD_{\bm \nu} \bm Q_1 \delta \bm Q_2 = \Tr_{12} [\bm Q_2,\bm r_{12} \cD_{\bm \nu} \bm Q_1] \delta \bm \phi_2\bm \phi^{-1}_2\,,
	\end{eqnarray*}
	to express $\delta \cD_{\bm \nu} \bm U(\bm\lambda,\bm\mu)$ on the basis of $\delta \bm\phi_1$, $\delta \cD_{\bm\mu} \bm\phi_1$ and $\delta \cD_{\bm\nu} \bm\phi_1$ (and similarly on the space $2$). Then, we collect the coefficients of $\delta \bm\phi_1$, $\delta \cD_{\bm\mu} \bm\phi_1$ and $\delta \cD_{\bm\nu} \bm\phi_1$ which provide the independent equations. From $\delta \cD_{\bm\nu} \bm K(\bm\lambda,\bm\mu) + \circlearrowleft$ we have
	\be
\Tr_1 ( - \cD_{\bm\mu} \bm\phi_1 \bm\phi^{-1}_1 \cD_{\bm\nu}\bm Q_1 + \cD_{\bm\nu}\bm \phi_1 \bm\phi^{-1}_1 \cD_{\bm\mu}\bm Q_1) \delta \bm\phi_1 \bm\phi^{-1}_1+ \Tr_1( - \cD_{\bm\mu} \delta \cD_{\bm\nu} \bm \phi_1 \bm\phi^{-1}_1 + \cD_{\bm\nu}\bm Q_1 \delta \cD_{\bm\mu} \bm\phi_1 \bm\phi^{-1}_1) + \circlearrowleft
	\ee
	and from $\delta \cD_{\bm \nu} \bm U(\bm\lambda,\bm\mu) + \circlearrowleft$, using the skew-symmetry of $r$, we obtain
	\bea
	&& \Tr_{12} [\bm Q_1,\bm r_{12} \bm Q_2] \delta \cD_{\bm \nu}\bm  \phi_1 \bm \phi^{-1}_1 - \Tr_{13}[\bm Q_1,\bm r_{13}\bm Q_3] \delta \cD_{\bm \mu} \bm \phi_1 \bm \phi^{-1}_1\nonumber\\
	&&+\Tr_{12}( [\cD_{\bm \nu} \bm Q_1, \bm r_{12}\bm Q_2] - \cD_{\bm \nu} \bm \phi_1 \bm \phi^{-1}_1 [\bm Q_1, \bm r_{12}\bm Q_2] + [\bm Q_1,\bm r_{12}\cD_{\bm \nu}\bm  Q_2])\delta \bm \phi_1 \bm \phi^{-1}_1\\
	&&+ \Tr_{13}(-[\cD_{\bm \mu}\bm  Q_1, \bm r_{13}\bm Q_3] + \cD_{\bm \mu} \bm \phi_1 \bm \phi^{-1}_1[\bm Q_1,\bm r_{13} \bm Q_3] -[\bm Q_1,\bm r_{13}\cD_{\bm \mu} \bm Q_3]) \delta \bm \phi_1 \bm \phi^{-1}_1+ \circlearrowleft\,.\nonumber
	\eea 
	The coefficients of $\delta \cD_{\bm\mu} \bm\phi_1$ and $\delta \cD_{\bm\nu} \bm\phi_1$ in $\delta \cD_{\bm \nu} \bm \Lag(\bm\lambda,\bm\mu) + \circlearrowleft=0$ give
	$$
	\cD_{\bm \mu} \bm Q_1 = \tfrac 12 [\Tr_2 \bm r_{12}\bm Q_2,\bm Q_1]\,, \quad \cD_{\bm \nu} \bm Q_1 = \tfrac 12 [\Tr_3 \bm r_{13}\bm Q_3,\bm Q_1]\,,
	$$
	\ie two equivalent copies of the same equation under the irrelevant change $2 \leftrightarrow 3$ and  $\mu \leftrightarrow \nu$. Explicitly, it reads
\be
		\label{Lax Q 1} 
\cD_{\bm \mu} \bm Q_1(\bm \lambda)= \tfrac 12 \big[ \Tr_2 \big( (\bm \iota_{\bm \lambda} \bm \iota_{\bm \mu} + \bm \iota_{\bm \mu} \bm \iota_{\bm \lambda}) r_{12}(\lambda,\mu) \bm Q_2(\bm \mu) \big),\bm Q_1(\bm \lambda) \big]\,,
\ee
which gives the desired result \eqref{Lax 1} upon recalling Lemma \ref{lem: iota reorder}.
		The coefficient of $\delta \bm\phi_1$ is just a consequence of this equation and of the commutativity of the flows: $[\cD_{\bm \mu}, \cD_{\bm \nu}]=0$. The coefficients of $\delta\bm \phi_2$, $\delta \bm\phi_3$ etc. contained in the cyclic permutations $\circlearrowleft$ give equivalent equations under the corresponding cyclic permutations of the spectral parameters and auxiliary spaces. 
	
\end{proof}	
\end{theorem}

\subsubsection{Generating closure relation}

\begin{theorem}
	\label{thm_closure}
	The generating closure relation
	\begin{equation} \label{closure in generating form}
	\cD_{\bm \nu} \bLag(\bm \lambda, \bm \mu) + \cD_{\bm \lambda} \bLag(\bm \mu, \bm \nu) + \cD_{\bm \mu} \bLag (\bm \nu, \bm\lambda) =0.
	\end{equation}
holds when \eqref{Lax 1} is satisfied. It is a consequence of the CYBE for $r$.
	
	\begin{proof}
		First consider the kinetic term \eqref{kinetic term}. We have
		\begin{align*}
		\cD_{\bm \nu} \bm K(\bm \lambda,\bm \mu) &= \Tr\big( \bm \phi(\bm \lambda)^{-1} \cD_{\bm \nu} \cD_{\bm \mu} \bm \phi(\bm \lambda) (\bm \iota_{\bm \lambda} F(\lambda))_- \big)\\
		&\qquad - \Tr \big(\bm \phi(\bm \mu)^{-1} \cD_{\bm \nu} \cD_{\bm \lambda} \bm \phi(\bm \mu) (\bm \iota_{\bm \mu} F(\mu))_- \big)\\
		&\qquad\quad - \Tr\big( \bm \phi(\bm \lambda)^{-1} \cD_{\bm \nu} \bm \phi(\bm \lambda) \bm \phi(\bm \lambda)^{-1} \cD_{\bm \mu} \bm \phi(\bm \lambda) (\bm \iota_{\bm \lambda} F(\lambda))_- \big)\\
		&\qquad\qquad + \Tr \big(\bm \phi(\bm \mu)^{-1} \cD_{\bm \nu} \bm \phi(\bm \mu) \bm \phi(\bm \mu)^{-1} \cD_{\bm \lambda} \bm \phi(\bm \mu) (\bm \iota_{\bm \mu} F(\mu))_- \big).
		\end{align*}
		It follows by adding the cyclic permutations of this expression in the variables $\lambda$, $\mu$ and $\nu$ that
		\begin{equation} \label{DK=0}
		\cD_{\bm \nu} \bm K(\bm \lambda, \bm \mu) + \cD_{\bm \lambda} \bm K(\bm \mu, \bm \nu) + \cD_{\bm \mu} \bm K(\bm \nu, \bm\lambda) =0.
		\end{equation}
		Consider now the potential term \eqref{potential term}. Using Theorem \ref{prop: eom Q} we find
		\begin{align*}
		\cD_{\bm \nu} \bm U(\bm \lambda,\bm \mu) &= \tfrac 12 \Tr_{12}\big( (\bm \iota_{\bm \lambda} \bm \iota_{\bm \mu} + \bm \iota_{\bm \mu} \bm \iota_{\bm \lambda})r_{12}(\lambda,\mu) \cD_{\bm \nu} \bm Q_1(\bm \lambda) \bm Q_2(\bm \mu)\big)\\
		&\qquad + \tfrac 12 \Tr_{12}\big( (\bm \iota_{\bm \lambda} \bm \iota_{\bm \mu} + \bm \iota_{\bm \mu} \bm \iota_{\bm \lambda})r_{12}(\lambda,\mu) \bm Q_1(\bm \lambda) \cD_{\bm \nu} \bm Q_2(\bm \mu)\big)\\
		&= \tfrac 12 \Tr_{123}\big( (\bm \iota_{\bm \lambda} \bm \iota_{\bm \mu} + \bm \iota_{\bm \mu} \bm \iota_{\bm \lambda})r_{12}(\lambda,\mu) \big[ \bm \iota_{\bm \lambda} \bm \iota_{\bm \nu} r_{13}(\lambda,\nu) \bm Q_3(\bm \nu),\bm Q_1(\bm \lambda) \big] \bm Q_2(\bm \mu)\big)\\
		&\quad + \tfrac 12 \Tr_{123}\big( (\bm \iota_{\bm \lambda} \bm \iota_{\bm \mu} + \bm \iota_{\bm \mu} \bm \iota_{\bm \lambda})r_{12}(\lambda,\mu) \bm Q_1(\bm \lambda) \big[ \bm \iota_{\bm \mu} \bm \iota_{\bm \nu} r_{23}(\mu,\nu) \bm Q_3(\bm \nu),\bm Q_2(\bm \mu) \big] \big).
		\end{align*}
		By using the cyclicity of the trace in space $1$ and $2$ in the first and second terms, respectively, we may write this as
		\begin{align*}
		\cD_{\bm \nu} \bm U(\bm \lambda,\bm \mu) &= - \tfrac 12 \Tr_{123}\big( \big[ (\bm \iota_{\bm \lambda} \bm \iota_{\bm \mu} + \bm \iota_{\bm \mu} \bm \iota_{\bm \lambda})r_{12}(\lambda,\mu) \bm Q_2(\bm \mu), \bm Q_1(\bm \lambda) \big] \bm \iota_{\bm \lambda} \bm \iota_{\bm \nu} r_{13}(\lambda,\nu) \bm Q_3(\bm \nu) \big)\\
		&\quad - \tfrac 12 \Tr_{123}\big( \big[ (\bm \iota_{\bm \lambda} \bm \iota_{\bm \mu} + \bm \iota_{\bm \mu} \bm \iota_{\bm \lambda})r_{12}(\lambda,\mu) \bm Q_1(\bm \lambda),\bm Q_2(\bm \mu) \big] \bm \iota_{\bm \mu} \bm \iota_{\bm \nu} r_{23}(\mu,\nu) \bm Q_3(\bm \nu) \big)\\
		&= - \Tr_{123}\big( \big[ \bm \iota_{\bm \lambda} \bm \iota_{\bm \mu}  r_{12}(\lambda,\mu) \bm Q_2(\bm \mu), \bm Q_1(\bm \lambda) \big] \bm \iota_{\bm \lambda} \bm \iota_{\bm \nu} r_{13}(\lambda,\nu) \bm Q_3(\bm \nu) \big)\\
		&\quad - \Tr_{123}\big( \big[ \bm \iota_{\bm \lambda} \bm \iota_{\bm \mu} r_{12}(\lambda,\mu) \bm Q_1(\bm \lambda),\bm Q_2(\bm \mu) \big] \bm \iota_{\bm \mu} \bm \iota_{\bm \nu} r_{23}(\mu,\nu) \bm Q_3(\bm \nu) \big)
		\end{align*}
		where in the second equality we used Lemma \ref{lem: iota reorder} in both terms.
		By using once again the cyclicity of the trace in space $1$ and $2$ in the first and second terms, respectively, we arrive at the expression
		\begin{subequations} \label{D potential}
			\begin{align}
			\cD_{\bm \nu} \bm U(\bm \lambda,\bm \mu)
			&= \Tr_{123}\big( \bm \iota_{\bm \lambda} \bm \iota_{\bm \mu} \bm \iota_{\bm \nu} \big( [r_{12}(\lambda,\mu), r_{13}(\lambda,\nu)] \notag\\
			&\qquad\qquad + [r_{12}(\lambda,\mu), r_{23}(\mu,\nu)] \big) \bm Q_1(\bm \lambda) \bm Q_2(\bm \mu) \bm Q_3(\bm \nu)\big).
			\end{align}
			Likewise, using the skew-symmetry of the $r$-matrix we find
			\begin{align*}
			\cD_{\bm \lambda} \bm U(\bm \mu, \bm \nu) 
			&= - \tfrac 12 \Tr_{123}\big( (\bm \iota_{\bm \mu} \bm \iota_{\bm \nu} + \bm \iota_{\bm \nu} \bm \iota_{\bm \mu})r_{23}(\mu, \nu) \big[ \bm \iota_{\bm \mu} \bm \iota_{\bm \lambda} r_{12}(\lambda, \mu) \bm Q_1(\bm \lambda),\bm Q_2(\bm \mu) \big] \bm Q_3(\bm \nu)\big)\\
			&\quad - \tfrac 12 \Tr_{123}\big( (\bm \iota_{\bm \mu} \bm \iota_{\bm \nu} + \bm \iota_{\bm \nu} \bm \iota_{\bm \mu})r_{23}(\mu,\nu) \bm Q_2(\bm \mu) \big[ \bm \iota_{\bm \nu} \bm \iota_{\bm \lambda} r_{13}(\lambda, \nu) \bm Q_1(\bm \lambda),\bm Q_3(\bm \nu) \big] \big).
			\end{align*}
			Then by following the same steps as above for $\cD_{\bm \nu} \bm U(\bm \lambda,\bm \mu)$ we deduce that
			\begin{align}
			\cD_{\bm \lambda} \bm U(\bm \mu, \bm \nu) 
			&= \Tr_{123}\big( \bm \iota_{\bm \lambda} \bm \iota_{\bm \mu} \bm \iota_{\bm \nu} \big( [r_{12}(\lambda,\mu), r_{23}(\mu,\nu)] \notag\\
			&\qquad\qquad + [r_{13}(\lambda,\nu), r_{23}(\mu,\nu)] \big) \bm Q_1(\bm \lambda) \bm Q_2(\bm \mu) \bm Q_3(\bm \nu)\big).
			\end{align}
			Similarly, we also find using the skew-symmetry of the $r$-matrix that
			\begin{align}
			\cD_{\bm \mu} \bm U(\bm \nu, \bm \lambda) &= \Tr_{123}\big( \bm \iota_{\bm \lambda} \bm \iota_{\bm \mu} \bm \iota_{\bm \nu} \big( [r_{13}(\lambda,\nu), r_{23}(\mu,\nu)] \notag\\
			&\qquad\qquad + [r_{12}(\lambda,\mu), r_{13}(\lambda,\nu)] \big) \bm Q_1(\bm \lambda) \bm Q_2(\bm \mu) \bm Q_3(\bm \nu)\big).
			\end{align}
		\end{subequations}
		It now follows from combining the three equations in \eqref{D potential} and using the classical Yang-Baxter equation for the skew-symmetry $r$-matrix that
		\begin{equation} \label{DU=0}
		\cD_{\bm \nu} \bm U(\bm \lambda, \bm \mu) + \cD_{\bm \lambda} \bm U(\bm \mu, \bm \nu) + \cD_{\bm \mu} \bm U(\bm \nu, \bm\lambda) = 0.
		\end{equation}
		The result now follows from \eqref{DK=0} and \eqref{DU=0} but together.
	\end{proof}
\end{theorem}

The rest of the paper is devoted to examples. To specify an example, the following ingredients need to be fixed:
\begin{itemize}
	\item[$(i)$] a skew-symmetric $r$-matrix as in Section \ref{sec: adeles} (rational or trigonometric in this work),
	\item[$(ii)$] an effective divisor $\mathcal D \coloneqq \sum_{a \in S} N_a a$, in particular with support given by a finite subset $S \subset \CP$ and with $N_a \in \ZZ_{\geq 1}$ for each $a \in S$, ($N_\infty\in \ZZ_{\geq 0}$ if $\infty \in S$),
	\item[$(ii)$] a Lie algebra $\g$ which for simplicity we take to be either $\gl_N$ or $\sl_N$,
	\item[$(iv)$] a $\g$-valued rational function $F(\lambda) \in R_\lambda(\g)$ with pole divisor $(F)_\infty = \mathcal D$, \emph{i.e.} with a pole of order $N_a \in \ZZ_{\geq 1}$ at each point $a \in S$, ($N_\infty\in \ZZ_{\geq 0}$ if $\infty \in S$).
\end{itemize}
Each section contains an example of a hierarchy for which the above formalism produces 
Lagrangian multiforms, Lax matrices and zero curvature equations. Some sections consist of known examples that we recover or cast in a new light, \emph{e.g.} AKNS and sine-Gordon. Other examples are new to the best of our knowledge and show the power of the formalism, \emph{e.g.} the trigonometric Zakharov-Mikhailov class of models or the examples where we couple different integrable field theories together.

\section{AKNS hierarchy} \label{sec: AKNS hierarchy}

We keep this section short as it is a matter of ``closing the loop'': we reproduce the motivating example of Section \ref{motivating_ex} which was dealt with in detail in \cite{CS3}) and the starting point of this whole project. The main objective is to illustrate how to use our machinery on the simplest and most well known example. We choose the rational $r$-matrix and we fix the required data as follows:
\be
\label{data_AKNS}
S=\{\infty\}\,,~~N_\infty=0\,,~~\g=\sl_2\,,~~F(\lambda)=-i\sigma_3\,.
\ee
The adjoint orbit description of Section \ref{sec: coadjoint} is implemented with 
\be
\label{form_phi}
\phi^\infty(\lambda_\infty)=\1+\sum_{n=1}^\infty \phi_n^\infty\lambda_{\infty}^n\,,
\ee
and gives 
\be
\label{form_Q}
Q^\infty(\lambda_\infty)=\sum_{n=0}^\infty Q_n^{\infty}\lambda_\infty^n\,,
\ee
with $Q_0^{\infty}=-i\sigma_3$ and $Q_1^{\infty}=i[\sigma_3,\phi_1^\infty]$, the familiar first two elements in the AKNS hierarchy. Since there is only one pole in this example, let us drop the subscripts and superscripts and simply write the fundamental objects in \eqref{form_phi} and \eqref{form_Q} as 
\be
\label{simpler}
\phi(\lambda)=\1+\sum_{n=1}^\infty \phi_n \lambda^{-n}\,,~~Q(\lambda)=\sum_{n=0}^\infty Q_n\lambda^{-n}\,.
\ee
Similarly, we will just write $t_n$ instead of $t_n^\infty$ for the times of the hierarchy.
The generating Lax equation \eqref{Lax 1} gives us,  
using the definitions \eqref{def_gen_deriv}, \eqref{V def} and \eqref{Vb expansion}, 
\be
\label{FNR}
\partial_{t_n}Q(\lambda)=\left[V_n(\lambda),Q(\lambda)\right]
\ee
where 
\be
V_n(\lambda)=\sum_{r=0}^n Q_r \lambda^{n-r}
\ee
are the Lax matrices of the hierarchy. Eqs \eqref{FNR} are the the central equations of \cite{FNR} where only Hamiltonian aspects of the theory were developed. The associated zero curvature equations read
\be
\label{ZC_NLS}
\partial_{t_k}V_n(\lda)-\partial_{t_n}V_k(\lda)+[V_n(\lda),V_k(\lda)]=0\,,~~n,k\ge 0\,,
\ee
and produce the equations of motion of the hierarchy. The famous (unreduced) NLS system corresponds to $n=1$ and $k=2$.  From our generating Lagrangian \eqref{Lagrangian mult}, we can of course reproduce the generating Lagrangian of \cite{CS3} and all the Lagrangians forming the Lagrangian multiform that gives these equations as its (multiform) EL equations. Since $S=\{\infty\}$ we only have $\Lag^{\infty,\infty}(\lambda_\infty, \mu_\infty)$ to consider. As above, let us simply denote it as $\Lag(\lambda, \mu)$. The coefficient $\Lag_{mn}$ of ${\lda^{-m-1}\mu^{-n-1}}$ in its expansion reads
\be
\Lag_{mn}=\sum_{i=1}^{m} \Tr\tilde \phi_{i} \partial_{t_{n}} \phi_{ m-i+1} X_0  -  \sum_{i=1}^{n} \Tr\tilde \phi_{i} \partial_{t_{m}} \phi_{ n-i+1} X_0 -U_{mn}
\ee
where we wrote $\displaystyle \phi^{-1}(\lambda)=\1+\sum_{n=1}^\infty \tilde\phi_n\lambda^{-n}$ for convenience and where $U_{mn}$ is given by
\be
U_{mn}=- \Tr \sum_{j=0}^{m} Q_{m+n+1-j} Q_j\,.
\ee
These are the coefficients of the AKNS Lagrangian multiform found in \cite{CS3} (up to an overall minus sign) to which we refer for more details. It was explained in \cite{CS3} that there exists a parametrization of $\phi(\lda)$ in terms of very nice coordinates $\displaystyle e(\lda)=\sum_{i=1}^\infty e_i\lda^{-i}$, $\displaystyle f(\lda)=\sum_{i=1}^\infty f_i\lda^{-i}$ as 
\be
\phi(\lda)=\frac{1}{\sqrt{2i}}\begin{pmatrix}
	\sqrt{2i-e(\lda)f(\lda)} & e(\lda)\\
	-f(\lda) & \sqrt{2i-e(\lda)f(\lda)}
\end{pmatrix}\,.
\ee
For the reader's convenience, let us give for instance
\be
\label{Lag_NLS}
\Lag_{12}=\frac{1}{2} (f_1 \partial_{t_2} e_{1} - e_1 \partial_{t_2} f_{1}) - \frac{1}{2}  \sum_{j=1}^{2}(f_j \partial_{t_1} e_{2-j+1} - e_j \partial_{t_1} f_{2-j+1})-2ie_2f_2-e_1^2f_1^2
\ee
and
\be
\Lag_{13}=\frac{1}{2} (f_1 \partial_{t_3} e_{1} - e_1 \partial_{t_3} f_{1}) - \frac{1}{2}  \sum_{j=1}^{3}(f_j \partial_{t_1} e_{3-j+1} - e_j \partial_{t_1} f_{3-j+1})-2i(e_2f_3+e_3f_2)-\frac{3}{2}e_1f_1(f_1e_2+f_2e_1)
\ee
Of course, one can check that the equations of motion for these Lagrangians give precisely the zero curvature equations \eqref{ZC_NLS} for $(k,n)=(1,2)$ and $(k,n)=(1,3)$ respectively. For instance, varying $\Lag_{12}$ with respect to $e_j$, $f_j$, $j=1,2$, we have
\bea
\label{NLS_syst1}
\partial_{t_1}e_1+2ie_2=0\,&,&~~\partial_{t_1}f_1-2if_2=0\,,\\
\label{NLS_syst2}\partial_{t_2}e_1-\partial_{t_1}e_2-2e_1^2f_1=0\,&,&~~\partial_{t_2}f_1-\partial_{t_1}f_2+2f_1^2e_1=0\,.
\eea
This is equivalent to \eqref{ZC_NLS} for $(k,n)=(1,2)$, upon recalling that 
$$\displaystyle Q_1=\begin{pmatrix}
0 & \sqrt{2i}e_1\\
\sqrt{2i}f_1 & 0
\end{pmatrix}\,,~~\displaystyle Q_2=\begin{pmatrix}
e_1f_1 & \sqrt{2i}e_2\\
\sqrt{2i}f_2 & -e_1f_1
\end{pmatrix}\,.$$

The top two equations can be used to eliminate $e_2,f_2$ in the bottom two equations. With $t_2=t$, $t_1=x$, $e_1=\frac{1}{\sqrt{2i}}q$, $f_1=\frac{1}{\sqrt{2i}}r$ we get 
\bea
i\partial_{t}q+\frac{1}{2}\partial_{x}^2q-q^2r=0\,,~~-i\partial_{t}r+\frac{1}{2}\partial_{x}^2r-r^2q=0\,,
\eea
and the reduction $r=\mp r^*$ yields the well-known (de)focusing NLS equation 
$$i\partial_{t}q+\frac{1}{2}\partial_{x}^2q\pm |q|^2q=0$$ 
for the complex field $q$. Similarly, $\Lag_{13}$ gives the complex modified KdV equation.

\section{Sine-Gordon hierarchy}\label{sec:sG}

For the example of the sine-Gordon equation
\be
\label{eq:sG}
u_{xy}+\sin u=0\,,
\ee
we choose the trigonometric $r$-matrix \eqref{trig kernel}. The required data is fixed as follows
\be
\label{choice_sG}
S=\{0,\infty\}\,,~~N_0=1=N_\infty\,,~~\g=\sl_2\,,~~F(\lambda)=\frac{i}{2}\left(\frac{1}{\lda}\sigma_++\sigma_--\sigma_+-\lda \sigma_-\right)\,,
\ee
and we work with the basis $\sigma_3$, $\sigma_+$, $\sigma_-$.
The adjoint orbit description of Section \ref{sec: coadjoint} is implemented with 
\begin{eqnarray}
\label{phi_sG}
&&\phi^0(\lambda)=\sum_{n=0}^\infty \phi_n^0\lambda^n\,,~~\phi_0^0=e^{i\frac{u}{4}\sigma_3}\,,\\
\label{phi_sG2}
&&\phi^\infty(\lambda_\infty)=\sum_{n=0}^\infty \phi_n^\infty\lambda_{\infty}^n\,,~~\phi_0^\infty=e^{-i\frac{u}{4}\sigma_3}\,.
\end{eqnarray}
The phase space coordinate $u$ will be the sine-Gordon field as will become clear soon. 
This gives, with $(\iota_{\lambda_0} F(\lda))^{\rm trig}_-=\frac{i}{2}\left(\frac{1}{\lda}\sigma_++\sigma_-\right)$ and $(\iota_{\lambda_\infty} F(\lda))^{\rm trig}_-=-\frac{i}{2}\left(\lda \sigma_-+\sigma_+\right)$,
\bea
\label{Q_sG1}
&&Q^0(\lambda_0)=\frac{i}{2}\phi^0(\lambda)\left(\frac{1}{\lda}\sigma_++\sigma_-\right) \phi^0(\lambda)^{-1}=\sum_{n=-1}^\infty
Q_n^{0}\lambda^n\,,\\
\label{Q_sG2} 
&&Q^\infty(\lambda_\infty)= -\frac{i}{2}\phi^\infty(\lambda_\infty)\left(\frac{1}{\lda_\infty} \sigma_-+\sigma_+\right) \phi^\infty(\lambda_\infty)^{-1}=\sum_{n=-1}^\infty Q_n^{\infty}\lambda_\infty^n\,,
\eea
with $Q_{-1}^0=\frac{i}{2}e^{\frac{iu}{2}}\sigma_+$ and $Q_{-1}^{\infty}=-\frac{i}{2}e^{\frac{iu}{2}}\sigma_-$. 
We now show how to use our formalism to recover the sine-Gordon equation (in light cone coordinates) as well as its first higher compatible flow which is nothing but the modified KdV equation, as presented in \cite{Su}. We take advantage of  this example to illustrate how our formalism also produces the Lagrangian multiform corresponding to these 3 times. In this context, our motivation is to show that 
the so-called ``alien derivatives'' problem that was discussed in \cite{V} does not appear with our approach. The problem only arises if one insists on using the variational equations we obtain to eliminate some of the phase space coordinates in favour of the sine-Gordon field $u$ and its derivatives with respect to a given time. In other words, we show in detail how our general discussion about the FNR procedure, when applied at the variational level, leads to this alien derivative problem. This is yet another reason in our opinion why it is preferable to work with the natural phase space coordinates that are provided by $\bm \phi(\bm \lambda) $.

It is convenient to parametrise \eqref{phi_sG}-\eqref{phi_sG2} as
\begin{eqnarray}
&&\phi^0(\lambda_0)=e^{i\frac{u}{4}\sigma_3}(\1+\psi^0(\lda))\,,~~\psi^0(\lda)=\sum_{n=1}^\infty \psi_n^0\lambda^n\,,\\
&&\phi^\infty(\lambda_\infty)=e^{-i\frac{u}{4}\sigma_3}(\1+\psi^\infty(\lda_\infty))\,,~~\psi^\infty(\lda_\infty)=\sum_{n=1}^\infty \psi_n^\infty\lambda_{\infty}^n\,,
\end{eqnarray}
where we recall that $\det \phi^0=1=\phi^\infty$ should hold. 
Using the gauge freedom of multiplying $\phi^0(\lambda_0)$ (resp. $\phi^\infty(\lambda_\infty)$) on the right by a matrix which commutes with 
$(\iota_{\lambda_0} F(\lda))^{\rm trig}_-$ (resp. $(\iota_{\lambda_\infty} F(\lda))^{\rm trig}_-$), we can work with 
\bea
\label{psi1}
&&\psi^0(\lda)=\sum_{n=1}^\infty \psi_n^0\lambda^n\,,~~
\psi_n^0=\begin{pmatrix}
	A_n^0 & 0\\
	C_n^0 & D_n^0
\end{pmatrix}\,,\\
\label{psi2}
&&\psi^\infty(\lda_\infty)=\sum_{n=1}^\infty \psi_n^\infty\lambda_{\infty}^n\,,~~\psi_n^\infty=\begin{pmatrix}
A_n^\infty & B_n^\infty\\
0 & D_n^\infty
\end{pmatrix}\,.
\eea
Note that one can show that there is a bijection between the group coordinates $A_n^0$, $C_n^0$, $D_n^0$ and $A_n^\infty$, $C_n^\infty$, $D_n^\infty$, and the algebra coordinates $a_n^0$, $b_n^0$ and $c_n^0$, which we would introduce via $Q_n^0=a_n^0\sigma_3+b_n^0\sigma_++c_n^0\sigma_-$ (and similarly at $\infty$). The reader familiar with the FNR construction or only interested in zero curvature equations would tend to use the algebra coordinates. However, since our Lagrangians are naturally expressed with group coordinates, we use the latter both for the zero curvature equations and the Lagrangians. It also facilitates comparison between the two ways of obtaining the equations of motion.

By our general results in Sections \ref{sec_gen_Lax} and \ref{sec:gen_ZC}, all the time flows commute and all the corresponding zero curvature equations of Proposition \ref{prop: ZC generating} hold, with the Lax matrices reading for $n\ge -1$, (see Proposition \ref{prop: Va explicit form})
\bea
&&V_n^0(\lambda)=-(P^- + \tfrac 12 P^0) Q^0_n -\frac{1}{\lda} Q^0_{n-1} -\cdots - \frac{1}{\lda^{n}} Q_{0}^0 - \frac{1}{\lda^{n+1}} Q_{-1}^0  \,,\\
&&V_n^\infty(\lambda)=(P^+ + \tfrac 12 P^0) Q^\infty_n +\lda Q^\infty_{n-1} +\cdots + \lda^n Q^\infty_0+\lda^{n+1} Q^\infty_{-1}\,.
\eea
The sine-Gordon equation is recovered by taking the pair of Lax matrices $(V^0_0(\lambda),V^\infty_0(\lambda))$ and the compatible higher flow attached to the pair $(V^\infty_0(\lambda),V^\infty_1(\lambda))$ gives the mKdV equation in potential form. The third possible Lax pair is $(V^0_0(\lambda),V^\infty_1(\lambda))$ and will be called the mixed equation. For convenience, let us label the corresponding times as follows $t^0_0=y$, $t^\infty_0=x$, $t^\infty_1=z$. 
Therefore, we focus on the following three zero curvature equations
\begin{enumerate}
	\item $\partial_{x} V_0^0(\lambda) -\partial_{y}V^\infty_0(\lambda)+\left[V_0^0(\lambda),V^\infty_0(\lambda)\right]=0$ (sG);

	\item $\partial_{z} V_0^\infty(\lambda) -\partial_{x}V^\infty_1(\lambda)+\left[V_0^\infty(\lambda),V^\infty_1(\lambda)\right]=0$ (mKdV);

	\item $\partial_{z} V_0^0(\lambda) -\partial_{y}V^\infty_1(\lambda)+\left[V_0^0(\lambda),V^\infty_1(\lambda)\right]=0$ (mixed).

\end{enumerate}
A direct calculation gives
\bea
&&Q_{-1}^0=\frac{i}{2}e^{\frac{iu}{2}}\sigma_+\,,~~Q_0^0=\frac{i}{2}\begin{pmatrix}
	-C_1^0	& 2 A_1^0e^{i\frac{u}{2}}\\
	e^{-i\frac{u}{2}} & C_1^0
\end{pmatrix}\,,\\
&&Q_1^0=\frac{i}{2}\begin{pmatrix}
	-C_2^0-A_1^0C_1^0	&  (2 A_2^0+(A_1^0)^2)e^{i\frac{u}{2}}\\
	(2 D_1^0-(C_1^0)^2)e^{-i\frac{u}{2}} & C_2^0+A_1^0C_1^0
\end{pmatrix}\,,\\
\label{form_Q0infty}
&&Q_{-1}^{\infty}=-\frac{i}{2}e^{\frac{iu}{2}}\sigma_-\,,~~Q_0^\infty=-\frac{i}{2}\begin{pmatrix}
	B_1^\infty	& e^{-i\frac{u}{2}}\\
	2D_1^\infty e^{i\frac{u}{2}} & -B_1^\infty
\end{pmatrix}\,,\\
&&Q_{1}^{\infty}=-\frac{i}{2}\begin{pmatrix}
	B_2^\infty+B_1^\infty D_1^\infty	& (2A_1^\infty-(B_1^\infty))^2e^{-i\frac{u}{2}}\\
	(2D_2^\infty+(D_1^\infty)^2)e^{i\frac{u}{2}} & -B_2^\infty-B_1^\infty D_1^\infty
\end{pmatrix}\,.
\eea
Hence,
\bea
&&V^0_0(\lambda)=-(P^- + \tfrac 12 P^0) Q^0_0 - \frac{1}{\lda}Q_{-1}^0=-\frac{i}{4} \begin{pmatrix}
	-C_1^0	& 2e^{i\frac{u}{2}}/\lda\\
	2e^{-i\frac{u}{2}} & C_1^0
\end{pmatrix}\,,\\
&&V^\infty_0(\lambda)=(P^+ + \tfrac 12 P^0) Q^\infty_0 + \lda Q_{-1}^\infty=-\frac{i}{4}\begin{pmatrix}
	B_1^\infty	& 2e^{-i\frac{u}{2}}\\
	2\lda e^{i\frac{u}{2}} & -B_1^\infty
\end{pmatrix}\,,\\
&&V^\infty_1(\lambda)=(P^+ + \tfrac 12 P^0) Q^\infty_1 + \lda Q_{0}^\infty+\lda^2 Q_{-1}^\infty\nonumber\\
&&\qquad\quad=-\frac{i}{4}\begin{pmatrix}
	2\lda B_1^\infty+B_2^\infty-A_1^\infty B_1^\infty	& 2(\lda+2A_1^\infty-B_1^\infty)e^{-i\frac{u}{2}}\\
	2\lda(\lda-2A_1^\infty) e^{i\frac{u}{2}} & -2\lda B_1^\infty-B_2^\infty+A_1^\infty B_1^\infty
\end{pmatrix}\,.
\eea
Therefore, we obtain the following equations of motion from the zero curvature equations:
\be
\label{ZC_sG}
\text{(sG)}~~\begin{cases}
	C_1^0=-u_y\,,\\
	B_1^\infty=-u_x\,,\\
	\partial_xC_1^0+\partial_yB_1^\infty-2\sin u=0\,\,.
\end{cases}
\ee
The first two equations show that the group coordinates $C_1^0$, $B_1^\infty$ can be thought of as auxiliary fields and can be eliminated from the dynamics to get \eqref{eq:sG}, as desired.
\be
\label{ZC_mKdV}
\text{(mKdV)}~~\begin{cases}
	B_1^\infty=-u_x\,,\\
	A_1^\infty=-\frac{i}{2}\partial_xB_1^\infty+\frac{1}{4}(B_1^\infty)^2\,,\\
	u_z-u_x(2A_1^\infty-(B_1^\infty)^2)-2i\partial_x(2A_1^\infty-(B_1^\infty)^2)-(B_1^\infty)^3+3 A_1^\infty B_1^\infty-B_2^\infty=0\,\\
	u_z+2u_xA_1^\infty-4i\partial_xA_1^\infty-A_1^\infty B_1^\infty-B_2^\infty=0\,,\\
	\partial_zB_1^\infty+\partial_x(A_1^\infty B_1^\infty-B_2^\infty)=0
\end{cases}
\ee
We see that both \eqref{ZC_sG} and \eqref{ZC_mKdV} contain the same equation for $B_1^\infty$ in terms of $u$, as it should be. 
A comment is in order. Under the first two equations, the third and fourth equation consistently give the same expression for $B_2^\infty$. In turn, replacing all the auxiliary fields into the last equation yields mKdV in potential form (\ie mKdV for $v=u_x$)
\be
\label{potential_mKdV}
u_{xz}+u_{xxxx}+\frac{3}{2}u_{xx}u_x^2=0\,.
\ee
\be
\label{ZC_mixed}
\text{(mixed)}~~\begin{cases}
	C_1^0=-u_y\,,\\
	A_1^\infty B_1^\infty-B_2^\infty=u_z\,,\\
		\partial_y B_1^\infty=\sin u\,,\\
	i\partial_y\left(e^{-iu/2}(2A_1^\infty-(B_1^\infty)^2)\right)+\frac{1}{2}C_1^0\left(2A_1^\infty-(B_1^\infty)^2 \right)e^{-iu/2}+B_1^\infty e^{iu/2}=0\,,\\
	2i\partial_y\left(A_1^\infty e^{iu}\right)+B_1^\infty e^{-iu/2}-A_1^\infty C_1^0e^{iu/2}=0\,,\\
	i\partial_zC_1^0-i\partial_y(A_1^\infty B_1^\infty-B_2^\infty)+2A_1^\infty (e^{iu}+e^{-iu})-(B_1^\infty)^2e^{-iu}=0\,.
\end{cases}
\ee
Using the first two equations to eliminate the auxiliary fields and noting that the fourth and fifth equations are equivalent (modulo the third equation), we obtain after simplification the following system of equations for the three fields $u$, $A_1^\infty$ and $B_1^\infty$,
\be
\label{mixed_bis}
\text{(mixed)}~~\begin{cases}
	\partial_y B_1^\infty=\sin u\,,\\
\partial_yA_1^\infty=\frac{i}{2}B_1^\infty e^{-iu}\,,\\
	-2iu_{yz}+2A_1^\infty (e^{iu}+e^{-iu})-(B_1^\infty)^2e^{-iu}=0\,.
\end{cases}
\ee
Note that this system of equations in $(y,z)$ can be perfectly studied on its own and is integrable. However, from our point of view, it should be included together with \eqref{ZC_sG} and \eqref{ZC_mKdV} into the sG hierarchy. This leads to interesting observations which are related to the Lagrangian multiform description we present below. First of all, using $B_1^\infty=-u_x$ and the sine-Gordon equation, we see that the first equation in \eqref{mixed_bis} is trivially satisfied. Similarly, the second equation in \eqref{mixed_bis} is a consequence of $B_1^\infty=-u_x$, the sine-Gordon equation and the second equation in \eqref{ZC_mKdV}. Perhaps more interesting is the fact that combining the first three equations in \eqref{ZC_mKdV} with the second equation in \eqref{ZC_mixed} yields
\be
\label{integrated_mKdV}
u_{z}+u_{xxx}+\frac{1}{2}u_x^3=0\,,
\ee
of which \eqref{potential_mKdV} is simply a differential consequence.

We now turn to the extraction of the coefficients of the Lagrangian multiform for the corresponding time flows. We need
$\Lag_{00}^{0,\infty}$ for (sG), $\Lag_{01}^{0,\infty}$ for (mixed) and $\Lag_{01}^{\infty,\infty}$ for (mKdV). 
We have
\begin{eqnarray*}
K^{0,\infty}(\lambda_0, \mu_\infty)&=&\Tr\left[\frac{i}{2}\left(\frac{1}{\lda}\sigma_++\sigma_-\right) \left( \phi^{(0)}_0 + \lambda \phi^{(0)}_1  +\lda^2\phi^{(0)}_2 +O(\lda^3)\right)^{-1}\right.\nonumber\\
&&\left. \times\left(\frac{1}{\mu_{\infty}} \partial_{t^\infty_{-1}}+\partial_{t^\infty_0} + \mu_{\infty}\partial_{t^\infty_1}+O(\mu_{\infty}^2)  \right)  \left( \phi^{(0)}_0 + \lambda \phi^{(0)}_1  + \lda^2 \phi^{(0)}_2\dots+O(\lda^3)\right) \right]\nonumber\\
&+&\Tr\left[\frac{i}{2}\left(\frac{1}{\mu_\infty}\sigma_-+\sigma_+\right) \left( \phi^{\infty}_0 + \mu_\infty \phi^{\infty}_1  +  \mu_\infty^2 \phi^{\infty}_2+O(\lda^3)\right)^{-1}\right.\nonumber\\
&&\left. \times\left(\frac{1}{\lda} \partial_{t^0_{-1}}+\partial_{t^0_0} + \lda\partial_{t^0_1}+O(\lda^2)  \right)  \left( \phi^{\infty}_0 + \mu_\infty \phi^{\infty}_1 + \mu_\infty^2 \phi^{\infty}_2\dots+O(\mu_\infty^3)\right) \right].
\end{eqnarray*}
Hence, dropping irrelevant total derivative terms and using again $t^0_0=y$, $t^\infty_0=x$, $t^\infty_1=z$ for convenience, we find
$$K_{00}^{0,\infty}=\frac{1}{4}C_1^0 u_x+\frac{1}{4}B_1^\infty u_y \,,~~K_{01}^{0,\infty}=\frac{1}{4}C_1^0u_z-\frac{1}{4}(A_1^\infty B_1^\infty-B_2^\infty)u_y-A_1^\infty\partial_{y}B_1^\infty+B_1^\infty\partial_{y}A_1^\infty\,.$$
To compute the potential terms, observe that for the trigonometric $r$-matrix, we have 
$$( \iota_{ \lambda_0}  \iota_{ \mu_\infty} + \iota_{ \mu_\infty} \iota_{ \lambda_0})r_{12}(\lambda,\mu)=-2P_{12}^--P_{12}^0+2P_{12}\sum_{n=0}^{\infty}\lda^n\mu_{\infty}^n\,.$$
Hence,
$$U_{00}^{0,\infty}=\Tr\left(Q_0^0(P^++\frac{1}{2}P^0)Q_0^\infty+Q_{-1}^0Q_{-1}^\infty\right)=\frac{1}{4}\left(e^{iu}+e^{-iu}-C_1^0B_1^\infty\right)\,,$$
\bea
U_{01}^{0,\infty}&=&\Tr\left(Q_0^0(P^++\frac{1}{2}P^0)Q_1^\infty+Q_{-1}^0Q_{0}^\infty\right)\nonumber\\
&=&\frac{1}{4}\left(2A_1^\infty-(B_1^\infty)^2\right)e^{-iu}+\frac{1}{4}C_1^0\left(A_1^\infty B_1^\infty-B_2^\infty\right)-\frac{1}{2}A_1^\infty e^{iu}\,.\nonumber
\eea
This gives us the desired Lagrangian densities for (sG) and (mixed) as
\be
\label{Lag_sG}
\Lag_{\text{sG}}\equiv \Lag_{00}^{0,\infty}= K_{00}^{0,\infty}-U_{00}^{0,\infty}\,,
\ee
and
\be
\Lag_{\text{mixed}}\equiv \Lag_{01}^{0,\infty}= K_{01}^{0,\infty} - U_{01}^{0,\infty}\,.
\ee
Similarly, we find
$$K_{01}^{\infty,\infty}=-\frac{1}{4}B_1^\infty u_z-\frac{1}{4}\left(A_1^\infty B_1^\infty-B_2^\infty\right)u_x-\frac{i}{2}A_1^\infty\partial_{x}B_1^\infty+\frac{i}{2}B_1^\infty\partial_{x}A_1^\infty$$
and, with
$$( \iota_{ \lambda_\infty}  \iota_{ \mu_\infty} + \iota_{ \mu_\infty} \iota_{ \lambda_\infty})r_{12}(\lambda,\mu)=-2P_{12}^--P_{12}^0+P_{12}\sum_{n=0}^{\infty}\frac{\mu_{\infty}^n}{\lda_\infty^n}-P_{12}\sum_{n=0}^{\infty}\frac{\lda_\infty^{n+1}}{\mu_{\infty}^{n+1}}\,,$$
we get
$$U_{01}^{\infty,\infty}=\Tr\left(Q_0^\infty(P^++\frac{1}{2}P^0)Q_1^\infty\right)=\frac{1}{4}B_1^\infty\left(A_1^\infty B_1^\infty-B_2^\infty\right)+\frac{1}{2}A_1^\infty\left(2A_1^\infty-(B_1^\infty)^2\right)  \,.$$
Thus, the Lagrangian density for (mKdV) is given by
\be
\Lag_{\text{mKdV}}\equiv \Lag_{01}^{\infty,\infty}=K_{01}^{\infty,\infty}-U_{01}^{\infty,\infty}\,.
\ee
It remains to derive the EL equations associated to each Lagrangian. For instance, by varying $B_1^\infty$, $C_1^0$ and $u$ in $\Lag_{\text{sG}}$ we find exactly the three equations in \eqref{ZC_sG}. Similarly, it can be checked that the E-L equations for $\Lag_{\text{mKdV}}$ and $\Lag_{\text{mixed}}$ reproduce \eqref{ZC_mKdV} and \eqref{ZC_mixed} respectively.

In particular, all the equations that determine the group coordinates in terms of $u$ and its (relevant) derivatives are reproduced variationally. This is an interesting feature that the FNR procedure is also obtained variationally with our construction. An important by-product is that the so-called problem of ``alien-derivatives'' is eliminated systematically. In the present context, the manifestation of this problem would be for instance that the Lagrangian $\Lag_{\text{mixed}}$ contains terms with derivatives of $u$ with respect to $x$, while this Lagrangian is supposed to produce equations of motion with respect to the variables $y$ and $z$ only. Clearly, our Lagrangians do not suffer from this problem since by construction, they always only involve the two times they are supposed to produce equations of motion for. The problem is an artefact of using some of the equations of motion to solve for some of the fields in terms of $u$ and its derivatives. In other words, it is an artefact of implementing the FNR procedure {\it a priori} to eliminate some of the group coordinates. If we do implement this procedure of elimination, we obtain Lagrangians which form a Lagrangian multiform equivalent to the one given originally in \cite{Su} and which suffers from this problem. Eliminating the auxiliary fields in favour of $u$ and its derivatives, we obtain
$$\Lag_{\text{sG}}=-\frac{1}{4}u_xu_y-\frac{1}{2}\cos u$$
which is a well-known Lagrangian for \eqref{eq:sG}, as well as
$$\Lag_{\text{mKdV}}=\frac{1}{4}u_xu_z+\frac{1}{16}u_x^4-\frac{1}{4}u_{xx}^2-\frac{i}{4}\partial_x\left(\frac{1}{6}u_x^3+iu_xu_{xx} \right)$$
and
$$\Lag_{\text{mixed}}=-\frac{1}{4}u_yu_z-\frac{1}{2}u_{xx}(u_{xy}+\sin u)+\frac{1}{4}u_x^2 \cos u-\frac{i}{4}\partial_y\left(\frac{1}{6}u_x^3+iu_xu_{xx}\right)\,.$$
Changing $x$ to $-x$, multiplying all our Lagrangian by $2$ and dropping the irrelevant total derivatives in $x$ and $y$, we recover exactly the three Lagrangian coefficients, eqs (31)-(33), in \cite{Su}. Our Lagrangian multiform expressed with the group coordinates (and restricted to the three times $x,y,z$) is thus equivalent to that in \cite{Su} but, as noted before, it does not suffer from the alien derivative problem.

The poles at $0$ and $\infty$ play a symmetric role in the construction so it would be natural to consider also the time $t_1^0$ and the associated Lax matrix $V^0_1(\lambda)$. This naturally leads to two additional zero curvature equations (denote $t^0_1=t$ and the other times as above) that can be combined with (sG)
\begin{enumerate}
	
	\item $\partial_{t} V_0^0(\lambda) -\partial_{y}V^0_1(\lambda)+\left[V_0^0(\lambda),V^0_1(\lambda)\right]=0$ (mKdV2):
		
	\item $\partial_{t} V_0^\infty(\lambda) -\partial_{x}V^0_1(\lambda)+\left[V_0^\infty(\lambda),V^0_1(\lambda)\right]=0$ (mixed 2):
		
\end{enumerate}
The first one is called (mKdV2) as it is another copy of the mKdV equation but in $(y,t)$ instead of $(x,t)$. It is a compatible flow 
with (sG) where we can think of the roles of $x$ and $y$ being swapped. Then, naturally (mixed 2) is the remaining compatible flow between the variables $x$ and $t$. All the expressions for the Lax matrices, the zero curvature equations and the corresponding Lagrangians are similar to the above ones with the appropriate changes and we omit them. To complete the picture related to the four times we have focussed on, it would remain to consider the zero curvature equation
$$\partial_{z} V_1^0(\lambda) -\partial_{t}V^\infty_1(\lambda)+\left[V_1^0(\lambda),V^\infty_1(\lambda)\right]=0\,.$$
The set of equations of motion is not particularly enlightening. When embedded in the hierarchy of the five zero curvature equations already discussed, this system is a consequence of them, as it should be. Our contruction gives us the means to derive the corresponding Lagrangian density $\Lag_{11}^{(0,\infty)}$ if required but again we omit its lengthy expression here.

\paragraph{FNR procedure for the sine-Gordon hierarchy.}

We have discussed the FNR procedure at the level of the EL equations above, using some of the equations to eliminate certain auxiliary coordinates/fields. Here, we discuss it at the level of the algebra coordinates using the Lax equation. This is more in line with the original work \cite{FNR} and with the explanation around Lemma \ref{lem: Q in terms of U} for which it provides an illustration in the sG case. We recall that our point of view is that the procedure is unnecessary. We show it in the sG case to make contact with a more traditional approach but also because to our knowledge, this is the first time that the FNR construction is obtained for a hierarchy other than AKNS. In the present sG case, it is based on the Lax equations \eqref{Lax_0}-\eqref{Lax_infty} below.

The generating Lax equation \eqref{Lax 1} gives the following equations, for $n\ge -1$,
\bea
\label{Lax_0}
\partial_{t_n^0} Q^0(\lambda)&=&\left[V^0_n(\lambda),Q^0(\lambda)\right]=\left[- \big( \lambda^{-n} Q^0(\lambda) \big)^{\rm trig}_-,Q^0(\lambda)\right]\,,~~\\
\label{Lax_infty}
\partial_{t_n^\infty} Q^\infty(\lambda_\infty)&=&\left[V^\infty_n(\lambda),Q^\infty(\lambda_\infty)\right]=\left[\big( \lambda_\infty^{-n} Q^\infty(\lambda_\infty) \big)^{\rm trig}_-,Q^\infty(\lambda_\infty)\right]\,.
\eea
We could use the Lax equations \eqref{Lax_0}-\eqref{Lax_infty} to derive the coefficients of $Q_n^0$ and $Q_n^\infty$ as differential polynomials in the coordinate $u$. Given the form of $F(\lda)$ here, we do not fall into the area of applicability of the argument given after Lemma \ref{lem: Q in terms of U}. Nevertheless, it is still possible to proceed. We illustrate this with \eqref{Lax_infty}, the other case being similar. 

Our choices \eqref{choice_sG} and \eqref{phi_sG} give $c_{-1}=-\tfrac i2 e^{iu/2}$, $a_{-1}=0=b_{-1}$. Then, consider \eqref{Lax_infty} for $n=0$ with $Q^\infty(\lda)=\begin{pmatrix}
a(\lda) & b(\lda)\\
c(\lda) & -a(\lda)
\end{pmatrix}$ (we drop the superscript for conciseness). Writing $t_0^\infty=x$ for convenience and projecting onto $\sigma_3$, $\sigma_+$ and $\sigma_-$, we obtain
\be
\label{syst1}
\begin{cases}
	\partial_{x}a(\lda)=b_0c(\lda)-\frac{1}{\lda}c_{-1}b(\lda)\,,\\
	\partial_{x}b(\lda)=a_0b(\lda)-2b_0a(\lda)\,,\\
	\partial_{x}c(\lda)=-a_0c(\lda)+\frac{2}{\lda}c_{-1}a(\lda)\,.
\end{cases}
\ee
Looking at the $\lda^j$ coefficient, this yields the following system
\be
\begin{cases}
	\partial_{x}a_j=b_0c_j-c_{-1}b_{j+1}\,,\\
	\partial_{x}b_j=a_0b_j-2b_0a_j\,,\\
	\partial_{x}c_j=-a_0c_j+2c_{-1}a_{j+1}\,,
\end{cases}
\ee
which we should use to determine the coefficients recursively. Suppose, we have determined $a_k$, $b_k$, $c_k$ for $k=1,\dots,n-1$ then the first equation gives us $b_{n}$ and hence the second equation yields $a_n$. However, we cannot deduce $c_n$ from the third equation since it would require the knowledge of $a_{n+1}$. It is possible to replace \eqref{syst1} by the following equivalent system
\be
\label{syst2}
\begin{cases}
	\partial_{x}a(\lda)=b_0c(\lda)-\frac{1}{\lda}c_{-1}b(\lda)\,,\\
	\partial_{x}b(\lda)=a_0b(\lda)-2b_0a(\lda)\,,\\
a^2(\lda)+b(\lda)c(\lda)=-\frac{\lda}{4}\,.
\end{cases}
\ee
To see this, note that \eqref{syst1} implies $\partial_{x}\Tr Q^\infty(\lda_\infty)^2=0$ so that 
$$\Tr Q^\infty(\lda_\infty)^2=cst=\Tr \big((\iota_{\lambda_\infty} F(\lda))^{\rm trig}_-\big)^2\,,$$ as it should by construction. Conversely, assume \eqref{syst2} holds. The third equation implies $2\partial_{x}a(\lda) a(\lda)+	\partial_{x}b(\lda)c(\lda)+b(\lda)\partial_{x}c(\lda)=0$. Using the first two equations to eliminate $\partial_{x}a(\lda)$ and $\partial_{x}b(\lda)$ yields $b(\lda)\left(	\partial_{x}c(\lda)+a_0c(\lda)-\frac{2}{\lda}c_{-1}a(\lda) \right)=0$, and the claim follows.
Now the advantage of system \eqref{syst2} is that the $j$-th term of the third equation gives the following relation:
\be
\sum_{i=0}^{j+1}\left(a_{i}a_{j-i}+b_{i}c_{j-i}\right)=-\frac{1}{4}\delta_{j,-1}\,.
\ee
Spelling it out, it can be seen that it can be used to determine $c_n$ from $a_k$, $b_k$, $c_k$, $k=1,\dots,n-1$ and $b_{n}$, $a_n$ obtained from the first two equations as explained before. Thus, \eqref{syst2} allow us to determine all $a_j$, $b_j$, $c_j$, $j\ge 0$ recursively.
We find the first few as
\begin{eqnarray}
\label{coeffs1}
&&a_0=\frac{i}{2}u_x\,,~~b_0=-\tfrac i2 e^{-iu/2}\,,~~c_0=\tfrac i2 e^{iu/2}\left(1+iu_{xx}+\tfrac 12u_x^2\right)\,,\\
\label{coeffs2}
&&a_1=-\tfrac i2 \left(u_x+u_{xxx}+\tfrac 12 u_x^3\right)\,,~~b_1=\tfrac i2 e^{-iu/2}\left(1-iu_{xx}+\tfrac 12u_x^2\right)\,,\\
\label{coeffs3}
&&c_1=-\tfrac i2 e^{iu/2}\left(\tfrac 12 u_x^2+\tfrac 38 u_x^4+iu_{xx}+u_xu_{xxx}-\tfrac 12 u_{xx}^2 +iu_{xxxx}+\tfrac{3i}{8} u_{xx}u_x^2\right)\,.
\end{eqnarray}
Now, for instance, the expression we find for $a_0$ is consistent with the fact that $a_0=-\tfrac i2 B_1^\infty$ from \eqref{form_Q0infty} and with the second equation in \eqref{ZC_sG}. This is what we mean when we say that the FNR procedure is automatically implemented with our Lagrangian approach. We reiterate that the advantage of not applying it is that the problem of alien derivatives disappears and that dependent variables are also treated on an equal footing, like the independent variables.

\section{Hierarchies of Zakharov-Mikhailov type } \label{ZM_examples}
In this section, we introduce a rather large class of models and their hierarchies by using the following data
\begin{gather}
\label{data1}
S = \{ a_1, \dots, a_{P}\} \subset \CC\,,~~ P >0\,,~~\g = \gl_N\,,\\
\label{data2}
F(\lambda) =- \sum_{i=1}^{P} \sum_{r=0}^{n_i} \frac{A_{ir}}{(\lambda-a_i)^{r+1}}\,.
\end{gather}
Each $A_{ir} \in \gl_N$ is a non-dynamical constant matrix and we have chosen to write the order $N_{a_i}$ of the pole $a_i$, $i =1,\dots, P$ as $N_{a_i}=  n_{i} +1$ for convenience. All the poles in $S$ are distinct. The $r$-matrix can be the rational or trigonometric one at this stage. 

The motivation behind such choices is that in the simplest setting (rational $r$-matrix and simple poles), our construction reproduces the Zakharov-Shabat Lax pair with simple poles whose equations of motion were cast in variational form in \cite{ZM1}. In fact, our construction automatically embeds this single Lax pair, its zero curvature equation and its Lagrangian into an integrable hierarchy. This point of view was first introduced in \cite{SNC} where the class of Zakharov-Mikhailov (ZM) models was cast into the formalism of Lagrangian multiforms. Allowing for higher order poles gives us the generalisation discussed in \cite[Chap. 20]{Dickey}. When we switch to the trigonometric $r$-matrix, we produce for the first time the trigonometric version of the large class of ZM models and their hierarchies. Finally, when specialising the construction via an appropriate reduction and choice of matrices $A_{ir}$, we obtain as a special case the class of models studied in \cite{ABW}. Their integrability is guaranteed by construction and they are naturally embedded in an integrable hierachy, a new feature for these models that were originally obtained as standalone models by a different method related to the $4d$ Chern-Simons construction (see conclusions for details and references). These examples are detailed in the next three subsections.

\subsection{Rational Zakharov-Mikhailov models}

We first describe in detail how to reproduce the class of Lax pairs and Lagrangians originally discussed in the pioneering paper \cite{ZM1}. The generalisation to higher order poles presented in \cite{Dickey} will be straightforward. The $r$-matrix is fixed to be the rational one in this subsection. We split the data \eqref{data1}-\eqref{data2} in the following way: $P=P_1+P_2$, $P_1,P_2 >0$, and
\begin{gather}
\label{data_ZM1}
S = \{ a_1, \dots, a_{P_1}, b_1, \dots, b_{P_2}\} \subset \CC\,,~~ \g = \gl_N\,,\\
\label{data_ZM2}
F(\lambda) =- \sum_{i=1}^{P_1} \sum_{r=0}^{n_i} \frac{A_{ir}}{(\lambda-a_i)^{r+1}} - \sum_{j=1}^{P_2} \sum_{r=0}^{m_j} \frac{B_{jr}}{(\lambda-b_j)^{r+1}}\,.  
\end{gather}
For notational convenience, we simply denoted $A_{j+P_1,r}=B_{jr}$ and $n_{j+P_1}=m_j$ for $j=1,\dots,P_2$.

\subsubsection{Case of simple poles}

Following \cite{ZM1}, let us consider a Lax pair of the form\footnote{In \cite{ZM1}, the authors include an additional term in $U$ and $V$ corresponding to a pole at $\infty$ but it can be gauged away.}
\begin{equation}
\label{ZM_Lax_pair}
U(\lambda) = \sum_{i=1}^{P_1}\frac{U_i}{\lambda - a_i},\qquad V(\lambda) = \sum_{j=1}^{P_2} \frac{V_j}{\lambda - b_j}\,.
\end{equation}
A prominent example of an integrable field theory that falls into this class is the Faddeev-Reshetikhin model \cite{FR} which was proposed as an ultralocal variant of the principal chiral model.
The main result of \cite{ZM1} is that the equations of motions encoded in the zero curvature equation $\partial_\eta U(\lambda)-\partial_\xi V(\lambda)  + [U(\lambda),V(\lambda)] =0$ associated to the auxiliary problem 
\be
\Psi_\xi=U\Psi\,,~~\Psi_\eta=V\Psi\,,
\ee
are variational and are obtained as the EL equations of the following Lagrangian density
\be
\label{ZM_Lagrangian}
\Lag_{ZM} = \Tr\left( \sum_{i=1}^{P_1} \phi_i^{-1}  \partial_{\eta}  \phi_i U^{(0)}_i - \sum_{j=1}^{P_2} \psi_j ^{-1}  \partial_{\xi} \psi_jV^{(0)}_j - \sum_{i=1}^{P_1} \sum_{j=1}^{P_2}\frac{\phi_i U^{(0)}_i\phi_i^{-1} \, \psi_jV^{(0)}_j\psi_j ^{-1}}{ a_i- b_j}\right)\,.
\ee
The key insight to obtain this result is to parametrise $U_i$ as $\varphi_i U^{(0)}_i \varphi_i^{-1}$ and $V_j$ as $\psi_j V^{(0)}_j \psi_j^{-1}$. The matrices $U^{(0)}_i$ and $V^{(0)}_j$ are constant and all the dynamical variables are contained in the fields $\varphi_i$ and $\psi_j$. 

We can reproduce \eqref{ZM_Lax_pair} by choosing $n_i=0$ and $m_j=0$ in our data \eqref{data_ZM2}. Since
$$
\left(\iota_{ \lambda_{a_i}}F(\lda)\right)^{\rm rat}_-=-\frac{A_{i0}}{\lda-a_i}\,,~~\left(\iota_{ \lambda_{b_j}}F(\lda)\right)^{\rm rat}_-=-\frac{B_{j0}}{\lda-b_j}\,,
$$
a direct calculation using Proposition \ref{prop: Va explicit form} gives
\be
\label{elem_Lax}
V_{-1}^{a_i}(\lda)=\frac{\phi_0^{a_i}A_{i0}(\phi_0^{a_i})^{-1}}{\lda-a_i}\,,~~V_{-1}^{b_j}(\lda)=\frac{\phi_0^{b_j}B_{j0}(\phi_0^{b_j})^{-1}}{\lda-b_j}\,.
\ee
Therefore, it remains to make the identifications $\phi^{a_i}_0 = \varphi_i$ and $A_{i0} = U^{(0)}_i$, and $\phi^{b_j}_0 = \psi_j$ and $B_{j0} = V^{(0)}_j$ and take linear combinations $\displaystyle\partial_\xi=\sum_{i=1}^{P_1}\partial_{t_{-1}^{a_i}}$, $\displaystyle\partial_{\eta}=\sum_{j=1}^{P_2}\partial_{t_{-1}^{b_j}}$ of the elementary time flows $\partial_{t_{-1}^{a_i}}$ and $\partial_{t_{-1}^{b_j}}$. The corresponding Lax matrices are simply the sum of the elementary Lax matrices \eqref{elem_Lax} which gives precisely \eqref{ZM_Lax_pair}.

To understand how to recover the Lagrangian \eqref{ZM_Lagrangian} with our method, note that the zero curvature equation associated to the elementary times $t_{-1}^{a_i}$ and $t_{-1}^{b_j}$ reads
\be
\label{elem_ZC}
\partial_{t_{-1}^{a_i}}V_{-1}^{b_j}(\lda)- \partial_{t_{-1}^{b_j}}V_{-1}^{a_i}(\lda)+\left[V_{-1}^{b_j}(\lda),V_{-1}^{a_i}(\lda) \right]=0\,.
\ee
Summing these elementary zero curvature equations over $i=1\,\dots,P_1$ and $j=1,\dots,P_2$ yields the desired $\partial_\eta U(\lambda)-\partial_\xi V(\lambda)  + [U(\lambda),V(\lambda)] =0$. Therefore, to find the Lagrangian $\Lag_{ZM}$ it suffices to sum the elementary Lagrangians $\Lag^{a_i,b_j}_{-1-1}$ (the coefficient of $\lambda_{a_i}^{-1}\mu_{b_j}^{-1}$ in $\Lag^{a_i,b_j}(\lambda_{a_i},\mu_{b_j})$ which yields the equations of motion in \eqref{elem_ZC}). A direct calculation gives 
\be
\label{ZM_Lag_elem}
\Lag^{a_i,b_j}_{-1-1}=-\Tr\left( (\phi_0^{a_i})^{-1}  \partial_{t_{-1}^{b_j}}  \phi_0^{a_i} A_{i0} - (\phi_0^{b_j})^{-1}  \partial_{t_{-1}^{a_i}}  \phi_0^{b_j} B_{j0} - \frac{\phi_0^{a_i} A_{i0}(\phi_0^{a_i})^{-1}\, \phi_0^{b_j} B_{j0}(\phi_0^{b_j})^{-1}}{ a_i- b_j}\right)
\ee
and the claim follows, \ie, with identifications made above, we derive $\Lag_{ZM}$(up to an irrelevant minus sign) as in \eqref{ZM_Lagrangian} by taking the double sum $\displaystyle \sum_{i=1}^{P_1}\sum_{j=1}^{P_2}\Lag^{a_i,b_j}_{-1-1}$.

It was shown for the first time in \cite{SNC} that the ZM Lagrangian can be incorporated into a Lagrangian multiform where each coefficient is a copy of the original ZM Lagrangian associated to the corresponding times. The explicit case of 3 times was considered. We now explain how to recover this multiform from our data. Instead of splitting the data \eqref{data1}-\eqref{data2} into two types of poles as in \eqref{data_ZM1}-\eqref{data_ZM2}, we split it into three types of poles by setting $P=P_1+P_2+P_3$ and restrict our attention to simple poles, \ie we set
\begin{gather}
\label{data_ZM_mult1}
S = \{ a_1, \dots, a_{P_1}, b_1, \dots, b_{P_2},c_1,\dots,c_{P_3}\} \subset \CC\,,~~ P_1,P_2,P_3 >0\,,~~\g = \gl_N\,,\\
\label{data_mult2}
F(\lambda) =- \sum_{i=1}^{P_1} \frac{A_{i}}{\lambda-a_i} - \sum_{j=1}^{P_2}  \frac{B_{j}}{\lambda-b_j}- \sum_{k=1}^{P_3}  \frac{C_{k}}{\lambda-c_k}\,.  
\end{gather}
As before, we take the linear combinations $\displaystyle\partial_\xi=\sum_{i=1}^{P_1}\partial_{t_{-1}^{a_i}}$, $\displaystyle\partial_{\eta}=\sum_{j=1}^{P_2}\partial_{t_{-1}^{b_j}}$ of the elementary time flows, as well as the new combinations 
$\displaystyle\partial_{\nu}=\sum_{k=1}^{P_3}\partial_{t_{-1}^{c_k}}$. The original ZM Lagrangian is now denoted by $\Lag_{\xi\eta}$ and is accompanied by two new copies 
\be
\Lag_{\eta\nu}=\sum_{j=1}^{P_2}\sum_{k=1}^{P_3}\Lag^{b_j,c_k}_{-1-1}\,,~~\Lag_{\nu\xi}=\sum_{k=1}^{P_3}\sum_{i=1}^{P_1}\Lag^{c_k,a_i}_{-1-1}\,.
\ee
The Lagrangian multiform in \cite[Section 2.4]{SNC} is precisely
\be
\Lag=\Lag_{\xi\eta}\,d\xi\wedge d\eta+\Lag_{\eta\nu} \, d\eta\wedge d\nu+ \Lag_{\nu\xi}\,d\nu\wedge d\xi\,.
\ee
The associated Lax matrices and zero curvature equations also reproduce those of \cite{SNC}.

\subsubsection{Case of higher poles}

The generalisation of the ZM result to Lax matrices with higher order poles of the form
\begin{equation}
\label{higher_Lax_pair}
U(\lambda) = \sum_{i=1}^{P_1} U_i(\lda), \qquad V(\lambda) = \sum_{j=1}^{P_2} V_j(\lda),
\end{equation}
where
\begin{equation}
U_i = \sum_{r=0}^{n_i} \frac{U_{ir}}{(\lambda - a_i)^{r+1}},\qquad V_j = \sum_{r=0}^{m_j} \frac{V_{jr}}{(\lambda - b_j)^{r+1}}
\end{equation}
was presented in \cite{Dickey}. We can reproduce it by simply allowing $n_i$ and $m_j$ in the data \eqref{data_ZM2} to be arbitrary positive integers and by following the same steps as for simple poles. In that case we find
\be
V^{a_i}_{-1}(\lambda) = - \sum_{r=0}^{n_i} \frac{Q^{a_i}_{ -r -1}}{(\lambda - a_i)^{r+1}} \equiv U_i(\lambda)\,,~~
V^{b_j}_{-1}(\lambda) =- \sum_{r=0}^{m_j} \frac{Q^{b_j}_{ -r -1}}{(\lambda - b_j)^{r+1}}\equiv  V_j(\lambda)
\ee
where the coefficients are identified as
\begin{align}
&Q^{a_i}_{-r-1} \eqqcolon-  U_{ir}\,,~~ i=1,\dots,P_1\,,~~ r=0,\dots,n_i\,,	 \\
& Q^{b_j}_{-r-1} \eqqcolon-  V_{jr} \,,~~ j=1,\dots,P_2\,,~~ r=0,\dots,m_j\,,
\end{align}
and calculated from the group coordinates using the following expansions
\begin{align}
& Q^{a_i}(\lambda_{a_i}) = -\phi^{a_i}(\lambda_{a_i}) \sum_{r=0}^{n_i} \frac{A_{ir}}{(\lambda - a_i)^{r+1}} \phi^{a_i}(\lambda_{a_i})^{-1}=\sum_{k=-n_i-1}^\infty Q^{a_i}_k (\lambda - {a_i})^k\,,\\
& Q^{b_j}(\lambda_{b_j}) = -\phi^{b_j}(\lambda_{b_j}) \sum_{r=0}^{m_j} \frac{B_{jr}}{(\lambda - b_j)^{r+1}} \phi^{b_j}(\lambda_{b_j})^{-1}= \sum_{k=-m_j-1}^\infty Q^{b_j}_k (\lambda - {b_j})^k\,.
\end{align}
As before, we simply assemble the elementary time flows into $\displaystyle\partial_\xi=\sum_{i=1}^{P_1}\partial_{t_{-1}^{a_i}}$ and $\displaystyle\partial_{\eta}=\sum_{j=1}^{P_2}\partial_{t_{-1}^{b_j}}$ which have the desired Lax pair \eqref{higher_Lax_pair}. This gives the corresponding equations of motion in zero curvature form $\partial_\eta U(\lambda)-\partial_\xi V(\lambda)  + [U(\lambda),V(\lambda)] =0$.
The Lagrangian producing these equations of motion is obtained by adding the elementary Lagrangians $\Lag^{a_i,b_j}_{-1-1}$. We give some details to show that we recover exactly \cite[Formula 20.2.12]{Dickey} (in the case of non coinciding poles which we consider here).

The kinetic part of $\Lag^{a_i,b_j}_{-1-1}$ reads
\begin{equation*}
\begin{split}
K^{a_i,b_j}_{-1-1}&=\lres_{\lambda = a_i} \lres_{\mu = b_j} \Tr ( - \phi^{a_i}(\lambda_{a_i}) ^{-1}  \cD_{\mu_{b_j}} \phi^{a_i}(\lambda_{a_i}) \sum_{r=0}^{n_i} \frac{A_{ir}}{(\lambda -a_i)^{r+1}} \\
&\qquad\qquad\qquad\qquad  +  \phi^{b_j}(\lambda_{b_j}) ^{-1}  \cD_{\lambda_{a_i}}\phi^{b_j}(\lambda_{b_j}) \sum_{r=0}^{m_j}\frac{B_{jr}}{(\lambda -b_j)^{r+1}}  )\\
&= \Tr (- \lres_{\lambda = a_i}  \phi^{a_i}(\lambda_{a_i}) ^{-1}  \partial_{t^{b_j}_{-1}} \phi^{a_i}(\lambda_{a_i})\sum_{r=0}^{n_i} \frac{A_{ir}}{(\lambda -a_i)^{r+1}}  \\
&\qquad\qquad\qquad  +  \lres_{\mu = b_j} \phi^{b_j}(\lambda_{b_j}) ^{-1}  \partial_{t^{a_i}_{-1}} \phi^{b_j}(\lambda_{b_j})\sum_{r=0}^{m_j}\frac{B_{jr}}{(\lambda -b_j)^{r+1}} )\\
&= \Tr (- \lres_{\lambda =a_j}  g_i^{-1} \partial_{t^{b_j}_{-1}}g_i A_i  + \lres_{\mu = b_j} h_j^{-1} \partial_{t^{a_i}_{-1}}h_j B_j)
\end{split}
\end{equation*}
where in the last equality, we introduced $g_i$ (resp. $h_j$) to denote the truncation of $\phi^{a_i}(\lambda_{a_i})$ (resp. $\phi^{b_j}(\lambda_{b_j})$) up to the order $n_i$ (resp. $m_j$), in order to help make the comparison with Dickey's formula. The equality holds since the truncation is possible under the residue. We also denoted $\displaystyle A_i \coloneqq \sum_{r=0}^{n_i} \frac{A_{ir}}{(\lambda -a_i)^{r+1}} $ and $\displaystyle B_j \coloneqq \sum_{r=0}^{m_j}\frac{B_{jr}}{(\lambda -b_j)^{r+1}}$ for conciseness. 

The potential term reads, noting that $\iota_{\lambda_{a_i}}\iota_{\mu_{b_j}} = \iota_{\mu_{b_j}}\iota_{\lambda_{a_i}}$ when $a_i\neq b_j$,
\begin{equation*}
\begin{split}
U^{a_i,b_j}_{-1-1}&= \Tr_{12} \left(\lres_{\lambda =a_i} \lres_{\mu = b_j}\iota_{\lambda_{a_i}}\iota_{\mu_{b_j}}\frac{P_{12}}{\mu - \lambda } (Q{a_i}( \lambda_{a_i}))_1 (Q{b_j} (\mu_{b_j}))_2\right)\\
& =- \lres_{\lambda =a_i} Q^{a_i}(\lambda_{a_i}) \left( Q^{b_j} (\lambda_{b_j}) \right)^{\rm rat}_- \\
& = - \lres_{\lambda =a_i} (Q^{a_i}(\lambda_{a_i}))^{\rm rat}_- ( Q^{b_j} (\lambda_{b_j}) )^{\rm rat}_-\\
&= - \lres_{\lambda =a_i} (g_i A_i g_i^{-1})^{\rm rat}_- ( h_j B_j h_j^{-1} )^{\rm rat}_-\,.
\end{split}
\end{equation*}
We obtain Dickey's Lagrangian, up to an overall sign and a relative sign due to a different convention in the zero-curvature equation, by taking the following sums
\begin{equation}
L_D = - \sum_{i=1}^{P_1}\sum_{j=1}^{P_2} \left(K^{a_ib_j}_{-1-1}-U^{a_ib_j}_{-1-1}\right)\,.
\end{equation}

\subsubsection{Interplay between hierarchies associated to simple and higher order poles}

Following Proposition \ref{prop: Va explicit form}, the Lax matrices read, for each $n \ge -n_i -1$ and $i=1,\dots,P_1$, and for each $m \ge -m_j -1$ and $j=1,\dots,P_2$:
\begin{align}
&V^{a_i}_n(\lambda) = - \left( \frac{Q^{a_i}(\lambda_{a_i})}{(\lambda - a_i)^{n+1}} \right)^{\rm rat}_-= - \sum_{r=0}^{n+n_i +1} \frac{Q^{a_i}_{n-r}}{(\lambda - a_i)^{r+1}},\\
&V^{b_j}_m(\lambda) = - \left( \frac{Q^{b_j}(\lambda_{b_j})}{(\lambda - b_j)^{m+1}} \right)^{\rm rat}_-= - \sum_{r=0}^{m+m_j +1} \frac{Q^{b_j}_{m-r}}{(\lambda - b_j)^{r+1}}.
\end{align}
At first glance, it is tempting to suggest that a Dickey hierarchy with certain fixed order $n_i$ and $m_j$ simply sits higher or lower in another Dickey hierarchy with different fixed $n_i$ and $m_j$. The situation is much more complicated in general. To illustrate what we mean and show that this is too naive, let us focus on the field content of a Lax matrix around a pole $a$ and compare the ZM case (where $a$ is a simple pole) with the Dickey case (where $a$ has order $n_1+1>1$). The corresponding Lax matrices are
\be
V^{ZM,a}_n(\lambda) = - \left( \frac{Q^{ZM,a}(\lambda_{a})}{(\lambda - a)^{n+1}} \right)^{\rm rat}_-= - \sum_{r=0}^{n +1} \frac{Q^{ZM,a}_{n-r}}{(\lambda - a)^{r+1}}\,,~~n\ge -1,
\ee
and 
\be
V^{D,a}_n(\lambda) = - \left( \frac{Q^{D,a}(\lambda_{a})}{(\lambda - a)^{n+1}} \right)^{\rm rat}_-= - \sum_{r=0}^{n+n_1 +1} \frac{Q^{D,a}_{n-r}}{(\lambda - a)^{r+1}}\,,~~n\ge -n_1-1\,.
\ee
In general, it is always the case that the Dickey hierarchy contains the ZM case as its lowest level. Indeed,
\be
V^{D,a}_{-n_1-1}(\lambda) =\left((\lda-a)^{n_1}\phi^a(\lda_a)\sum_{r=0}^{n_1} \frac{A^D_{r}}{(\lambda -a)^{r+1}}(\phi^a(\lda_a))^{-1}  \right)_-^{\rm rat}= \frac{\phi^{a}_0 A^D_{n_1} (\phi^{a}_0)^{-1}}{\lambda - a}
\ee
and it suffices to choose $A^D_{n_1}=A^{ZM}_{0}$ to see that this is equal to 
\be
V^{ZM,a}_{-1}(\lambda) =  \left(\phi^a(\lda_a)\frac{A^{ZM}_{0}}{\lambda -a}(\phi^a(\lda_a))^{-1}  \right)_-^{\rm rat}=\frac{\phi^{a}_0 A^{ZM}_{0} (\phi^{a}_0)^{-1}}{\lambda - a}.
\ee
However, the crucial point is that $Q^{ZM,a}(\lambda_{a})$ and $Q^{D,a}(\lambda_{a})$ are constructed as orbits around different elements in general so the phase space is different in general. This means that the previous identification only gives {\it some} of the fields of the Dickey case which happen to be identifiable with the full phase space for ZM. The ``converse'' is not true in general. The Dickey case can only be seen as a higher flow in the ZM hierarchy if we construct it around a special element of the form $\displaystyle
\sum_{r=0}^{n_1} \frac{A^D_{r}}{(\lambda -a)^{r+1}}$ with $A^D_{r}=0$ for $r= 1,\dots,n_1$ and $A^D_{n_1}=A^{ZM}_{0}$. In that case, we see that 
\be
V^{D,a}_{n}(\lambda) = V^{ZM,a}_{n+n_1}(\lambda) 
\ee
so that the two hierarchies simply correspond to shifting the starting point in the elementary times $t_j^a$. This discussion was local in the sense that we looked at a typical pole $a$. Of course, similar conclusions hold around the other poles. If one assembles them to obtain compound times, then the situation is similar but technically more complicated. The summary is that in general, the Dickey case is a genuine generalisation of the ZM case unless it is constructed as an orbit around a specific element dictated by the ZM element. Of course, this comparison extends to the corresponding Lagrangians since the building blocks are the same as for the Lax matrices.

\subsection{Trigonometric Zakharov-Mikhailov models}\label{sec:trig_models}

We can repeat the construction of the previous subsection but with the rational $r$-matrix replaced by the trigonometric one. To the best of our knowledge, this produces for the first time a new class of models which we call {\it trigonometric Zakharov-Mikhailov models}. 

For conciseness, we 
simply illustrate this on the simplest example of simple poles in the data \eqref{data_ZM2}. To derive the elementary Lax matrices, we need to use the trigonometric formula in Proposition \ref{prop: Va explicit form} which brings interesting differences compared to the rational case, already for the lowest times $t_{-1}^{a_i}$ and $t_{-1}^{b_j}$. With $Q_{-1}^{a_i}=\phi_0^{a_i}A_{i0}(\phi_0^{a_i})^{-1}$ and $Q_{-1}^{b_j}=\phi_0^{b_j}B_{j0}(\phi_0^{b_j})^{-1}$, the corresponding elementary Lax matrices read
\bea
\label{trig_elem_Lax}
&&V_{-1}^{a_i}(\lda)=-\frac{a_iQ_{-1}^{a_i}}{\lda-a_i}-\left(P^{-}+\frac{1}{2}P^0\right)Q_{-1}^{a_i}\,,\\
&&V_{-1}^{b_j}(\lda)=-\frac{b_jQ_{-1}^{b_j}}{\lda-b_j}-\left(P^{-}+\frac{1}{2}P^0\right)Q_{-1}^{b_j}\,.
\eea
It will be convenient to introduce the following notations, for $M\in\gl_N$:
\be
\left(P^{+}+\frac{1}{2}P^0\right)M=M^>\,,~~\left(P^{-}+\frac{1}{2}P^0\right)M=M^<\,.
\ee
In particular $M=M^>+M^<$.
We derive from our general formula the following elementary Lagrangian: 
\bea
\label{trig_Lag_elem}
&&\qquad\qquad\Lag^{a_i,b_j}_{-1-1}=\Tr\left((\phi_0^{a_i})^{-1}\partial_{t^{b_j}_{-1}}\phi_0^{a_i}A_{i0}-(\phi_0^{b_j})^{-1}\partial_{t^{a_i}_{-1}}\phi_0^{b_j}B_{j0}\right)\\
&&-\frac{b_j}{b_j-a_i}\Tr\left(\phi_0^{a_i}A_{i0}(\phi_0^{a_i})^{-1}\phi_0^{b_j}B_{j0}(\phi_0^{b_j})^{-1}\right)+\Tr\left[\phi_0^{a_i}A_{i0}(\phi_0^{a_i})^{-1}\left(\phi_0^{b_j}B_{j0}(\phi_0^{b_j})^{-1}\right)^<\right]\nonumber\,.
\eea
The last term represents the main difference with the rational case, see \eqref{ZM_Lag_elem}. 

We now show that the so-called anisotropic chiral model presented in Section 6 of \cite{FR} can be obtained as a particular case of our trigonometric ZM Lagrangians and ZS Lax matrices. We will refer to it as {\it anisotropic Faddeev-Reshetikhin model} to avoid the confusion with the ``anisotropic chiral model'' terminology used in \cite{FR} which would assume that we parametrise the currents differently from our coadjoint parametrization, see \eqref{param_currents}. 

We proceed in two steps. First, we specialise our data as follows: in \eqref{data_ZM1}, we take $P_1=P_2=1$ and write $a_1=a$ and $b_1=b$; in \eqref{data_ZM2}, we simply write
$$F(\lda)=\frac{A}{\lda-a}+\frac{B}{\lda-b}\,.$$
We also restrict $\g$ to $\sl_2$\footnote{Thus, it would be more accurate to say that we derive the $\sl_2$ anisotropic FR model, as opposed to the $su(2)$ version of \cite{FR}. This is not important for our considerations here.}.
Second, we apply the automorphism discussed in Appendix \ref{comparison} to make the connection with \cite{FR} easier. 
Let us denote for convenience $t_{-1}^{a}=\xi$, $t_{-1}^{b}=\eta$,
\be
\label{param_currents}
Q_{-1}^{a}=\phi_0^{a}A(\phi_0^{a})^{-1}\equiv J_0\,,~~Q_{-1}^{b}=\phi_0^{b}B(\phi_0^{b})^{-1}\equiv J_1\,,
\ee
and the Lax pair \eqref{trig_elem_Lax},
\bea
\label{LPU}&&V_{-1}^{a}(\lda)\equiv U(\lda)=-\frac{aJ_0}{\lda-a}-J_0^<\,,\\
\label{LPV}&&V_{-1}^{b}(\lda)\equiv V(\lda)=-\frac{bJ_1}{\lda-b}-J_1^<\,.
\eea
The Lagrangian \eqref{trig_Lag_elem} becomes
\be
\label{Lag_aFR}
\Lag_{\rm aFR}=
\Tr\left((\phi_0^{a})^{-1}\partial_{\eta}\phi_0^{a}A-(\phi_0^{b})^{-1}\partial_{\xi}\phi_0^{b}B-\frac{b}{b-a}J_0J_1+J_0J_1^<\right)\,.
\ee
Varying with respect to $\phi_0^{a}$ and $\phi_0^{b}$, the EL equations read\footnote{The property $\Tr\left(J_0J_1^<\right)=\Tr\left(J_0^>J_1\right)$ is useful in deriving the EL equations.}
\be
\label{eq:aFR}
\partial_\eta J_0=\left[-\frac{bJ_1}{a-b}-J_1^<,J_0\right]\,,~~\partial_\xi J_1=\left[-\frac{aJ_0}{b-a}-J_0^<,J_1\right]\,.
\ee
Projecting on the basis $J_{0,1}=J_{0,1}^+\sigma_++J_{0,1}^-\sigma_-+J_{0,1}^3\sigma_3$, we get
\bea
\label{EL_aFR}
\quad\quad
\begin{cases}
\partial_\eta J_0^+=\frac{2b}{a-b}J_1^+J_0^3-\frac{a+b}{a-b}J_1^3J_0^+\,,\\
\partial_\eta J_0^-=\frac{2a}{b-a}J_1^-J_0^3+\frac{a+b}{a-b}J_1^3J_0^-\,,\\
\partial_\eta J_0^3=\frac{b}{b-a}J_1^+J_0^-+\frac{a}{a-b}J_1^-J_0^+\,,
\end{cases}
\begin{cases}
	\partial_\xi J_1^+=\frac{2a}{b-a}J_0^+J_1^3+\frac{a+b}{a-b}J_0^3J_1^+\,,\\
	\partial_\xi J_1^-=\frac{2b}{a-b}J_0^-J_1^3-\frac{a+b}{a-b}J_0^3J_1^-\,,\\
	\partial_\xi J_1^3=\frac{a}{a-b}J_0^+J_1^-+\frac{b}{b-a}J_0^-J_1^+\,.
\end{cases}
\eea
The residue at infinity of the zero curvature equation for the Lax pair \eqref{LPU}-\eqref{LPV} yields the equation $-\partial_\eta J_0^<+ \partial_\xi J_1^<+\left[J_0^<,J_1^<\right]=0$ in addition to \eqref{eq:aFR}. However, when projecting, one can see that this is a consequence of the system \eqref{EL_aFR}.

To make the comparison with the equations for the fields $S_{1,2,3}$ and $T_{1,2,3}$ used in \cite{FR}, we use the automorphism mentioned above and express the final answer using the Pauli matrices $\sigma_{1,2,3}$. We also implement the changes $\lda\to e^{2\lda}$, $a\to e^{2a}$, $b\to e^{-2a}$ to go from rational to hyperbolic parametrisation. We find 
\bea
&&\qquad\qquad e^{\lda/2\sigma_3}U(e^{2\lda})e^{-\lda/2\sigma_3}=\nonumber\\
&&-\frac{1}{2}\left[  w_1(\lda-a)\frac{1}{2}\left(e^aJ_0^++e^{-a}J_0^-\right)\sigma_1+ w_2(\lda-a)\frac{i}{2}\left(e^aJ_0^+-e^{-a}J_0^-\right)\sigma_2+w_3(\lda-a)J_0^3 \sigma_3 \right]\,,\nonumber
\eea
and
\bea
&&\qquad\qquad e^{\lda/2\sigma_3}V(e^{2\lda})e^{-\lda/2\sigma_3}=\nonumber\\
&&-\frac{1}{2}\left[  w_1(\lda+a)\frac{1}{2}\left(e^{-a}J_1^++e^{a}J_1^-\right)\sigma_1+ w_2(\lda+a)\frac{i}{2}\left(e^{-a}J_1^+-e^{a}J_1^-\right)\sigma_2+w_3(\lda+a)J_1^3 \sigma_3\right]\,,\nonumber
\eea
where $w_1(\lda)=w_2(\lda)=\frac{1}{\sinh \lda}$, $w_3(\lda)=\coth \lda$. It remains to compare with the Lax operator (6.22) in \cite{FR} and remember that they work with $x$ and $t$ instead of the light-cone coordinates $\xi$ and $\eta$. This leads to the identifications
\be
\label{JST_rel}
\begin{cases}
	S_1=-\frac{1}{4}\left(e^aJ_0^++e^{-a}J_0^-\right)\,,\\
	S_2=-\frac{i}{4}\left(e^aJ_0^+-e^{-a}J_0^-\right)\,,\\
	S_3=-\frac{1}{2}J_0^3\,,
\end{cases}\qquad 
\begin{cases}
	T_1=-\frac{1}{4}\left(e^{-a}J_1^++e^{a}J_1^-\right)\,,\\
	T_2=-\frac{i}{4}\left(e^{-a}J_1^+-e^{a}J_1^-\right)\,,\\
	T_3=-\frac{1}{2}J_1^3\,.
\end{cases}
\ee
Using \eqref{JST_rel}, eqs \eqref{EL_aFR} become
\bea
&&\partial_{\eta}S_a=2i\sum_{b,c}\epsilon^{abc}w_b(2a)T_bS_c\,,\\
&&\partial_{\xi}T_a=-2i\sum_{b,c}\epsilon^{abc}w_b(2a)S_bT_c\,,
\eea
which are of the same form as (6.26)-(6.27) in \cite{FR} when moving from the light-cone coordinates $\xi,\eta$ to the coordinates $x,t$.

\subsection{Deformed Gross-Neveu models}\label{deformed_sigma}

Here, we show how to produce the Lax pair and Lagrangian for the deformed Gross-Neveu model discussed in \cite[Section 16.2]{ABW} (see also \cite{By} and references therein for the particular case of rank $M=1$) as a particular case of our construction. The deformation is controlled by the $r$-matrix in the potential term which appears naturally in our construction. In fact, more than just the single Lagrangian and its Lax pair, we can in principle generate all the elementary Lagrangians in the whole Lagrangian multiform and all the elementary Lax pairs for the hierarchy containing this model as its main representative. This explains the origin of the integrability of such a class of models observed in \cite{ABW,By} and is seen to be a particular case of our construction.

The idea is to apply a {\it reduction}, in the spirit of \cite{Mikh}, to a Zakharov-Mikhailov model. The $r$-matrix could in principle be any skew-symmetric solution of the CYBE as we have already mentioned. Of course, if we want to resort to our explicit formulas for Lagrangians or Lax matrices, then it will be either the rational or trigonometric one since we have given an explicit construction only in those cases. Nevertheless, we will write most results without specifying the $r$-matrix to emphasize this observation. 

Choose the data in \eqref{data_ZM1}-\eqref{data_ZM2} as follows
\begin{gather}
S = \{ a,a^*\}\,,~~a\notin \RR\,,~~\g = \gl_N\,,\\
F(\lambda) = \frac{A}{\lambda-a} -\frac{A^\dagger}{\lambda-a^*}\,.  
\end{gather}
In particular, we chose $N_a=1=N_{a^*}$.
As mentioned, we want to use the idea of reduction which we implement as a reality condition on the objects of the theory. Writing
\be
Q^a(\lambda_a)=\sum_{k=-1}^\infty Q_k^{a}(\lambda-a)^k
\ee
and 
\be
Q^{a^*}(\lambda_{a^*})=\sum_{k=-1}^\infty Q_k^{a^*}(\lambda-{a^*})^k
\ee
we require $Q_k^{{a^*}}=-\left(Q_k^{a}\right)^\dagger$ for all $k\ge -1$. 
Accordingly, at the group level, we require that when writing
\be
\varphi^{-1}_a(\lambda_a)=\sum_{k=0}^\infty \widetilde{\varphi}_k^{a}(\lambda-a)^k
\ee
and 
\be
\varphi_{a^*}(\lambda_{a^*})=\sum_{k=0}^\infty \varphi_k^{{a^*}}(\lambda-{a^*})^k
\ee
we must have $\widetilde{\varphi}_k^{a}=\left(\varphi_k^{{a^*}}\right)^\dagger$ for all $k\ge 0$. Then, for any skew-symmetric $r$-matrix which is well-defined at $\lda=a$ and $\mu=a^*$, a direct computation gives
\bea
\label{Lag_Bykov1}
\Lag^{a,a^*}_{-1-1}&=&\Tr\left( \left(\varphi_0^{{a^*}}\right)^\dagger \partial_{t_{-1}^{a^*}}\varphi_0^{{a}} A+ \left(\varphi_0^{{a}}\right)^\dagger \partial_{t_{-1}^{a}}\varphi_0^{{a^*}}A^\dagger \right) \nonumber\\
&&+\Tr_{12}\left( r_{12}(a,a^*)\left(\varphi_0^{{a}} A\left(\varphi_0^{{a^*}}\right)^\dagger\right)_1 \left(\varphi_0^{{a}} A\left(\varphi_0^{{a^*}}\right)^\dagger\right)_2^\dagger\right)\,.
\eea
It remains to choose $A$ as a rank $M$ matrix and parametrize it as $A=(uv)^\dagger$ where $u$ is a constant $N\times M$ matrix and $v$ is a constant $M\times N$ matrix ($M\le N$). Then, setting $U=\varphi_0^{{a^*}}u$, $V=v\left(\varphi_0^{{a}}\right)^\dagger$, $t_{-1}^a=\bar{z}$ and $t_{-1}^{a^*}=z$, we get
\be
\label{Lag_Bykov2}
\Lag^{a,a^*}_{-1-1}=\Tr\left( V\partial_{\bar{z}}U+U^\dagger\partial_z V^\dagger \right)+\Tr_{12}\left( r_{12}(a,a^*)(UV)_1\left(UV\right)_2^\dagger\right)\,.
\ee
This is the Lagrangian given in \cite{ABW} (without the covariant derivative), with the relation to their notation being $r_a(A)_1=\Tr_{2}(r_{12}(a,a^*)A_2)$ so that the potential term reads
$$\Tr_{12}\left( r_{12}(a,a^*)\left(UV\right)_1^\dagger(UV)_2\right)=\Tr\left( r_{a}(UV)\left(UV\right)^\dagger\right)\,.$$
The interpretation of the parameter appearing in the $r$-matrix ($a$ here, $s$ in \cite{ABW}) is clear in our context: it corresponds to the pole structure of the constant matrix in our data \eqref{data_ZM2}. 

The corresponding Lax pair is derived from \eqref{V def} and reads, with $K=UV$,
\be
V_{-1}^{a}(\lda)=\Tr_2 \big(  r_{12}(\lambda,a)K^\dagger_2\big)\,,~~V_{-1}^{a^*}(\lda)=-\Tr_2 \big(  r_{12}(\lambda,a^*)K_2\big)\,,
\ee
and coincides with the Lax connection (16.7) in \cite{ABW}. Hence, the zero curvature equation yields
\bea
&&\partial_z \Tr_2 \big( \res_\lambda^a r_{12}(\lambda,a)K^\dagger_2\big)=\left[\Tr_2 \big( \res_\lambda^a r_{12}(\lambda,a)K^\dagger_2\big) ,  \Tr_2 \big(  r_{12}(a,a^*)K_2\big)\right]\,,\nonumber\\
&&\partial_{\bar{z}} \Tr_2 \big( \res_\lambda^{a^*} r_{12}(\lambda,a^*)K_2\big)=\left[ \Tr_2 \big(  r_{12}(\lambda,a)K^\dagger_2\big) ,  \Tr_2 \big( \res_\lambda^{a^*} r_{12}(\lambda,a^*)K_2\big)  \right]\,,\nonumber
\eea
which reduces to\footnote{This is true for $r$-matrices whose singular part at $\lda=\mu$ is of the form $\frac{f(\mu)}{\mu-\lda}P_{12}$, which is the case for the rational and trigonometric matrices we work with here.}
\bea
&&\partial_z K^\dagger=\left[K^\dagger ,  \Tr_2 \big(  r_{12}(a,a^*)K_2\big)\right]\,,\nonumber\\
&&\partial_{\bar{z}} K_2=\left[ \Tr_2 \big(  r_{12}(\lambda,a)K^\dagger_2\big) ,  K  \right]\,.\nonumber
\eea
In our opinion, it is rather beautiful that our generating Lagrangian multiform produces this class of models which was originally obtained via a completely different method, related to $4d$ Chern-Simons theory (see the conclusion for details and references). Unlike the latter method which necessarily focuses on a single Lagrangian at a time, we can also obtain all the Lagrangians corresponding to the higher commuting flows of the hierarchy, if desired.

\section{Coupling integrable hierarchies together}

	To show the flexibility of the construction, we explain by way of two examples how we can couple integrable field theories together in a simple way. The reader familiar with integrable hierarchies will recognize the procedure of assembling elementary time flows and the corresponding Lax matrices into linear combinations. What we gain here is the possibility to derive the corresponding Lagrangian (multiform) systematically for the new model as well. The procedure is an analog in the ultralocal case of the construction presented in \cite{DMLV1} for a class of non ultralocal field theories. Unlike the latter, the coupling here is at the level of an entire hierarchy. We give an example in the rational class and one in the trigonometric class of models. In the rational class, we couple together the AKNS hierarchy with the hierarchy of the Faddeev-Reshetikhin model (the simplest instance of a ZM model). In the trigonometric class, we couple the sine-Gordon hierarchy as discussed in Section \ref {sec:sG} with the hierarchy of the anisotropic Faddeev-Reshetikhin model as presented in Section \ref{sec:trig_models}. In each case, for conciseness, we present all the details for the lowest levels of the hierarchy but it should be clear by now that one can extract higher levels (Lagrangians and Lax matrices) systematically if desired. 

\subsection{AKNS-FR hierarchy}

To couple models in the AKNS hierarchy with models in the simplest ZM hierarchy (with two poles), we assemble the corresponding data as
\be
S=\{a,-a,\infty\}\,,~~a\in \CC^\times\,,N_a=N_b=1\,,~~N_\infty=0\,,~~\g=\sl_2\,,
\ee
and we choose 
\be
F(\lambda)=-i\alpha\sigma_3+\frac{A}{\lambda-a}+\frac{B}{\lambda+a}\equiv \alpha F^{AKNS}(\lambda)+F^{FR}(\lda)\,,
\ee
where $A,B$ are constant $\sl_2$ matrices. The parameter $\alpha$ is the coupling between the two theories: $\alpha=0$ gives a pure FR theory while sending $\alpha$ to infinity produces a pure AKNS hierarchy. The effect of multiplying $F^{AKNS}(\lambda)=-i\sigma_3$ by $\alpha$ is to yield $Q^\infty(\lda_\infty)=\alpha Q(\lambda)$ where $Q(\lambda)$ is the AKNS series \eqref{simpler}. Hence the Lax matrix $V_n^\infty(\lda)$ is equal to the AKNS Lax matrix $V_n(\lda)$ multiplied by $\alpha$. With this in mind, we have for instance $V_1^\infty(\lda)=-i\alpha \lda \sigma_3+\alpha Q_1$.

For simplicity, we illustrate the coupling by looking at the two main models in each hierarchy( NLS in AKNS and FR in ZM), \ie by considering the Lax pair
\be
V_{-1}^a(\lda)+V_1^\infty(\lda)\equiv U(\lda)\,,~~
V_{-1}^b(\lda)+V_2^\infty(\lda)\equiv V(\lda)\,,
\ee
with associated times $\xi$ and $\eta$ respectively. This choice of Lax pair corresponds to assembling the four flows $t_1^\infty$, $t_2^\infty$ (AKNS) and $t_{-1}^a$, $t_{-1}^{-a}$ (FR) such that $\partial_\xi=\partial_{t^{a}_{-1}}+\partial_{t_1^\infty}$ and $\partial_{\eta}=\partial_{t^{-a}_{-1}}+\partial_{t^\infty_2}$. 

Denoting  $Q_{-1}^a=J_0$ and $Q_{-1}^{-a}=J_1$ and recalling the above comments on the effect of multiplying by $\alpha$, we have
\be
U(\lda)=\frac{J_0}{\lda-a}  -i\alpha \lda \sigma_3+\alpha Q_1 \equiv U_{FR}(\lda)+\alpha U_{NLS}(\lda)\,,
\ee
\be
V(\lda)= \frac{J_1}{\lda-b} -i\lda^2 \alpha\sigma_3+\lda \alpha Q_1+\alpha Q_2 \equiv V_{FR}(\lda)+\alpha V_{NLS}(\lda)\,.
\ee
The zero curvature equation $\partial_\eta U(\lambda)-\partial_\xi V(\lambda)  + [U(\lambda),V(\lambda)] =0$ yields the following four (matrix) equations by looking at the residue at $\lda=a$, $\lda=-a$, $\lda=\infty$ and at the constant term in the $1/\lda$ expansion  respectively,
\be
\label{NLS_FR1}
\partial_\eta J_0+\frac{1}{2a}\left[J_0,J_1\right]+\alpha\left[J_0,V_{NLS}(a)\right]=0\,,
\ee
\be
\label{NLS_FR2}
\partial_\xi J_1+\frac{1}{2a}\left[J_0,J_1\right]-\alpha\left[U_{NLS}(-a),J_1\right]=0\,,
\ee
\be
\label{NLS_FR3}
\alpha\partial_{\xi}Q_1+i\alpha^2[\sigma_3,Q_2]+i\alpha[J_0,\sigma_3]=0\,,
\ee
\be
\label{NLS_FR4}
\alpha\partial_{\eta}Q_1-\alpha\partial_{\xi} Q_2+\alpha^2[Q_1,Q_2]  
-ia\alpha[J_0,\sigma_3]-i\alpha[\sigma_3,J_1]+\alpha[J_0,Q_1]=0\,.
\ee
Setting $\alpha=0$, \eqref{NLS_FR1} and \eqref{NLS_FR2} gives the FR version of the principal chiral model \cite{ZM2,FR} which is usually written as
\be
\partial_{\eta} J_0+\partial_{\xi} J_1+\frac{1}{a}\left[J_0,J_1\right]=0\,,~~\partial_{\eta} J_0-\partial_{\xi} J_1=0\,.
\ee
In the limit $\alpha\to\infty$ (recall that $\partial_{\xi}$ scales like $\alpha \partial_{t_1}$ and $\partial_{\eta}$ scales like $\alpha \partial_{t_2}$, with $t_1$, $t_2$ the NLS times), we see that \eqref{NLS_FR3}-\eqref{NLS_FR4} yield the NLS system \eqref{NLS_syst1}-\eqref{NLS_syst2}
\be
	\partial_{t_1}Q_1+i[\sigma_3,Q_2]=0\,,~~
	\partial_{t_2}Q_1-\partial_{t_1} Q_2+[Q_1,Q_2]=0\,.
\ee

The Lagrangian of this coupled model is obtained by adding the NLS Lagrangian $\Lag_{12}^{\infty\infty}$ (which is $\Lag_{12}$ in \eqref{Lag_NLS} properly rescaled)
\be
\Lag_{12}^{\infty\infty}=\frac{\alpha}{2} (f_1 \partial_{t_2^\infty} e_{1} - e_1 \partial_{t_2^\infty} f_{1}) - \frac{\alpha}{2}  \sum_{j=1}^{2}(f_j \partial_{t_1^\infty} e_{2-j+1} - e_j \partial_{t_1^\infty} f_{2-j+1})-\alpha^2\left(2ie_2f_2+e_1^2f_1^2\right)\,,
\ee
the FR Lagrangian $\Lag_{-1-1}^{a,-a}$
\be
\Lag_{-1-1}^{a,-a}=\Tr\left[ (\phi_0^{a})^{-1} \partial_{t_{-1}^{-a}} \phi_0^{a}A- (\phi_0^{-a})^{-1} \partial_{t_{-1}^{a}} \phi_0^{-a}B-\frac{J_0J_1}{2a}\right]\,,
\ee
 and the following two mixed elementary Lagrangians (discarding some irrelevant total derivatives),
\be
\Lag_{-12}^{a,\infty}=\Tr\left[ (\phi_0^{a})^{-1} \partial_{t_{2}^{\infty}} \phi_0^{a}A\right]- \frac{\alpha}{2}  \sum_{j=1}^{2}(f_j \partial_{t_{-1}^{a}} e_{2-j+1} - e_j \partial_{t_{-1}^{a}} f_{2-j+1})-\alpha \Tr\left[J_0V_{NLS}(a)\right]\,,
\ee
\be
\Lag_{1-1}^{\infty,-a}=\frac{\alpha}{2} (f_1 \partial_{t_{-1}^{-a}} e_{1} - e_1 \partial_{t_{-1}^{-a}} f_{1})-\Tr\left[(\phi_0^{-a})^{-1} \partial_{t_{1}^{\infty}} \phi_0^{-a}B+\alpha J_1U_{NLS}(-a)\right]\,.
\ee
Summing we get our Lagrangian for the coupled model
\bea
&&\Lag_{\rm NLS-FR}=\Tr\left[(\phi_0^{a})^{-1} \partial_{\eta} \phi_0^{a}A-(\phi_0^{-a})^{-1} \partial_{\xi} \phi_0^{-a}B\right]\\
&&+\frac{\alpha}{2} (f_1 \partial_{\eta} e_{1} - e_1 \partial_{\eta} f_{1}) - \frac{\alpha}{2}  \sum_{j=1}^{2}(f_j \partial_{\xi} e_{2-j+1} - e_j \partial_{\xi} f_{2-j+1})\nonumber\\
&&-\alpha^2\left(2ie_2f_2+e_1^2f_1^2\right)-\Tr\left[ \frac{J_0J_1}{2a} +\alpha J_0 V_{NLS}(a)-\alpha J_1 U_{NLS}(-a) \right]\nonumber\,.
\eea
It can be checked directly that the variations with respect to $\phi_0^{a}$, $\phi_0^{-a}$, $e_{2},f_2$ and $e_1,f_1$ gives \eqref{NLS_FR1},  \eqref{NLS_FR2}, \eqref{NLS_FR3} and \eqref{NLS_FR4} respectively.

\subsection{sG-aFR hierarchy}

The same strategy can of course be applied in the trigonometric case and we illustrate this by assembling the data of the sine-Gordon (sG) hierarchy as in Section \ref{sec:sG} with that of the anisotropic Faddeev-Reshetikhin (aFR) model as in Section \ref{sec:trig_models}, in the following way
\be
S=\{0,a,b,\infty\}\,,~~a,b\in \CC^\times\,,N_0=N_a=N_b=N_\infty=1\,,~~\g=\sl_2\,,
\ee
and we choose 
\be
F(\lambda)=\frac{i\beta}{2}\left(\frac{1}{\lda}\sigma_++\sigma_--\sigma_+-\lda \sigma_-\right)+\frac{A}{\lambda-a}+\frac{B}{\lambda-b}\,,
\ee
where $A,B$ are constant $\sl_2$ matrices and it is understood that $b=1/a$. We keep $b$ instead of $1/a$ as it makes notations lighter but all calculations are done with $b=1/a$. The parameter $\beta$ is the coupling between the two theories: $\beta=0$ gives a pure aFR theory while sending $\beta$ to infinity produces a pure sG model. 

To illustrate the procedure on the easiest case, we choose the main representative of each hierarchy, \ie we consider the Lax pair (recall from Section \ref{sec:trig_models} that we set $J_0=Q_{-1}^a=\phi_0^{a}A(\phi_0^{a})^{-1}$ and $J_1=Q_{-1}^b=\phi_0^{b}B(\phi_0^{b})^{-1}$)
\be
U(\lda)=V_{-1}^a(\lda)+V_0^0(\lda)=\frac{-aJ_0^>-\lda J_0^<}{\lda-a}   -\frac{i\beta}{4} \begin{pmatrix}
	-C_1^0	& 2e^{i\frac{u}{2}}/\lda\\
	2e^{-i\frac{u}{2}} & C_1^0
\end{pmatrix}\equiv U_{\rm aFR}(\lda)+\beta U_{\rm sG}(\lda)\,,
\ee
\be
V(\lda)=V_{-1}^b(\lda)+V_0^\infty(\lda)= \frac{-b J_1^>-\lda J_1^<}{\lda-b}   -\frac{i\beta}{4}\begin{pmatrix}
	B_1^\infty	& 2e^{-i\frac{u}{2}}\\
	2\lda e^{i\frac{u}{2}} & -B_1^\infty
\end{pmatrix}\equiv V_{\rm aFR}(\lda)+\beta V_{\rm sG}(\lda)\,,
\ee
with associated times $\xi$ and $\eta$ respectively. This corresponds to assembling the two sG times $t_0^0$, $t_0^\infty$ with the two aFR times $t_{-1}^a$, $t_{-1}^b$ such that $\partial_\xi=\partial_{t^{a}_{-1}}+\partial_{t_0^0}$ and $\partial_{\eta}=\partial_{t^{b}_{-1}}+\partial_{t^\infty_0}$. The zero curvature equation $\partial_\eta U(\lambda)-\partial_\xi V(\lambda)  + [U(\lambda),V(\lambda)] =0$ yields the following four equations by looking at the residue at $\lda=0$, $\lda=\infty$, $\lda=a$, $\lda=b$ respectively,
\be
\label{sG_FR1}
u_\eta+\beta B_1^\infty=-2iJ_1^3\,,
\ee
\be
\label{sG_FR2}
u_\xi+\beta C_1^0=-2iJ_0^3\,,
\ee
\be
\label{sG_FR3}
\partial_\eta J_0=\left[V_{\rm aFR}(a),J_0\right]-\beta\left[J_0,V_{\rm sG}(a)\right]\,,
\ee
\be
\label{sG_FR4}
\partial_\xi J_1=\left[ U_{\rm aFR}(b),J_1\right]+\beta\left[U_{\rm sG}(b),J_1\right]\,.
\ee
Equations \eqref{sG_FR1}-\eqref{sG_FR2} should be compared with the first two equations in \eqref{ZC_sG} and \eqref{sG_FR3}-\eqref{sG_FR4} should be compared with \eqref{eq:aFR}. The last independent equation contained in the zero curvature can be obtained for instance by setting $\lambda=1$. It can be shown that only the component $\sigma_3$ gives an equation that is not a consequence of those already written. It takes the form
\bea
\label{sG_FR5}
&&\frac{i\beta}{4}\left(\partial_{\eta}C_1^0+\partial_{\xi}B_1^\infty\right)-\frac{i\beta^2}{2}\sin u\nonumber\\
&&+\frac{1}{2}\frac{a+1}{a-1}\left(\partial_{\eta}J_0^3+\partial_{\xi}J_1^3\right)+\frac{1}{(a-1)(b-1)}\left(aJ_0^+J_1^--bJ_0^-J_1^+\right)\nonumber\\
&&+\frac{i\beta}{2(a-1)}\left(J_0^-e^{-i\frac{u}{2}}-aJ_0^+e^{i\frac{u}{2}}-J_1^+e^{-i\frac{u}{2}}+aJ_1^-e^{i\frac{u}{2}}\right)=0\,.
\eea
We can use \eqref{sG_FR1}-\eqref{sG_FR4} to cast this equation in the following more suggestive form which shows the coupling between sG and the aFR currents
\be
\label{sG_FR6}
i\left(\partial_{\eta}J_0^3+\partial_{\xi}J_1^3\right)+u_{\eta\xi}+\beta^2\sin u+\frac{\beta}{2}\left((J_0^- -J_1^+)e^{-i\frac{u}{2}}+a(J_1^--J_0^+)e^{i\frac{u}{2}}\right)=0.
\ee

We can derive the Lagrangian producing \eqref{sG_FR1}-\eqref{sG_FR4} and \eqref{sG_FR6} by adding the sine-Gordon Lagrangian \eqref{Lag_sG} (with appropriate inclusion of $\beta$)
\be
\Lag_{\rm sG}=\frac{\beta}{4}C_1^0 \partial_{t^\infty_0}u+\frac{\beta}{4}B_1^\infty \partial_{t^0_0}u-\frac{\beta^2}{4}\left(e^{iu}+e^{-iu}-C_1^0B_1^\infty\right)\,,
\ee
the anisotropic FR Lagrangian \eqref{Lag_aFR}
\be
\Lag_{\rm aFR}=
\Tr\left((\phi_0^{a})^{-1}\partial_{t_{-1}^b}\phi_0^{a}A-(\phi_0^{b})^{-1}\partial_{t_{-1}^a}\phi_0^{b}B-\frac{b}{b-a}J_0J_1+J_0J_1^<\right)\,,
\ee
 and the following two mixed elementary Lagrangians
\be
\Lag_{-10}^{a,\infty}=\Tr\left((\phi_0^{a})^{-1}\partial_{t^\infty_0}\phi_0^{a}A+\frac{\beta}{4}C_1^0 \partial_{t_{-1}^a}u  -\beta J_0V_{\rm sG}(a)\right)\,,
\ee
\be
\Lag_{0-1}^{0,b}=\Tr\left(\frac{\beta}{4}C_1^0 \partial_{t_{-1}^b}u-(\phi_0^{b})^{-1}\partial_{t^0_0}\phi_0^{b}B  +\beta J_1U_{\rm sG}(b)\right)\,.
\ee
We obtain
\bea
&&\qquad\Lag_{\rm sG-aFR}=\frac{\beta}{4}C_1^0 \partial_{\eta}u+\frac{\beta}{4}B_1^\infty \partial_{\xi}u+\Tr\left((\phi_0^{a})^{-1}\partial_{\eta}\phi_0^{a}A-(\phi_0^{b})^{-1}\partial_{\xi}\phi_0^{b}B\right)\\
&&-\frac{\beta^2}{4}\left(e^{iu}+e^{-iu}-C_1^0B_1^\infty\right)-\Tr\left(\frac{b}{b-a}J_0J_1+J_0J_1^< + \beta J_0V_{\rm sG}(a) -\beta J_1U_{\rm sG}(b)\right)\nonumber\,.
\eea
The variation with respect to $C_1^0$, $B_1^\infty$, $\phi_0^{a}$, $\phi_0^{b}$, and $u$ gives \eqref{sG_FR1}, \eqref{sG_FR2}, \eqref{sG_FR3}, \eqref{sG_FR4}, and \eqref{sG_FR5} respectively.

\section{Discussion and conclusion}\label{conclusion}

By introducing a certain generating Lagrangian multiform, we were able to relate two important but so far separate aspects of integrable systems: the well established theory of the classical $r$-matrix and the comparatively much newer framework of Lagrangian multiforms. In doing so, we bring closer together the vast amount of results in the Hamiltonian approach to integrable systems and the Lagrangian approach in the form advocated in the seminal paper \cite{LN}. A rich byproduct of this effort is that the generating Lagrangian multiform and its accompanying generating Lax equation and zero curvature equation provide a systematic framework to construct integrable hierarchies of field theories, both in terms of Lagrangians and of Lax matrices. This was illustrated at length over many examples, both known and new. As already emphasised in the introduction, this versatility to accommodate a very large class of examples stems from the fact that we work in the ad\`elic framework. 

\medskip

The most immediate open question that comes to mind relates to the restrictions imposed on the classical $r$-matrix appearing in the generating Lagrangian multiform. Certain aspects of our construction appear to remain true under only the assumption that $r$ is a solution of the CYBE \eqref{CYBE}. In particular, the restriction to the rational or trigonometric case that we studied in detail only played a role in the explicit construction of the projectors associated to the decomposition of the Lie algebra of $\g$-valued ad\`eles. It is easy to imagine that one could use a more general skew-symmetric $r$-matrix provided similar technicalities can be dealt with. Specifically, given a solution $r$ of the CYBE, one would like to establish results along the following lines:

- Define a pair of linear operators on the Lie algebra of $\g$-valued ad\`eles $\bm\A_{\bm\lambda}(\g)$ as
\begin{equation} \label{pi pm def}
\pi_\pm : \bm\A_{\bm\lambda}(\g) \longrightarrow \bm\A_{\bm\lambda}(\g), \qquad
\bm X(\bm \lambda) \longmapsto \big( (\pi_\pm X)_a(\lambda_a) \big)_{a \in \CP}
\end{equation}
with formulas similar to e.g. \eqref{P+ P- expressions}.

- Show that the linear maps $\pi_\pm$ so defined are projection operators onto complementary subspaces of $\bm\A_{\bm\lambda}(\g)$, \ie $(\pi_\pm)^2 = \pi_\pm$ and $\pi_+ + \pi^k_- = \id$.

- Show that the images $\pi_\pm \bm\A_{\bm\lambda}(\g)$ of the projection operators $\pi_\pm$ are both Lie subalgebras of $\bm\A_{\bm\lambda}(\g)$ and are isotropic with respect to the bilinear form analogous to that defined in \eqref{bilinear form}.

If one could accomplish this then it would follow that one would  have a direct sum decomposition of $\bm\A_{\bm\lambda}(\g)$ into complementary Lagrangian Lie subalgebras
\begin{equation*}
\bm\A_{\bm\lambda}(\g) = \pi_+ \bm\A_{\bm\lambda}(\g) \dotplus \pi_- \bm\A_{\bm\lambda}(\g).
\end{equation*}
The corresponding $r$-matrix would be defined as $r \coloneqq \pi_+ - \pi_- \in \End \bm\A_{\bm\lambda}(\g)$ and would presumably have a kernel of the form $\big( (\iota_{\mu_b} \iota_{\lambda_a} + \iota_{\lambda_a} \iota_{\mu_b}) r_{12}(\lambda, \mu) \big)_{a, b \in \CP}$. We could then use this kernel into our generating Lagrangian multiform and construct integrable hierarchies by the same method as we have done. One candidate to see if such a programme can be realised is the elliptic $r$-matrix \cite{Sk,Belavin-elliptic}.

The other obvious restriction of the present work is the condition that $r$ be skew-symmetric. In fact, we wrote the CYBE \eqref{CYBE} in its non-skew-symmetric form on purpose. Once again, some of our results appear to hold without this assumption. This is the case for the commutativity of the vector fields \eqref{def_gen_vf} as can be seen from the proof of Proposition \ref{commuting_vf}. The extension of our construction to the non-skew-symmetric case, hence to non-ultralocal field theories, appears rather challenging as the current form of our generating Lagrangian multiform simply does not allow for such an extension. We are currently investigating this exciting issue which promises to have connection with the framework of classical affine Gaudin models, developed in \cite{Vicedo:2017cge, Delduc:2019bcl}, that provides a unifying formalism for constructing and studying a very broad class of non-ultralocal classical integrable field theories. A first step in that direction was achieved recently \cite{CDS} where it was shown how to incorporate the non-skew symmetric case naturally in the context of finite-dimensional integrable hierarchies.

It was shown in \cite{Vicedo:2019dej} that classical affine Gaudin models are closely related to $4d$ mixed topological-holomorphic Chern-Simons theory introduced and studied in \cite{Costello:2013zra, Costello:2013sla, Costello:2017dso, Costello:2018gyb, Costello:2019tri}, see also \cite{Delduc:2019whp, Benini:2020skc, Lacroix:2020flf}. In fact, $4d$ Chern-Simons theory also naturally provides a framework for constructing a very broad class of ultralocal integrable field theories (see also  \cite{Zotov:2010kb} for a description of ultralocal integrable field theories as affine Gaudin models). In this context, it was shown in \cite{CSV}, see also \cite{Fukushima:2020tqv}, that the rational Zakharov-Mikhailov models, one of the main classes of examples that we reproduced here, could be obtained from $4d$ Chern-Simons theory with certain line defects. However, the construction of \cite{CSV} is, by design, able to produce only the action of a single Zakharov-Mikhailov model, as opposed to its entire hierarchy, starting from that of $4d$ Chern-Simons theory. It seems natural to wonder if such a construction, and in fact the whole $4d$ Chern-Simons approach, could be adapted to our generating Lagrangian multiform framework in order to derive entire integrable hierarchies and not just single models from this point of view.

\medskip

In the simplest case of the AKNS hierarchy, the concept of Hamiltonian multiform, initially introduced in \cite{CS2}, was illustrated in \cite{CS3}. The main idea is that it is possible to apply a version of the covariant Legendre transformation to an entire Lagrangian multiform to obtain the Hamiltonian analog of a multiform. Each coefficient of the resulting Hamiltonian multiform can be seen as a covariant Hamiltonian for the field theory described by the associated Lagrangian coefficient in the Lagrangian multiform. Important accompanying objects are the symplectic multiform and the multitime Poisson bracket which generalise to an entire hierarchy the concepts of multisymplectic form and of covariant Poisson bracket respectively. The latter are essential ingredients of the framework generally called covariant Hamiltonian field theory, see e.g. \cite{Gi} and references therein for a very useful recent review of the many facets of this rich topic. We believe it is important to try and obtain the generating Hamiltonian multiform and related structures corresponding to our generating Lagrangian multiform. Indeed, historically, one of the driving motivations of the above mentioned covariant Hamiltonian approach to field theory has been to allow for a (canonical) quantization of field theories that removes from the start the breaking of covariance associated to the standard Hamiltonian approach. The idea of covariant Hamiltonian field theory is to use a Poisson bracket that does not suffer from the lack of covariance of the traditional Poisson bracket: a {\it covariant} Poisson bracket. The results of \cite{CS2,CS3} show that one can extend this idea to a whole integrable hierarchy and that the classical $r$-matrix plays a key role in this ``covariant'' context, see also \cite{CS1,CSV}. The hope is that this could allow one to use the nice features of integrability encoded in the passage from the classical $r$-matrix to the quantum $R$-matrix, to fully implement the idea of covariant canonical quantization for such field theories.

Finally, our work also opens the possibility for quantization using another route: combining Feynman's path integral ideas with a Lagrangian multiform, thus taking advantage again of integrability features now encoded in a Lagrangian object entering the path integral. This tantalising idea was first put forward and explored in \cite{KN} but is still very much in its infancy.

\section*{Acknowledgments}

V.C and M.S. would like to acknowledge the vibrant atmosphere of the Leeds research group on Lagrangian multiforms involving F. Nijhoff, J. Richardson, D. Sleigh and M. Vermeeren.

\section*{Data Availability}

Data sharing not applicable to this article as no datasets were generated or analysed in this work.

\appendix

\section{ Comparison of trigonometric $r$-matrices for sine-Gordon}\label{comparison}

For the reader's convenience, we make the connection between the trigonometric $r$-matrix we used in this paper and the perhaps more familiar one usually used for treating the sine-Gordon model. The former reads
\be
r_{12}(\lambda,\mu)=\sigma_+\otimes\sigma_--\sigma_-\otimes\sigma_++\frac{\mu+\lambda}{\mu-\lambda}P_{12}\,,
\ee
with
\be
P_{12}=\frac{1}{2}\left(\1\otimes\1+\sigma_1\otimes\sigma_1+\sigma_2\otimes\sigma_2+\sigma_3\otimes\sigma_3\right)\,,
\ee
while the latter, which can be found for instance in \cite[pp. 432-433]{FT}, reads
\be
\tilde{r}_{12}(\lambda,\mu)=\frac{\gamma}{2}\frac{\mu^2+\lambda^2}{\lambda^2-\mu^2}\left(\1\otimes\1-\sigma_3\otimes\sigma_3\right)-\frac{\gamma\mu\lambda}{\lambda^2-\mu^2}\left(\sigma_1\otimes\sigma_1+\sigma_2\otimes\sigma_2\right)\,,
\ee
where $\gamma$ is related to the coupling constant of the sine-Gordon model and is set to $1$ as it is not relevant here.
We relate the two matrices by showing that they both give rise to the same matrix in trigonometric form. 
Set $\lambda=e^{2i\alpha}$, $\mu=e^{2i\beta}$ and define
\be
M(\alpha)=e^{\frac{i\alpha}{2}\sigma_3}
\ee
with property $M(\alpha)\sigma_\pm M^{-1}(\alpha)=e^{\pm i\alpha}\sigma_\pm$. Then we have
\bea
&&M(\alpha)\otimes M(\beta) r_{12}(\alpha,\beta)M^{-1}(\alpha)\otimes M^{-1}(\beta)\nonumber\\
&&\quad\qquad=-\frac{1}{2i\sin(\alpha-\beta)}   \left[2\left(\sigma_+
\otimes\sigma_-+\sigma_-\otimes\sigma_+\right) + \cos(\alpha-\beta) \left(\sigma_3\otimes\sigma_3+\1\otimes\1\right)  \right] .
\eea
Now set instead $\lambda=e^{i\alpha}$, $\mu=e^{i\beta}$,
and recall the relation $\sigma_1\otimes\sigma_1+\sigma_2\otimes\sigma_2=2\left(\sigma_+\otimes\sigma_-+\sigma_-\otimes\sigma_+\right)$,
to get
\be
\tilde{r}_{12}(\alpha,\beta)=-\frac{1}{2i\sin(\alpha-\beta)}   \left[2\left(\sigma_+
\otimes\sigma_-+\sigma_-\otimes\sigma_+\right) + \cos(\alpha-\beta) \left(\sigma_3\otimes\sigma_3-\1\otimes\1\right)  \right] .
\ee
The term proportional to $\1\otimes\1$ is irrelevant as it plays no role in the Sklyanin bracket or in the CYBE. 

There is a deeper reason for this connection which has to do with the fact that the twist of a loop algebra by an inner automorphism is isomorphic to the loop algebra. In simple terms here, the version of sine-Gordon considered in \cite{FT} is built on the twisted loop algebra $\cL^\theta(\g)$ where $\theta$ is an automorphism of order $2$ of $\g$, defined by $\theta X=\sigma_3 X \sigma_3$ for all $X\in \g$, and extended to the algebra 
\be
\cL(\g)=
\bigoplus_{n\in\ZZ}\g\otimes \lambda^n
\ee
by setting $\theta(X\lambda^n)=(-1)^n\theta(X)\lambda^n$ for all $X\in\g$. With $\g=\n_-\oplus\h\oplus\n_+$, $\cL^\theta(\g)$ can be decomposed as
\be
\cL^\theta(\g)=
\left(\bigoplus_{n\in\ZZ}\n_-\otimes \lambda^{2n+1}\right)\oplus \left(\bigoplus_{n\in\ZZ}\h\otimes \lambda^{2n}\right)\oplus\left(\bigoplus_{n\in\ZZ}\n_+\otimes \lambda^{2n+1}\right).
\ee
Now we apply the map
\be
\sigma_+\lambda^{2n+1}\mapsto \sigma_+\lambda^{2n}\,,~~\sigma_-\lambda^{2n+1}\mapsto \sigma_-\lambda^{2n+2}\,,~~\sigma_3\lambda^{2n}\mapsto \sigma_3\lambda^{2n}
\ee
which amounts to the transformation $X(e^{2i\alpha})\mapsto M(-\alpha)X(e^{2i\alpha})M^{-1}(-\alpha)$ to obtain that $\cL^\theta(\g)$ is isomorphic to 
\be
\left(\bigoplus_{n\in\ZZ}\n_-\otimes \lambda^{2n+2}\right)\oplus\left(\bigoplus_{n\in\ZZ}\h\otimes \lambda^{2n}\right)\oplus\left(\bigoplus_{n\in\ZZ}\n_+\otimes \lambda^{2n}\right)\,.
\ee
We now apply a second map
\be
\sigma_i\lambda^{2n}\mapsto \sigma_i\lambda^{n}\,,~~i=3,\pm\,,
\ee
to obtain that $\cL^\theta(\g)$ is isomorphic to 
\be
\left(\bigoplus_{n\in\ZZ}\n_-\otimes \lambda^{n}\right)\oplus\left(\bigoplus_{n\in\ZZ}\h\otimes \lambda^{n}\right)\oplus\left(\bigoplus_{n\in\ZZ}\n_+\otimes \lambda^{n}\right)=\cL(\g)\,.
\ee

\end{document}